\documentclass[aps,prd,superscriptaddress,twocolumn,amsmath,amssymb,floatfix,nofootinbib,10pt]{revtex4-2}

\usepackage{physics}
\usepackage{bm}% bold math

\usepackage{graphicx}   % Include figure files
\usepackage[caption=false]{subfig}
\usepackage[colorlinks=true,citecolor=royalBlue,linkcolor=bordeauxRed,urlcolor=oliveGreen]{hyperref}   % references with hyperlinks
\usepackage{cleveref}   % references that recognize the object type (must be imported after hyperref for compatibility)
\usepackage{dcolumn}% Align table columns on decimal point

\usepackage[english]{babel} % language
\usepackage[T1]{fontenc} % > and < fontsetting 
\usepackage[utf8]{inputenc}% use inputenc when using pdflatex
\usepackage{dsfont} % use mathds
\usepackage[mathlines]{lineno}% Enable numbering of text and display math
\usepackage{braket}
\usepackage{nicefrac} % compact symbols for 1/2, etc.

\usepackage[export]{adjustbox} % adjust relative position of subfloats
\usepackage{xcolor, etoolbox}

\usepackage[normalem]{ulem} % For crossing out 

\newcommand{\amax}{a_{\text{max}}}
\newcommand{\amin}{a_{\text{min}}}

\newcommand{\im}{\text{i}}
\newcommand{\e}{\text{e}}
\newcommand{\Vr}{V_\text{r}}
\newcommand{\Vs}{V_\text{s}}
\newcommand{\af}{a_\text{f}}
\newcommand{\ai}{a_\text{i}}
\newcommand{\atwo}{a_{\text{II}}}
\newcommand{\ti}{t_{\text{i}}}
\newcommand{\tf}{t_{\text{f}}}
\newcommand{\etai}{\eta_{\text{i}}}
\newcommand{\etaf}{\eta_{\text{f}}}
\newcommand{\cs}{c_{\text{s}}}
\newcommand{\Hi}{\mathcal{H}_\text{i}}
\newcommand{\Hf}{\mathcal{H}_\text{f}}

\newcommand{\as}{a_{\text{s}}}

\newcolumntype{P}[1]{>{\centering\arraybackslash}p{#1}}
\newcolumntype{M}[1]{>{\centering\arraybackslash}m{#1}}

\definecolor{marius}{rgb}{0.7,0.65,0.0}

\newcommand\varpm{\mathbin{\vcenter{\hbox{%
  \oalign{\hfil$\scriptstyle+$\hfil\cr
          \noalign{\kern-.3ex}
          $\scriptscriptstyle({-})$\cr}%
}}}}
\newcommand\varmp{\mathbin{\vcenter{\hbox{%
  \oalign{$\scriptstyle({+})$\cr
          \noalign{\kern-.3ex}
          \hfil$\scriptscriptstyle-$\hfil\cr}%
}}}}

\makeatletter
\def\@ssect@ltx#1#2#3#4#5#6[#7]#8{%
  \def\H@svsec{\phantomsection}%
  \@tempskipa #5\relax
  \@ifdim{\@tempskipa>\z@}{%
    \begingroup
      \interlinepenalty \@M
      #6{%
       \@ifundefined{@hangfroms@#1}{\@hang@froms}{\csname @hangfroms@#1\endcsname}%
       {\hskip#3\relax\H@svsec}{#8}%
      }%
      \@@par
    \endgroup
    \@ifundefined{#1smark}{\@gobble}{\csname #1smark\endcsname}{#7}%
  }{%
    \def\@svsechd{%
      #6{%
       \@ifundefined{@runin@tos@#1}{\@runin@tos}{\csname @runin@tos@#1\endcsname}%
       {\hskip#3\relax\H@svsec}{#8}%
      }%
      \@ifundefined{#1smark}{\@gobble}{\csname #1smark\endcsname}{#7}%
      \addcontentsline{toc}{#1}{\protect\numberline{}#8}%
    }%
  }%
  \@xsect{#5}%
}%
\makeatother

\begin{document}
\title{Cosmological particle production in a quantum field simulator as a quantum mechanical scattering problem}

\date{\today}

\author{Christian F. Schmidt}
\email{scatteringanalogy@matterwave.de}
\affiliation{Theoretisch-Physikalisches Institut, Friedrich-Schiller-Universität Jena, Max-Wien-Platz 1, 07743 Jena, Germany}

\author{Álvaro Parra-López}
\email{scatteringanalogy@matterwave.de}
\affiliation{Departamento de Física Teórica and IPARCOS, Facultad de Ciencias Físicas, Universidad Complutense de Madrid, Ciudad Universitaria, 28040 Madrid, Spain}

\author{Mireia Tolosa-Simeón}
\email{scatteringanalogy@matterwave.de}
\affiliation{Institut für Theoretische Physik III, Ruhr-Universität Bochum, Bochum, Germany}

\author{Marius Sparn}
\email{scatteringanalogy@matterwave.de}
\affiliation{Kirchhoff-Institut für Physik, Universität Heidelberg,
Im Neuenheimer Feld 227, 69120 Heidelberg, Germany}

\author{Elinor Kath}
\email{scatteringanalogy@matterwave.de}
\affiliation{Kirchhoff-Institut für Physik, Universität Heidelberg,
Im Neuenheimer Feld 227, 69120 Heidelberg, Germany}

\author{Nikolas Liebster}
\email{scatteringanalogy@matterwave.de}
\affiliation{Kirchhoff-Institut für Physik, Universität Heidelberg,
Im Neuenheimer Feld 227, 69120 Heidelberg, Germany}

\author{Jelte Duchene}
\email{scatteringanalogy@matterwave.de}
\affiliation{Kirchhoff-Institut für Physik, Universität Heidelberg,
Im Neuenheimer Feld 227, 69120 Heidelberg, Germany}

\author{Helmut Strobel}
\email{scatteringanalogy@matterwave.de}
\affiliation{Kirchhoff-Institut für Physik, Universität Heidelberg,
Im Neuenheimer Feld 227, 69120 Heidelberg, Germany}

\author{Markus K. Oberthaler}
\email{scatteringanalogy@matterwave.de}
\affiliation{Kirchhoff-Institut für Physik, Universität Heidelberg,
Im Neuenheimer Feld 227, 69120 Heidelberg, Germany}

\author{Stefan Floerchinger}
\email{scatteringanalogy@matterwave.de}
\affiliation{Theoretisch-Physikalisches Institut, Friedrich-Schiller-Universität Jena, Max-Wien-Platz 1, 07743 Jena, Germany}

\begin{abstract}
The production of quantum field excitations or particles in cosmological spacetimes is a hallmark prediction of curved quantum field theory. The generation of cosmological perturbations from quantum fluctuations in the early universe constitutes an important application. The problem can be quantum-simulated in terms of structure formation in an interacting Bose-Einstein condensate (BEC) with time-dependent s-wave scattering length.
Here, we explore a mapping between cosmological particle production in general (D+1)-dimensional spacetimes and scattering problems described by the non-relativistic stationary Schrödinger equation in one dimension.
Through this mapping, intuitive explanations for emergent spatial structures in both the BEC and the cosmological system can be obtained for analogue cosmological scenarios that range from power-law expansions to periodic modulations. 
The investigated cosmologies and their scattering analogues are tuned to be implemented in a (2+1)-dimensional quantum field simulator.
\end{abstract}

% Define your custom highlight color
\definecolor{bordeauxRed}{RGB}{178, 34, 52} % bordeaux red
\definecolor{royalBlue}{RGB}{65,105,225} % royal blue
\definecolor{oliveGreen}{RGB}{128,128,0} % olive green

\maketitle

\makeatletter
\def\l@subsubsection#1#2{}
\makeatother
\tableofcontents

\section{Introduction}

In this work, we want to address three different but interconnected problems that offer valuable insights into quantum field theory in curved spacetime \cite{mukhanov_winitzki_2007,birrell_davies_1982,Fulling1989,Wald1995}: 
\begin{enumerate}
\item The production of excitations of a real, relativistic, scalar quantum field from an evolving spacetime geometry, such as an expanding or contracting cosmological solution of the gravitational field equations.

\item A quantum mechanical scattering problem described by the stationary Schrödinger equation in one spatial dimension with a potential that features a critical bound state with vanishing binding energy.

\item The formation of spatial structure in a Bose-Einstein condensate with time-dependent scattering length that can be manipulated through a magnetic field in the vicinity of a Feshbach resonance.
\end{enumerate}

All three problems are of considerable interest by themselves. For example, cosmological particle production in time-dependent spacetimes is a hallmark prediction from quantum field theory in curved spacetime~\cite{Parker1969}, with some parallels to Hawking radiation from black holes~\cite{Hawking1975}. Moreover, the formation of structure in gravitational field perturbations during an inflationary period in the early universe, triggered by a scalar field, can be mapped to this problem. These perturbations are conserved by causality arguments afterwards for some time, and later constitute, just after horizon re-entry, the seeds for the dynamical formation of large-scale structure through gravitational attraction.

Quantum mechanical scattering on a potential in one spatial dimension is a textbook problem due to its conceptual clarity \cite{LandauLifshitzQuantum,Sakurai_Napolitano_2020,griffiths_schroeter_2018,Schwabl2007,Flügge1999,Boya2008}. It is suitable to describe a broad range of physical phenomena where negative frequency waves are generated; in particular, its connection to cosmological particle production was already mentioned in~\cite{mukhanov_winitzki_2007}. Furthermore, several quantum tunneling problems can be reduced to this setup.
However, it is not easy to investigate this problem in detail experimentally and specifically to find an experimental setup that would allow to design arbitrary forms of the scattering potential.

Finally, structure formation in a Bose-Einstein condensate with time-dependent scattering length is a formidable non-equilibrium problem in many-body quantum mechanics, or non-relativistic quantum field theory~\cite{CalzettaHu2008}, which has the nice feature that it can be well realized and investigated experimentally \cite{Viermann2022,Tolosa2022,Sanchez2022}.

The idea to use ultracold atomic quantum gases to realize quantum simulators for interesting problems in condensed matter physics or quantum field theory has a well-established history \cite{Unruh1981,Unruh1995,UnruhSchuetzhold2007,Volovik2009,Feynman1982,Jaksch1998,Bloch2012,Cirac2012,Georgescu2014,Aidelsburger2021,Mil2020,Yang2020,Schweizer2019,Goerg2019}.
Precise experimental control and high tunability render table-top experiments with ultracold atoms ideal quantum simulation platforms. Particular success has been achieved in the context of the analogue gravity program (consider \cite{Barcelo2011,Visser2002,Novello2002} as an introduction and \cite{Jacquet2020} for an overview of experimental development), where acoustic excitations of the condensate background experience an emergent spacetime geometry shaped by the latter. This allows to simulate kinematics of quantum field theory in curved space. In particular, the background can be engineered to correspond to a cosmological spacetime \cite{Barcelo2003c,Fedichev2003,Fedichev2004,Fischer2004,Fischer2004b,Uhlmann2005,Calzetta2005,Liberati2006,Liberati2006b,Weinfurtner2007,Prain2010,Bilic2013} that can be continued to a dispersive rainbow spacetime in the transphononic regime \cite{Weinfurtner2006,Weinfurtner2007BoseNova,Weinfurtner2009}. Proposals to use dipolar Bose-Einstein condensates to simulate cosmological particle production exist \cite{Fischer2017}. 
Experimental studies of cosmological particle production were performed on cold-atom-platforms in $(1+1)$ dimensions \cite{Eckel2018,Tajik2023} and in $(2+1)$-dimensions \cite{Hung2013,Chen2021,Viermann2022}. The $(1+1)$-dimensional effect has also been realized with trapped ions \cite{Wittemer2019} and an optical system \cite{Steinhauer2022}.
Outside of the cosmological context, cold atoms were also used in theoretical and experimental studies of
Hawking radiation \cite{Horstmann2010,Weinfurtner2011,Steinhauer2016,MunozdeNova2019,Ribeiro2022}, the Casimir effect \cite{Jaskula2012}, Unruh radiation \cite{Rodriguez-Laguna2017,Hu2019,Gooding2020} and false vacuum decay \cite{Zenesini2024,Jenkins2024}.
Furthermore, new simulation platforms to simulate kinematics of fermionic fields in curved spacetime have been proposed \cite{Tolosa2023,HallerMeng2023}.
Recently, impressive developments have been made with microcavity-polaritons \cite{Falque2023,Jacquet2022,Jacquet2023} and superfluid helium-4 \cite{Svancara2024} in the context of simulating (rotating) black hole spacetimes. Finally it should be noted that the scattering aspects of the mentioned analogue gravity systems are already used in the analysis, as for example in the emerging topic of black hole superradiance \cite{Giacomelli2021,Delhom2024}. 

In this work, we will discuss how the essential elements of all three problems can be mapped to each other. This allows to transfer insights, be it intuition or precise quantitative statements, from one problem to another. 
Specifically, we shall show that the third problem, i.e.\ structure formation in an interacting Bose-Einstein condensate, 
which lies at the core of the quantum field simulator (QFS) established in \cite{Viermann2022,Tolosa2022,Sanchez2022}, shares its salient features with the conceptually well-understood problem of quantum mechanical scattering in one dimension. 
Our investigations furthermore highlight that the non-trivial transition amplitudes in the analogue cosmological system possess fundamentally quantum properties, similar to reflection at an attractive or tunneling through a repulsive landscape. These quantum-physical aspects are manifested, for example, in the appearance of discrete jumps in a relative phase of the particle spectrum. 

\emph{The remainder of this paper is organized as follows.}
The rest of this section is dedicated to briefly recalling how a scalar field in a curved background can be simulated with Bose-Einstein condensates, focusing on the case of a curved $(D+1)$-dimensional Friedmann-Lemaître-Robertson-Walker (FLRW) spacetime. In \cref{sec:CosmPartProd}, we showcase the equivalence between the mode evolution and a quantum mechanical scattering problem in one dimension, and describe the scattering analogy of cosmological particle production, providing direct relations between the spectrum of quantum field excitations and the scattering amplitudes of the analogue quantum mechanical problem. We particularize to the case of a (2+1)-dimensional BEC-quantum field simulator in \cref{sec:ScatteringPotentialsInTheQuantumSimulation}, and analyze specific examples, highlighting their connection to analogue cosmology. 
In \cref{sec:AnalyticalSolutionOfElementaryScenarios}, we give explicit solutions to the problems introduced in \cref{sec:ScatteringPotentialsInTheQuantumSimulation} and summarize their physical properties. 
Periodic scattering landscapes, which correspond to oscillating cosmological spacetimes, are investigated in \cref{sec:OscillatingSpacetimes}. A minimal lattice-model which admits an analytic solution is designed and provides clear explanations for emerging structures in the quasiparticle power spectrum.
In \cref{sec:ZeroEnergySolutions}, we show that a large class of analogue scattering potentials allows for zero-energy-resonances (given in terms of the scale factor) and investigate their impact on the particle spectrum.  
\Cref{subsec:SmoothSwitches} constitutes an analysis of the impact of smoothed transitions between stasis and dynamics in the experiment.
Finally, we summarize our findings and provide an outlook in \cref{Sec:Conclusions}.

\emph{Notation.} In this paper, we work in SI-units unless stated otherwise. Vectors are denoted by bold symbols. Greek indices $\mu,\nu$ run from $0$ to $D$, while latin indices $i,j$ run from $1$ to $D$. The metric signature is $(-,+,\ldots,+)$. 

\subsection{Quantum simulation of a scalar field in curved spacetime with Bose-Einstein condensates}
\label{subsec:RecapBEC}

Weakly interacting Bose-Einstein condensates in $D+1$ dimensions can be described in terms of non-relativistic complex scalar quantum fields $\Phi(t, \mathbf{x})$ with the effective action \cite{Floerchinger2008}
\begin{equation}
\begin{split}
&\Gamma[\Phi] = \int \text{d}t \, \text{d}^D \mathbf{x} \Bigg\{ \hbar \Phi^*(t,\mathbf x) \left[\im \frac{\partial}{\partial t} - V_\mathrm{ext}(t,\mathbf x) \right]  \Phi(t,\mathbf x) \\
&- \frac{\hbar^2}{2 m} \boldsymbol{\nabla}\Phi^*(t, \mathbf x) \boldsymbol{\nabla} \Phi(t, \mathbf x) - \frac{\lambda(t)}{2} \left[ \Phi^*(t,\mathbf x) \Phi(t,\mathbf x) \right]^2 \Bigg\}.
\end{split} \label{eq:ActionBEC}
\end{equation}
Here, $m$ denotes the mass of the atoms, $V_\mathrm{ext}(t,\mathbf x)$ is a configurable external trapping potential and $\lambda(t)$ is a time-dependent interaction strength between the atoms of the BEC, which can be expressed in terms of the s-wave scattering length $a_\text{s}$. 
For a quasi-(2+1)-dimensional BEC, in the regime of the first Born approximation, their relation is given by \cite{Pitaevskii2016}
\begin{equation}
    \lambda(t) = \sqrt{\frac{8\pi \omega_z \hbar^3}{m}} a_\text{s}(t),
    \label{eq:ScatteringLength}
\end{equation}
where $\omega_z$ is the trapping frequency in the $z$-direction and $a_\mathrm{s}$ is the s-wave scattering length. 
The interaction strength can be dynamically controlled in the laboratory by changing an external magnetic field in the vicinity of a Feshbach resonance \cite{Feshbach1958,Viermann2022}.

We consider small excitations around a background mean field solution and decompose the field as \cite{Floerchinger2008,Heyen2020}
\begin{equation}
\begin{split}
    \Phi(t, \mathbf{x}) = & \, \e^{\im S_0(t, \mathbf{x})} \left\{  \sqrt{n_0(t, \mathbf{x})} \right. \\ & \left.+ \frac{1}{\sqrt{2}} \left[ \phi_1(t, \mathbf{x}) + \im \phi_2(t, \mathbf{x}) \right]\right\},
\end{split}
\end{equation}
where $n_0(t, \mathbf{x})$ is the background superfluid density and the phase $S_0(t, \mathbf{x})$ is a potential for the background superfluid velocity, $\mathbf{v}_S(t, \mathbf{x})=(\hbar/m) \boldsymbol{\nabla} S_0(t, \mathbf{x})$.
The perturbation fields  $\phi_1(t,\mathbf{x})$ and $\phi_2(t, \mathbf{x})$ are real and they can be seen as a superfluid density fluctuation and a phase fluctuation, respectively.

We focus on the regime where perturbations are small enough so that only terms up to quadratic order in fluctuations are taken into account. 
Moreover, we assume the background superfluid velocity to be spatially constant, i.e. $\boldsymbol{\nabla}^2 S_0(t, \mathbf{x})=0$. In the low momentum regime, it is convenient to integrate out the density perturbation field $\phi_1(t, \mathbf{x})$ by evaluating it on its equation of motion. The remaining quadratic action for the phonon field~${\phi(t, \mathbf{x}) \equiv \phi_2(t, \mathbf{x})/\sqrt{2m}}$ can then be brought to the form of an effective action for a free massless scalar field in a curved spacetime determined by the acoustic metric~$g_{\mu\nu}(x)$~\cite{Tolosa2022}
\begin{equation}
    \Gamma_2 [\phi] = -\frac{\hbar^2}{2} \int  \dd t\, \text{d}^D \mathbf{x}\, \sqrt{g(x)} \, g^{\mu\nu}(x) \partial_\mu \phi(x) \partial_\nu \phi(x),
    \label{eq:QEACurvedSpacetime}
\end{equation}
where $x^\mu = (t,\mathbf{x})$ and $\sqrt{g(x)}\equiv\sqrt{-\det(g_{\mu\nu}(x))}$ with
\iffalse
\begin{equation}
    \sqrt{g(\mathbf{x})} g^{\mu\nu}(x) = \frac{1}{c^2(x)} \begin{pmatrix}
        -1 & v_j(x) \\ v_i(x) & c^2(x) \delta_{ij} - v_i(x) v_j(x)
    \end{pmatrix}.
\end{equation}
By taking the determinants on both sides on finds $g(\mathbf{x}) = 1/c(\mathbf{x})^{4/(D-1)}$ and then
\fi
\begin{equation}
    g_{\mu\nu}(x) = \frac{1}{c^\frac{2}{D-1}(x)} \begin{pmatrix} -c^2(x) + \mathbf{v}_S^2(x) & \mathbf{v}_S^T(x) \\ \mathbf{v}_S(x) & \mathds{1} \end{pmatrix},
\end{equation}
where $c(x)$ is the space- and time-dependent speed of sound.

In the following, we consider the case of vanishing background superfluid velocity, $\mathbf{v}_S(x)=0$, yielding a stationary background density $n_0(\mathbf{x})$, which can be manipulated through the spatial dependence of the external trapping potential $V_\mathrm{ext}(t,\mathbf x)$, and we allow a time-dependent interaction strength $\lambda(t)$, leading to the line element
\begin{equation}
    \dd s^2 = - c^{\frac{2D-4}{D-1}}(t,\mathbf{x}) \dd t^2 + \frac{1}{c^{\frac{2}{D-1}}(t, \mathbf{x})} \dd \mathbf{x}^2,
    \label{eq:Metric}
\end{equation}
where the speed of sound is introduced as
\begin{equation}
    c(t, \mathbf{x}) = \sqrt{\frac{\lambda(t) n_0(\mathbf{x})}{m}}.
    \label{eq:SpeedOfSoundDefinition}
\end{equation}
We restrict to rotation invariant density profiles of the form
\begin{equation}
    n_0(\mathbf{x}) = \bar n_0 \left(1+\frac{\kappa}{4} r^2 \right)^2,
\end{equation}
where $\kappa$ is a real parameter that can take the values ${\kappa=0}$, $\kappa=4/R^2$ or $\kappa=-4/R^2$, with $R$ being the radius of the condensate, $r$ the radial coordinate and $\Bar{n}_0=n(r=0)$ is the constant background density at the center of the trap.
Therefore, in this case, the acoustic line element \eqref{eq:Metric} takes the form
\begin{equation}
    \text{d} s^2 = - \text{d}t^2 + a^2(t) \left(1 + \frac{\kappa}{4} r^2 \right)^{-1} (\text{d}r^2 + r^2 \text{d} \Omega_D^2),
\label{eq:LineElement}
\end{equation}
with $\text{d}\Omega_{D}$ being the solid angle element in $D$ spatial dimensions and where we introduced the time-dependent scale factor as
\begin{equation}
    a^2(t) \equiv \frac{m \, }{\bar{n}_0} \frac{1}{\lambda(t)}.
\label{eq:ScaleFactorDefinition}
\end{equation}
Then, performing a radial coordinate transformation
\begin{equation}
    u(r) = \frac{r}{1 +\kappa r^2/4},
\end{equation}
the acoustic line element becomes
\begin{equation}
    \text{d} s^2 = - \text{d}t^2 + a^2(t)  \left(\frac{\text{d}u^2}{1 - \kappa u^2} + u^2 \text{d} \Omega_D^2 \right),
    \label{eq:FLRWLineElement}
\end{equation}
which corresponds to the standard line element of a FLRW spacetime with spatial curvature $\kappa$ and scale factor $a(t)$, which is assumed to be a real, continuous and positive function of time  with the limit $a(t)\to 0$ corresponding to a big bang singularity.

\subsection{Real relativistic scalar field in cosmological spacetimes}

Let us now discuss the physics of a real, relativistic scalar field in an expanding or contracting universe. We work here, in the context of cosmology, in units where~${\hbar = c = 1}$.

We will consider a real, massive scalar field $\phi$ that is non-minimally coupled to gravity in a (D+1)-dimensional FLRW spacetime~\eqref{eq:FLRWLineElement}. The corresponding action~\eqref{eq:QEACurvedSpacetime} takes the form~\cite{birrell_davies_1982} 
\begin{equation}
    \Gamma[\phi] = - \frac{1}{2} \int \dd t\, \text{d}^D \mathbf{x}\sqrt{g}\left[\partial_{\mu} \phi \partial^{\mu} \phi + \left(m_\phi^2 + \xi \mathcal{R} \right)\phi^2 \right],
    \label{eq:EffAKG}
\end{equation}
where $\xi$ is the coupling constant to the Ricci scalar $\mathcal{R}$, $m_\phi$ the mass of the scalar field $\phi$ and $D$ the number of spatial dimensions.

It is convenient to introduce the conformal time $\eta$ defined by $\dd \eta = \dd t/a(t)$ as well as the rescaled field
\begin{equation}
\chi(x) = a^\frac{D-1}{2}(\eta) \phi(x), 
\label{eq:RescaledField}
\end{equation}
such that the action becomes
\begin{equation}
    \Gamma[\chi] = - \frac{1}{2} \int \dd \eta \  \dd^{D}\mathbf{x} \sqrt{\gamma} \  \chi \left[ \dv[2]{\eta} -  \Delta + m_{\text{eff}}^2(\eta)  \right] \chi,
\label{eq:ActionRescaledField}
\end{equation}
where $\sqrt{\gamma}\equiv  \sqrt{\det(\gamma_{ij})}$ is the determinant of the time-independent spatial metric, with line element
\begin{equation}
    \gamma_{ij} \dd \text{x}^i \dd \text{x}^j = \frac{\dd u^2}{1- \kappa u^2} + u^2 \dd \Omega_{D}^2,   
\end{equation}
and $\Delta$ is the Laplace-Beltrami operator on hypersurfaces of constant cosmic time  \cite{Cohl2011,donnelly2002}. Note that the field $\chi$ acquires an effective, time-dependent mass given by
\begin{equation}
    \begin{aligned}
        m_{\text{eff}}^2(\eta) =& \, a^2(\eta) \left[m_\phi^2+\xi \mathcal{R}(\eta)\right] \\
        &+ \frac{1-D}{2}\left[\frac{a''(\eta)}{a(\eta)} - \frac{3-D}{2}\left( \frac{a'(\eta)}{a(\eta)} \right)^2\right],
        \label{eq:EffectiveMass}
    \end{aligned}
\end{equation}
where the prime notation denotes differentiation with respect to the conformal time $\eta$, and the Ricci scalar is given by
\begin{equation}
\begin{split}
 \mathcal{R} (\eta) =& \frac{D}{a^{2}(\eta)} \Bigg[(D-1) \kappa + (D-3) \left( \frac{a'(\eta)}{a(\eta)} \right)^2+2 \frac{a''(\eta)}{a(\eta)} \Bigg].
 \label{eq:Ricci}
\end{split}
\end{equation}
Note that for $\xi=(D-1)/(4D)$, which corresponds to the conformally coupled case, only the bare mass term in \eqref{eq:EffectiveMass} survives. Therefore, when $m_\phi=0$, the action is invariant under conformal transformations. As a consequence, the effective mass in \cref{eq:EffectiveMass} becomes independent of conformal time, $m_{\text{eff}}^2 = (D-1)^2 \kappa / 4$, such that the quantum field is not affected by the time-dependent background.

Taking the functional derivative of the action in \cref{eq:ActionRescaledField} with respect to $\chi$ yields the field equation of motion,
\begin{equation}
    \chi''(\eta,\mathbf{x}) - \Delta \chi(\eta,\mathbf{x})  + m_{\text{eff}}^2(\eta) \chi(\eta,\mathbf{x}) = 0,
    \label{eq:FieldEquation_RescaledField}
\end{equation}
in the form of a Klein-Gordon equation with time-dependent mass.

We expand the field $\chi$ in terms of the eigenfunctions of the Laplace-Beltrami operator, $\mathcal{H}_{\mathbf{k}}(\mathbf{x})$, together with the time-dependent mode functions,  $\psi_k(\eta)$, %\eqref{eq:LaplaceBeltrami}
\begin{equation}
    \begin{split}
        \chi(\eta, \mathbf{x}) = \int_{\mathbf{k}}\left[\hat a_{\mathbf{k}} \mathcal{H}_{\mathbf{k}} (\mathbf{x}) \psi_k (\eta) + \hat a^\dagger_{\mathbf{k}} \mathcal{H}_{\mathbf{k}}^{*} (\mathbf{x}) \psi_k^*(\eta) \right], 
    \end{split}
    \label{eq:ModeExpansion}
\end{equation}
where the momentum integral $\int_{\mathbf{k}}$, the eigenfunctions $\mathcal{H}_{\mathbf{k}}(\mathbf{x})$ and the (creation) annihilation operators $\hat{a}_\mathbf{k}^{(\dagger)}$ that follow bosonic commutation relations and are discussed in further detail for a (2+1)-dimensional spacetime in our previous works in Refs.~\cite{Sanchez2022,Tolosa2022}. 
Inserting the mode expansion of the field into the Klein-Gordon equation~\eqref{eq:FieldEquation_RescaledField} leads to the mode equation 
\begin{equation}
     \left[\dv[2]{}{\eta} -h(k) + m_\text{eff}^2 (\eta) \right] \psi_k(\eta) = 0,
    \label{eq:ModeEvolutionSchrödinger}
\end{equation}
where $h(k)$ are the eigenvalues of the Laplace-Beltrami operator,
\begin{equation}
    h(k) = \begin{cases} -k\left[k + (D-1) \sqrt{\abs{\kappa}} \right] &\text{for} \quad \kappa > 0, \\
        -k^2 &\text{for} \quad \kappa = 0, \\
        - \left[k^2 + \left(\frac{D-1}{2}\right)^2 \abs{\kappa} \right] &\text{for} \quad \kappa < 0. \end{cases}
        \label{eq:LaplacianEigenvalue}
\end{equation}

It is clear now how the dynamics of phononic excitations in a BEC can be mapped to that of a scalar field in an expanding geometry, and that the latter is determined by the mode equation \eqref{eq:ModeEvolutionSchrödinger}, in which the effect of the background is encoded in the time-dependent effective mass. We will examine in the next section how particle production by a dynamical background takes place.

\section{Cosmological particle production as a scattering problem}
\label{sec:CosmPartProd}
The key observation here is that the mode equation~\eqref{eq:ModeEvolutionSchrödinger} has the form of a stationary Schrödinger equation 
%(in which the prefactor $\hbar/(2M)$ has been set to one)
\begin{equation}
    \left[ - \dv[2]{}{\eta} + V(\eta) \right] \psi_k(\eta) = E_k \psi_k(\eta),
    \label{eq:SchrodingerEq}
\end{equation}
with energy eigenvalues $E_k = - h(k)$ given by the eigenvalues of the Laplace-Beltrami operator (see eq.\ \eqref{eq:LaplacianEigenvalue}), and scattering potential given by 
\begin{equation}
V(\eta) = - m_{\text{eff}}^2(\eta),
\label{eq:GeneralScatteringPotential}
\end{equation}
defined in eq.\ \eqref{eq:EffectiveMass}. 
Therefore, solving the mode equation for the dynamical evolution of a mode with wave number $k$ is equivalent to finding a solution to the stationary Schrödinger equation with energy $E_k=-h(k)$ and a potential $V(\eta)$ determined by the cosmological history in terms of the scale factor $a(\eta)$. 

In the following, the framework developed in our previous work \cite{Sanchez2022,Tolosa2022,Viermann2022} shall be formulated as a quantum mechanical scattering problem (leading to \cref{fig:ScatteringAnalogy}).
To that end, we consider a cosmological evolution during a finite interval of time $[t_{\text{i}},t_{\text{f}}]$ specified as region II, as well as static regimes before and after the expansion, referred to as regions I and III, respectively.

The mode function $\psi_k(\eta)$  is a solution to the mode evolution equation \eqref{eq:ModeEvolutionSchrödinger} in all three regions of time; in particular, the form of the solution in regions I and III is that of plane waves, since in these regimes we recover Minkowski spacetime.
During the time evolution in region II, the mode functions $\psi_k(\eta)$ are excited through their interaction with the time-dependent background, leading to particle production. 
Through a Bogoliubov transformation~\cite{mukhanov_winitzki_2007}, one can express the mode solution~$\psi_k(\eta)$ as a superposition of the mode functions $u_k(\eta)$ and $u_k^*(\eta)$, which are positive and negative frequency plane waves in region~III, respectively,
\begin{equation}
    u_k = \alpha_k \psi_k + \beta_k \psi_k^*, \quad \psi_k = \alpha_k^* u_k - \beta_k u_k^*
    \label{eq:BogoliubovTrafo}
\end{equation}
and the corresponding ladder operators associated to the mode functions $u_k(\eta)$ are $\hat{b}^{(\dagger)}_{\mathbf{k}}$ , whose relation with the ladder operators $\hat{a}^{(\dagger)}_{\mathbf{k}}$ associated to $\psi_k(\eta)$, is given here for completeness, 
\begin{equation}
    \hat b_{\mathbf{k}} = \alpha_k^* \hat a_{\mathbf{k}} - \beta_k^* \hat a^\dagger_{\mathbf{-k}}.
    \label{eq:BogoliubovTrafoOperators}
\end{equation}
Then, the Bogoliubov coefficients are
\begin{equation}
    \begin{aligned}
        \alpha_k &= \frac{\text{Wr}[u_k,\psi_k^*]}{\sqrt{\text{Wr}[u_k,u_k^*]} \sqrt{\text{Wr}[\psi_k,\psi_k^*]}} 
    \end{aligned}
    \label{eq:BogoliubovAlphaGeneralExpr}
\end{equation}
and
\begin{equation}
    \begin{aligned}
        \beta_k &= -\frac{\text{Wr}[u_k,\psi_k]}{\sqrt{\text{Wr}[u_k,u_k^*]} \sqrt{\text{Wr}[\psi_k,\psi_k^*]}}, 
    \end{aligned}
    \label{eq:BogoliubovBetaGeneralExpr}
\end{equation}
where the Wronskian is defined as 
\begin{equation}
    \mathrm{Wr}[u_k,\psi_k] = u_k \psi_k' - u_k' \psi_k.
\end{equation}

In particular, the coefficient $\beta_k \neq 0$ represents a finite overlap between negative and positive frequency modes in region III (according to the positive-frequency notion employed in region I), indicating that the notion of the vacuum disagrees between region I and region III such that one can speak of particle production. 
An alternative viewpoint is to realize a finite overlap between particle annihilation in region III and particle creation in region I, as can be deduced from \cref{eq:BogoliubovTrafoOperators}. A more detailed discussion can be found in \cite{Tolosa2022,Sanchez2022}. 

\subsection{Description of the scattering analogy}
\label{subsec:DescriptionScatteringAnalogy}

\begin{figure}
   \makebox[\columnwidth][c]
   {
    \includegraphics[width=1\columnwidth]{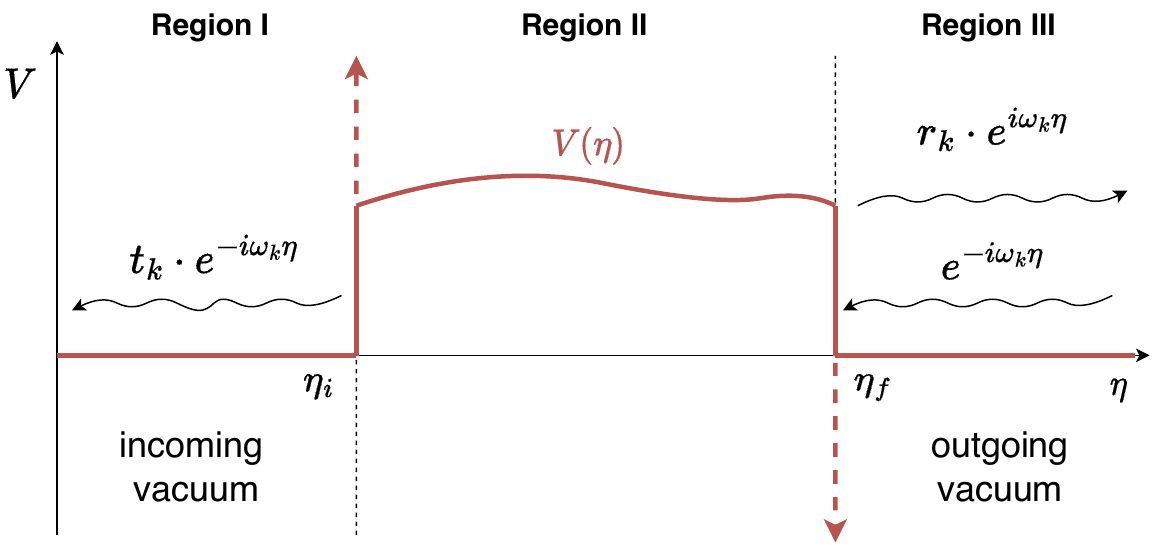}
   }
    \caption{ Graphical illustration of the scattering analogy. For a clearer illustration in terms of $r_k$ and $t_k$, the coefficient $a_k$ is normalized to unity here.}
    \label{fig:ScatteringAnalogy}
\end{figure}

Within the scattering analogy, the evolution of the mode functions is described as follows (see \cref{fig:ScatteringAnalogy} for a graphical illustration): 

1. In region I, $\eta < \eta_{\text{i}}$, the expansion has not yet taken place. In the incoming vacuum state, the mode functions are plane waves $\psi_k^\mathrm{I}(\eta)$ with positive frequency $\omega_k$ such that the relation
    \begin{equation}
        \dv{\psi_k^\mathrm{I}(\eta)}{\eta} = - \im \omega_k \psi_k^\mathrm{I}(\eta) \quad \text{ with } \quad \omega_k > 0
        \label{eq:ConformalFrequencyDefinition}
    \end{equation}
     holds \cite{Carroll2019}.
    Therefore, we can write the mode solution in region I as
    \begin{equation}
    \psi_k^\mathrm{I}(\eta) = c_k \e^{-\im \omega_k \eta}.
    \label{eq:PlaneWaveRegionI}
    \end{equation}
In the scattering analogy, the mode corresponds to a plane wave solution of the free Schrödinger equation, which was transmitted by a scattering potential and now propagates freely towards smaller $\eta$ (to the left in \cref{fig:ScatteringAnalogy}). 
The dispersion relation for the mode is $\omega_k = \sqrt{-h(k)}$ with $h(k)$ given in \cref{eq:LaplacianEigenvalue}.

2. In region II, $\eta_{\text{i}} \leq \eta \leq \eta_{\text{f}}$, the field is subjected to a dynamic spacetime.
The Klein-Gordon evolution is mapped to a Schrödinger equation with a non-vanishing scattering potential defined in \eqref{eq:GeneralScatteringPotential}. 
      
3. In region III, $\eta > \eta_{\text{f}}$, the expansion has ceased and the modes are again of plane wave form. However, now they are a superposition of positive and negative frequencies 
 \begin{equation}
    \psi_k^\mathrm{III}(\eta) = a_k \e^{-\im \omega_k \eta} + b_k  \e^{\im \omega_k \eta},
    \label{eq:PlaneWaveRegionIII}
    \end{equation}
    where the presence of negative frequency modes with amplitude $b_k$  indicates the non-emptiness of the vacuum state after the expansion has taken place, equivalent to a non-vanishing Bogoliubov coefficient $\beta_k$.
    The negative frequency modes with amplitude $\sim b_k$  correspond in the scattering analogy to a wave that is reflected at the potential $V(\eta)$,
    whereas the incoming wave of the scattering problem with amplitude $\sim a_k$  is the positive frequency part of the vacuum after the expansion has ceased (propagating from right to left in \cref{fig:ScatteringAnalogy}).

In the context of one-dimensional scattering problems in quantum mechanics, it is common to define the reflection and transmission amplitudes $r_k$ and $t_k$, which in this context are given by
\begin{equation}
    r_k = \frac{b_k}{a_k} \quad \text{ and } \quad t_k = \frac{c_k}{a_k}.
    \label{eq:ScatteringAmplitudes}
\end{equation}
These amplitudes satisfy 
\begin{equation}
    1=\abs{r_k}^2 + \abs{t_k}^2  \quad \text{ or } \quad \abs{a_k}^2 = \abs{b_k}^2 + \abs{c_k}^2,
    \label{eq:ProbabilityConservation} 
\end{equation}
which reflect the probability conserving condition of unitarity. 

To complete the shift of perspective, it is useful to express the Bogoliubov coefficients in \cref{eq:BogoliubovAlphaGeneralExpr,eq:BogoliubovBetaGeneralExpr} in terms of the early and late time solutions in \cref{eq:PlaneWaveRegionI,eq:PlaneWaveRegionIII}, respectively. 
A straightforward calculation leads to
\begin{equation}
    \begin{aligned}
        \alpha_k &= \frac{a_k^*}{c_k^*} \quad \text{and} \quad \beta_k = - \frac{b_k}{c_k^*}.
    \end{aligned}
\label{eq:BogoliubovCoeffScattering}
\end{equation} 
Note that the condition
\begin{equation}
    \abs{\alpha_k}^2 - \abs{\beta_k}^2 = 1,
    \label{eq:BogoliubovCofficientsNormalization}
\end{equation}
which is necessary to preserve the Wronskian under the Bogoliubov transformation, $\text{Wr}[u_k,u_k^*] = \text{Wr}[\psi_k,\psi_k^*]$, is provided by \cref{eq:ProbabilityConservation}. 

The described analogy between Bogoliubov coefficients and scattering amplitudes was used in \cite{Visser1999} to derive general bounds on the process of one-dimensional scattering.

\subsection{Spectral properties of produced quanta}

In this section, we will discuss the spectral properties of the produced quanta. 
The power spectrum of quantum field excitations is defined via a generalized Fourier transform of the equal-time statistical function of the canonically conjugate momentum field $\dot{\phi}$  \cite{Sanchez2022,Tolosa2022}
\begin{equation}
    \frac{1}{2} \bra{0} \lbrace \dot{\phi}(t,\mathbf{x}), \dot{\phi}(t,\mathbf{y}) \rbrace \ket{0}_c = \int_k \mathcal{F}(k,L) \frac{\sqrt{-h(k)}}{a_{\text{f}}^3} S_k(t),
    \label{eq:SpectrumDefinition}
\end{equation}
where the dot notation denotes differentiation with respect to the cosmic time $t$ and the vacuum state $\ket{0}$  is taken to be the natural vacuum state for an observer in region I.
The kernel $\mathcal{F}(k,L)$ represents the geometry of the homogeneous and isotropic hypersurfaces of constant time (consider \cite{Tolosa2022,Sanchez2022} for more details), which depends on the comoving distance between two spatial points $L = \abs{\mathbf{x}-\mathbf{y}}$. 
We can compute the expectation value \eqref{eq:SpectrumDefinition} in region III, where $t > t_{\text{f}}$, through a Bogoliubov transformation on the modes and can identify the spectrum $S_k$ with the general form 
\begin{equation}
    S_k(\eta(t)) = \frac{1}{2} + N_k + \Delta N_k(\eta(t))
\end{equation}
for $t > t_{\text{f}}$. Relating the Bogoliubov coefficients to the scattering amplitudes via \cref{eq:BogoliubovCoeffScattering}, one finds for the mean particle number 
\begin{equation}
    N_k = \abs{\beta_k}^2 = \abs{\frac{b_k}{c_k}}^2 = \abs{\frac{r_k}{t_k}}^2.
    \label{eq:SakharovOffset}
\end{equation}
Furthermore, for $\eta(t) > \eta_{\text{f}}$, we have the coherent term
\begin{equation}
    \begin{aligned}
    \Delta N_k(\eta(t)) &= \Re \left[ \alpha_k \beta_k \e^{2\im \omega_k [\eta(t) - \Delta \eta]} \right] \\[5pt]
    &= \Delta N_k^0 \cos \left[ 2  \omega_k \eta(t) + \vartheta_k \right],
    \end{aligned}
    \label{eq:SakharovOscillations}
\end{equation}
where we introduced the amplitude
\begin{equation}
    \Delta N_k^0 = \abs{\alpha_k \beta_k} = \frac{\abs{a_k b_k}}{\abs{c_k}^2} = \abs{\frac{r_k}{t_k^2}}
    \label{eq:SakharovAmplitude}
\end{equation}
and the phase shift
\begin{equation}
    \begin{aligned}
    \vartheta_k &= \arg (\alpha_k \beta_k \e^{2 \im \omega_k \Delta \eta } ) \\
    & = \arg \left( -\frac{c_k}{c_k^*} r_k \e^{2 \im \omega_k \Delta \eta  } \right).%= \arg \left( - \frac{a_k^*}{c_k^*} \frac{b_k}{c_k^*} \e^{-2 \im \omega_k \eta_{\text{f}}}  \right )
    \end{aligned}
    \label{eq:SakharovPhaseShift}
\end{equation}
with $\Delta \eta = \eta_\mathrm{f} - \eta_\mathrm{i}$. 

Proper quantum mechanical observables should be invariant under a global $\mathrm{U}(1)$-transformation of the wave function.
In the scattering analogy, the wave function is a mode of a quantum field and the observable is the occupation number of this mode, defined in \cref{eq:SakharovOffset,eq:SakharovOscillations}.
We will now show that global $\mathrm{U}(1)$-symmetry is manifested in these observables. Consider the global unitary transformation $\psi_k(\eta) \to \e^{\im \lambda_k} \psi_k(\eta)$ and ${u_k(\eta) \to \e^{\im \mu_k} u_k(\eta)}$, which entails
$ {a_k \to \e^{\im \mu_k} a_k }$, ${b_k \to \e^{\im \mu_k} b_k}$ and ${c_k \to \e^{\im \lambda_k} c_k}$ such that $N_k$ is left invariant. Furthermore, $\Delta N_k(\eta)$ is invariant as well, taking into account that ${\e^{2\im \omega_k \eta} \to \e^{-2\im \mu_k} \e^{2\im \omega_k \eta}}$, since the initial value of conformal time can be chosen arbitrary and, by \cref{eq:ConformalFrequencyDefinition}, the mode frequency $\omega_k$ is defined up to a global phase of the mode $u_k(\eta)$. 

At this point it should be noted that, to respect the (bosonic) canonical commutation relations \cite{mukhanov_winitzki_2007,Tolosa2022,Sanchez2022}, the mode $\psi_k^\mathrm{I}$ has to be normalized according to $\text{Wr}[\psi_k^\mathrm{I},(\psi_k^\mathrm{I})^*]=\mathrm{i}$. This condition is fulfilled by~${c_k = 1/\sqrt{2 \omega_k}}$, such that
\begin{equation}
    \vartheta_k = \arg(-r_k \e^{2 \im \omega_k \Delta \eta }),
    \label{eq:PhaseShiftInTermsOfROnly}
\end{equation}
which means that the phase-shift of the spectrum $S_k$ is equivalent to the phase-shift of the reflected wave relative to the incoming left-mover of the analogue scattering problem.

Interestingly, the spectrum $S_k$ can be deduced solely from the amplitude of the mode function in region III, i.e.\
\begin{equation}
\begin{aligned}
        \lvert \psi_k^\mathrm{III}(\eta(t)) \rvert^2 = 2 \lvert c_k \rvert^2 \left[\frac{1}{2} + N_k + \Delta N_k(\eta(t))\right],
        \label{eq:SpectrumWavefunction}
\end{aligned}
\end{equation}
showing that the coherent population $\Delta N_k$ stems from the interference of the incoming and reflected wave in region III. 

Let us furthermore highlight that properties of the scattering amplitudes $r_k$ and $t_k$ can be well understood from the shape of the scattering landscape, which in turn enables intuitive interpretations about the attributes of the excitation power spectrum $S_k$ in both the cold-atom as well as the cosmological context.  
For example, the broadness of the potential landscape generally corresponds to the sound-horizon at final times i.e. 
\begin{equation}
    \Delta \eta \equiv \etaf - \etai = \int_{t_\mathrm{i}}^{t_\mathrm{f}} \frac{\mathrm{d}t}{a(t)} = \int_{t_\mathrm{i}}^{t_\mathrm{f}} \mathrm{d}t \, \bar c(t),
\end{equation}
where we introduced $\bar c(t)$ as the speed of sound in the center of the trap (cref.\@ \cref{eq:SpeedOfSoundDefinition}) and used \cref{eq:ScaleFactorDefinition} together with $\mathrm{d} \eta  = \mathrm{d}t/a(t)$. It is therefore expected that acoustic maxima and minima in the fluctuation power spectrum of a BEC with time-dependent scattering length will occur and can be linked, via the quantum-mechanical analogy, to modes whose wavelengths satisfy extremal conditions that are formulated relative to the sound-horizon $\Delta \eta$ (similar to baryonic acoustic oscillations in the cosmic microwave background \cite{Dodelson2003}). In the cosmological context, these oscillations are known as Sakharov oscillations \cite{Sakharov1966,Hung2013,Viermann2022,Steinhauer2022}. 
Let us also appreciate that, in the cosmological context, conformal time $\eta(t)$ is equivalent to the so-called particle horizon in comoving coordinates \cite{Hobson2006,Weinberg2008}. At a given time $t$, it defines a region in space that could have been explored by a light-like particle within time $t$. Therefore one has a direct correspondence between the sound-horizon for phonons in the BEC and the particle horizon for excitations of a massless quantum field in a cosmological spacetime.

\subsubsection*{Examples for scattering potential}

In the following, let us discuss a couple of examples. The simplest possible case is a constant scale factor $a(t)=a_0$, implemented in a BEC through a time-independent scattering length, such that ${\lambda(t)= m/(\bar n_0 a_0^2)}$. 
In this case, the conformal time $\eta$ is related to the standard time $t$ by a simple affine transformation, ${t=a_0 (\eta-\eta_0)}$. From eq.~\eqref{eq:EffectiveMass} one can find that the potential is also constant
\begin{equation}
    V(\eta) = - a_0^2 m_\phi^2 - \xi D(D-1) \kappa.
\end{equation}
In the BEC, the effective mass corresponds to the case  ${m_\phi=\xi=0}$, such that $V(\eta)=0$ in this situation. More generally, one can consider a vanishing potential ${V(\eta)=0}$ and obtain the corresponding scale factor. For ${m_\phi=\xi=0}$, this implies
\begin{equation}
    0 = \left[\frac{D-1}{2}-q(t)\right] \dot a^2(t),
\label{eq:DiffEqnV0}
\end{equation}
where $q(t)$ is the deceleration parameter
\begin{equation}
    q(t) = - \frac{\ddot a(t) a(t)}{\dot a^2(t)},
    \label{eq:DecelerationParameter}
\end{equation}
which characterizes the cosmic acceleration or deceleration of the expansion of an FLRW universe. 

One solution of eq.~\eqref{eq:DiffEqnV0} is a constant scale factor, $\dot a(t) =0$, as we just saw.
Another possibility is the constant deceleration parameter $q(t)=(D-1)/2$, yielding a power-law scale factor,
\begin{equation}
    a(t) = a_0 \left(1+\frac{D+1}{2}H_0t\right)^\frac{2}{D+1},
\end{equation}
which corresponds to a radiation-dominated FLRW universe \cite{Chen2014}.

\section{Hallmark potential landscapes}
\label{sec:ScatteringPotentialsInTheQuantumSimulation}

In this section, we discuss the analogy between the time evolution of momentum modes of a real scalar field in a $(2+1)$-dimensional cosmology or, equivalently, of density perturbations in an interacting Bose-Einstein condensate, and the scattering problem in quantum mechanics in further detail. 
We explore the form of the effective scattering potential $V(\eta)$ for different classes of scale factors $a(\eta)$.
In cosmology, the scale factor is determined as a solution to the Friedmann equations, which depend on the specific energy composition of the universe. However, for our purposes it is convenient to take a more general approach and study classes of scale factors determined by a few parameters.

Let us stress that one can actually design the time-dependent scale factor $a(\eta)$ in the QFS, and thereby the scattering potential $V(\eta)$ to a large extent. Indeed, using  eq.\ \eqref{eq:ScaleFactorDefinition} in eq.\ \eqref{eq:EffectiveMass} yields, for $m_\phi=\xi=0$ and $D=2$,
\begin{equation}
    V(\eta) = \frac{m}{4 \bar n_0} \left[ \frac{7}{4} \frac{\dot \lambda^2(t(\eta))}{\lambda^3(t(\eta))} - \frac{\ddot \lambda(t(\eta))}{\lambda^2(t(\eta))} \right],
\end{equation}
where $\lambda(t)$ is the interatomic coupling strength. 
This correspondence allows experimental realizations of one-dimensional scattering problems for a wide class of (possibly idealized) potentials $V(\eta)$ using a Bose-Einstein condensate with variable scattering length.

\subsection{Emergence of singular contributions}

More generally, let us study the scattering potentials corresponding to the framework developed in Refs.~\cite{Sanchez2022,Tolosa2022,Viermann2022}, considering a cosmological expansion that starts at $t_{\text{i}}$ and ends at $t_{\text{f}}$, as discussed above. For an arbitrary scale factor in region II, we have from eq.~\eqref{eq:EffectiveMass} that the scattering potential reads
\begin{equation}
    \begin{aligned}
        V(\eta)  &=   \frac{1}{2} \frac{a^{\prime\prime}(\eta)}{a(\eta)} - \frac{1}{4} \left( \frac{a^{\prime}(\eta)}{a(\eta)} \right)^2 \\
        &= \frac{1}{4} \dot{a}^2(t(\eta)) + \frac{1}{2} \ddot{a}(t(\eta)) a(t(\eta)).
    \end{aligned}
    \label{eq:ScatteringPotential} 
\end{equation}

Let us first take the perspective of choosing a specific scale factor $a(t)$ and calculate the corresponding scattering potential.
In the QFS, the scale factor is a continuous time-dependent function 
\begin{equation}
    a(t) = \begin{cases}
        \ai & \text{for } t \leq t_\text{i}, \\
        \atwo(t) & \text{for }  t_\text{i} \leq t \leq t_\text{f}, \\
        \af & \text{for }  t \geq t_\text{f},
    \end{cases}
    \label{eq:ScaleFactorForm}
\end{equation}
where $\atwo(t)$ can be engineered to have arbitrary functional shape with boundary values $\atwo(\ti) = \ai$ and $\atwo(\tf) = \af$. 
The derivative of the scale factor has the shape
\begin{equation}
    \dot{a}(t) = \begin{cases}
        0 &\text{for }  t < \ti, \\
        \dot{a}_\text{II}(t) & \text{for } \ti \leq t \leq \tf, \\
        0& \text{for } t > \tf,
    \end{cases}
    \label{eq:ScaleFactorDerivativeForm}
\end{equation}
and, consequently, is not continuously differentiable at the boundaries, except if we choose $\atwo$ such that ${\dot{a}_\text{II}(\ti)= \dot{a}_\text{II}(\tf) = 0}$. These possible discontinuous transitions from stasis to dynamics 
lead to singular contributions to the potential landscape due to the last term in eq.~\eqref{eq:ScatteringPotential}.

Indeed, inserting \cref{eq:ScaleFactorForm,eq:ScaleFactorDerivativeForm} into \cref{eq:ScatteringPotential}, we find
\begin{equation}
    \begin{aligned}
        &V(\eta) = \left[ \frac{1}{4} \dot{a}_\text{II}^2(t(\eta)) + \frac{1}{2} \ddot{a}_\text{II}(t(\eta)) a(t(\eta)) \right] \\
        &\phantom{V(\eta) = [ }{ \times \Theta(t(\eta) - \ti) \Theta(\tf - t(\eta))} \\
        &+ \frac{1}{2} \dot{a}_\text{II}(t(\eta)) \dv{}{t} \left[\Theta(t(\eta) - \ti) \Theta(\tf - t(\eta)) \right] \atwo(t(\eta)),
    \end{aligned}
\end{equation}
where $\Theta$ is the Heaviside function. 
Therefore, it is convenient to split the scattering potential into regular and singular terms,
\begin{equation}
V(\eta) = \Vr(\eta) +  \Vs(\eta),
\label{eq:ScatteringPot_Split}
\end{equation}
with 
\begin{equation}
\begin{aligned}
    \Vr(\eta) &= \left[ \frac{1}{4} \dot{a}_\text{II}^2(t(\eta)) + \frac{1}{2} \ddot{a}_\text{II}(t(\eta)) \atwo(t(\eta)) \right] \\
    & \quad \times \Theta(\eta - \etai) \Theta(\etaf - \eta)
    \label{eq:RegularTerms_General}
\end{aligned}
\end{equation}
and
\begin{equation}
\begin{aligned}
    \Vs(\eta) 
    &= \frac{1}{2} \dot{a}_\text{II}(t(\eta)) \atwo(t(\eta)) \left[\delta(t(\eta) -\ti) - \delta(t(\eta) - \tf) \right] \\
    &= \frac{\mathcal{H}(\eta)}{2} \left[  \delta(\eta-\eta_{\text{i}}) -\delta{(\eta - \eta_{\text{f}})} \right],
\end{aligned}
\label{eq:IrregularTerms_General}
\end{equation}
where we introduced $\eta_{\mathrm{i},\mathrm{f}} = \eta(t_{\mathrm{i},\mathrm{f}})$ as well as the conformal Hubble rate ${\mathcal{H}(\eta) = a'(\eta) / a(\eta)}$. We also used that ${\delta(t(\eta)-t_*) = \delta(\eta-\eta_*)/a_*}$ with $a_*=a_\text{II}(\eta_*)$.

For the solution of the Schrödinger equation \eqref{eq:SchrodingerEq}, the singular terms of the scattering potential \cref{eq:IrregularTerms_General} at the boundaries imply discontinuities in $\psi_k'(\eta)$, i.e.
% (cf. \cref{Appendix:BoundaryConditions}) 
\begin{equation}
\lim_{\epsilon \to 0} \, [\psi_k'(\eta_{\text{i(f)}}+ \epsilon) -\psi_k'(\eta_{\text{i(f)}}-\epsilon)] = \varpm \frac{\mathcal{H}_{\text{i(f)}}}{2}\psi_k(\eta_{\text{i(f)}}),
\end{equation}
but not in $\psi_k(\eta)$, which is continuous in all the regions.
In this sense, the discontinuous transitions of the cosmological evolution at the boundaries of region II modify the boundary conditions for the scattering problem. The extent of this modification is determined by the respective conformal Hubble rate, that represents the speed towards (or away from) the discontinuous transitions at initial (final) time. 

\subsection{Exemplification}

As a first example, let us consider power-law expansions
\begin{equation}
    a_\mathrm{II}(t) = \left[1+ (q+1)H_0 t \right]^\frac{1}{q+1},
    \label{eq:PowerLawScaleFactorII}
\end{equation}
with constant deceleration parameter $q$ defined in \cref{eq:DecelerationParameter} and the analogue Hubble parameter $H_0$ set in accordance with the boundary conditions in \cref{eq:ScaleFactorForm} (see \cref{Appendix:HubbleinLab} for explicit examples details).

\begin{figure*}
\raisebox{-0.4cm}{\subfloat{\includegraphics[width=0.33\textwidth]{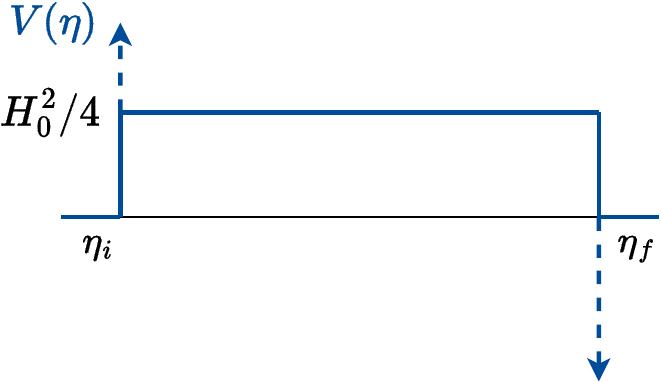}}}
\subfloat{\includegraphics[width=0.33\textwidth]{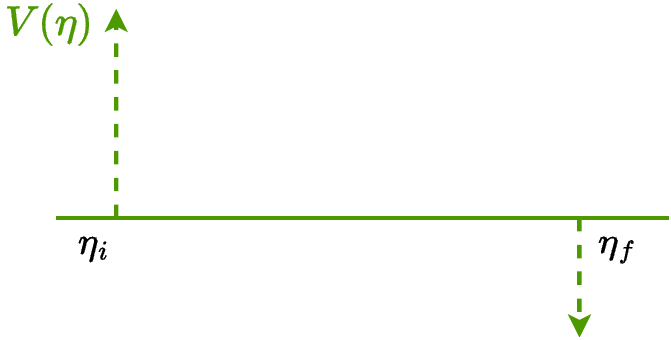}}
\raisebox{-0.06cm}{\subfloat{\includegraphics[width=0.33\textwidth]{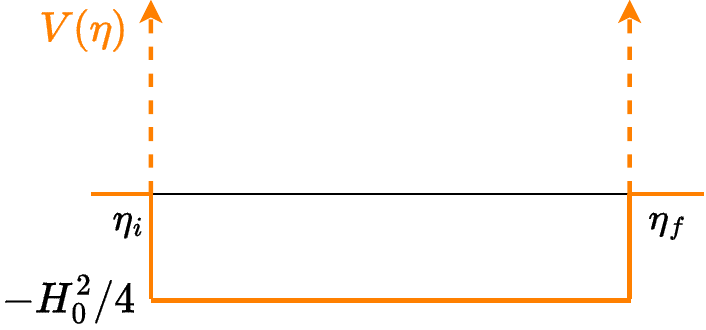}}}
\caption{Scattering potential landscapes that can be realized in a  quantum field simulation, which correspond to cosmological power-law expansions in $2+1$ dimensions ($q=0$ in the left image, $q=1/2$ in the central image) and, in the right image, the anti-bouncing scale-factor \eqref{eq:SymmetricExpansionContraction_ScaleFactor}.
The cosmological scenarios are also depicted in \cref{fig:ExpansionScenarios}.}
\label{fig:ScatteringPotentialsQuantumSimulator}
\end{figure*}

\begin{figure}
    \includegraphics[width=0.9\columnwidth]{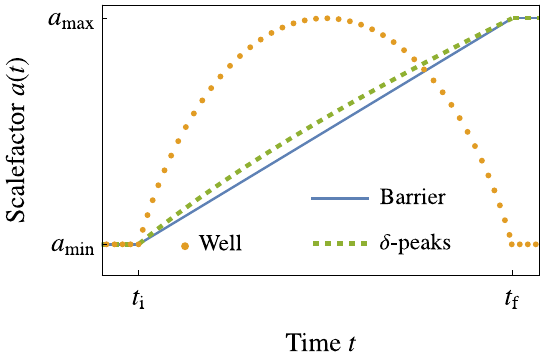}
    \caption{Cosmological expansion scenarios according to their analogue elementary scattering potentials.}
    \label{fig:ExpansionScenarios}
\end{figure}

For $q \neq 0$, one finds the scattering potential 
\begin{equation} 
\begin{aligned}
V(\eta)
=& \left( \frac{1}{4q^2} - \frac{1}{2q} \right) \frac{1}{\eta^2} \Theta(\eta - \eta_{\text{i}}) \Theta(\eta_{\text{f}} - \eta) \\
&+ \frac{H_0}{ 2 \left[a(\eta)\right]^q}\left[\delta(\eta - \eta_{\text{i}}) - \delta(\eta-\eta_{\text{f}})\right],
    \end{aligned}
\label{eq:ScatteringPotential_PowerLawExpansion_ConformalTimeQneq0}
\end{equation}
while for $q=0$ one has
\begin{equation} 
    \begin{aligned}
V(\eta)
=& \frac{H_0^2}{4} \Theta(\eta - \eta_{\text{i}}) \Theta(\eta_{\text{f}} - \eta) \\
& + \frac{H_0}{ 2}\left[\delta(\eta - \eta_{\text{i}}) - \delta(\eta-\eta_{\text{f}})\right],
    \end{aligned}
\label{eq:ScatteringPotential_PowerLawExpansion_ConformalTimeQeq0}
\end{equation}
which corresponds to the interesting case of a rectangular-barrier scattering potential; while $q=1/2$ corresponds to a pair of $\delta$-peaks (consider the left and central panel of \cref{fig:ScatteringPotentialsQuantumSimulator} together with \cref{fig:ExpansionScenarios}). More details on this class of cosmological evolutions and their scattering-analogous properties are to be found in \cref{subsec:PowerLaw}.

Let us now reverse the perspective by choosing a specific quantum mechanical scattering problem of interest and finding the corresponding cosmological scenario.
In that agenda, it is convenient to introduce
\begin{equation}
    y(\eta) \propto \sqrt{a(\eta)} 
\label{eq:Substitution_ParametricOscillator}
\end{equation} 
as a solution to the zero-energy Schrödinger equation
\begin{equation}
0=y''(\eta) - V(\eta) y(\eta) .  
\label{eq:ParametricOscillator}
\end{equation}
This solution is a zero-energy resonant state and follows directly from \cref{eq:ScatteringPotential}. In fact, it also exists in the general case of $D$ spatial dimensions, as we show in \cref{sec:ZeroEnergySolutions}. 
As \cref{eq:ParametricOscillator} is a well-studied differential equation (that can be converted into a Riccati equation), a scale factor $a(\eta)$ corresponding to a specific potential of interest is most likely to be found via this approach.
However, transforming $a(\eta)$ into an analytic expression for the scale factor as a function of cosmic time, $a(t)$, is non-trivial since the coordinate transformation between $t$ and $\eta$ may be transcendental. 
Indeed, this already occurs in elementary situations as we will see in the next section.

In summary, the strategy to design a specific potential landscape is to:
\begin{enumerate}
    \item Solve the differential equation \eqref{eq:ParametricOscillator} with ${V(\eta) = V_\text{r}(\eta)}$ for $y(\eta)$ to infer the scale factor $a(\eta)$.
    \item Compute the time derivatives of $a(\eta)$ at the boundaries of region II and infer the singular part $V_\text{s}(\eta)$ according to \cref{eq:IrregularTerms_General}. 
    \item Since the differential equation \eqref{eq:ParametricOscillator} is of second order, two integration constants are available to give the cosmological scenario desired properties like symmetry, extrema, acceleration, etc.
    In particular, the $\delta$-peaks at each boundary of region II can be tuned. Depending on the functional form of $a(\eta)$, one may even be able to set both singular contributions to zero. 
\end{enumerate}
Let us showcase with an instructive example the use of this strategy and the occurrence of the aforementioned features by designing a rectangular-well potential as the regular part of the scattering potential landscape (cf. right panel of \cref{fig:ScatteringPotentialsQuantumSimulator})
\begin{equation}
    \Vr(\eta) = - \frac{H_0^2}{4} \Theta(\eta - \eta_{\text{i}}) \Theta(\eta_{\text{f}} - \eta), 
    \label{eq:SquarePotentialWell_Definition}
\end{equation}
which leads to the solution to \cref{eq:ParametricOscillator}
\begin{equation}
    a(\eta) = a_{\text{max}} \cos^2 \left[ \frac{H_0}{2} \left( \eta + \varphi \right) \right],
    \label{eq:SquareWellGeneralScaleFactor}
\end{equation}
and is displayed in \cref{fig:ExpansionScenarios}.
The analogue cosmological scenario has singularities when $H_0 \left( \eta + \varphi \right) =\pi $. Therefore, we have to restrict the potential landscape of interest by choosing a suitable value of $\varphi$. Here, we choose symmetric anti-bounce, i.e. the decelerating half-cycle of \cref{eq:SquareWellGeneralScaleFactor}, as it maximizes the dynamical range accessible to the cosmological expansion under regularity constraints. With this choice, the scale factor takes the form
\begin{equation}
    a(\eta) = \frac{\amax}{2} \left\lbrace 1 + \cos \left [ H_0 \left(\eta - \frac{\eta_{\text{f}}+\etai}{2}\right) \right] \right \rbrace 
    \label{eq:SymmetricExpansionContraction_ScaleFactor}, 
\end{equation}
and the expansion duration in cosmic time 
\begin{equation}
        \Delta t =  \frac{\amax}{2} \bigg \lbrace \eta_{\text{f}}  -  \eta_{\text{i}} + \frac{2}{H_0} \sin \left[ \frac{H_0}{2} \left(\eta_{\text{f}}-\etai\right) \right] \bigg \rbrace, 
\label{eq:SymmetricExpansionContraction_Deltat}  
\end{equation}
where the product between depth and width of the potential is constrained by the expansion ratio
\begin{equation}
    \frac{H_0}{2} \left(\eta_{\text{f}}-\etai\right) = \arccos \left( 2 \frac{\amin}{\amax} - 1 \right),
    \label{eq:SquareWellEtaF}
\end{equation}
which follows directly from parametrizing $\amin = a(\eta_\text{i})$. 
As from \cref{eq:SquareWellEtaF}, $\tfrac{H_0}{2} \left(\eta_{\text{f}}-\etai\right) < \pi$, the regularity of the analogue cosmological scenario entails a trade-off between the broadness and depth of the potential landscape.

Using \cref{eq:SymmetricExpansionContraction_ScaleFactor,eq:SymmetricExpansionContraction_Deltat,eq:SquareWellEtaF}, we find the singular contributions to the scattering potential
\begin{equation}
    \begin{aligned}
    \Vs(\eta) 
    &=  \frac{H_0}{2} \sqrt{\frac{\amax}{\amin} - 1} \, [\delta(\eta-\eta_{\text{i}}) + \delta(\eta-\eta_{\text{f}})],
    \end{aligned}
    \label{eq:WellSingular}
\end{equation}
according to \cref{eq:IrregularTerms_General}.
\begin{center}
\begin{table*}[t]
\centering
\begin{tabular}{  M{8cm}  M
{8cm}  } 
\hline \hline
  
  \\ \textbf{Expansion scenario in the quantum field simulator} & \textbf{Scattering potential landscape} \\ \\
  \hline
 \\  Power-law scale factor ($q=0$)  \vbox{\begin{equation}\begin{split}
      a(t) &= 1+ H_0 t \\
      H_0 \left( \eta_\text{f} - \etai \right) &= \ln \left( \frac{\amax}{\amin} \right)
  \end{split}\nonumber\end{equation}} & Rectangular-barrier bounded by $\delta$-peaks  \vbox{\begin{equation} \begin{split}
      V_\text{r}(\eta) &= \frac{H_0^2}{4} \Theta(\eta-\eta_{\text{i}}) \Theta(\eta_{\text{f}} - \eta)\\
      V_\text{s}(\eta)&=\frac{H_0}{2}[\delta(\eta-\eta_{\text{i}}) - \delta(\eta-\eta_{\text{f}})]
  \end{split}\nonumber\end{equation}}
 \\ 
  \hline
\\  Power-law scale factor ($q=1/2$) \vbox{\begin{equation}
      \begin{split}
           a(t) &= \left(1+ \frac{3}{2}H_0 t \right)^\frac{2}{3}\\
          \frac{H_0}{2} (\etaf - \etai) &= \sqrt{\amax} - \sqrt{\amin}
          \nonumber
      \end{split}
  \end{equation}} &Double $\delta$-peak  \vbox{ \vspace{0.3cm} \begin{equation*}
  \begin{split} V_\text{r}(\eta) &= 0 \\
V_\text{s}(\eta) &= \frac{H_0}{2\sqrt{a_\text{{min}}}} \delta(\eta - \etai) - \frac{H_0}{2\sqrt{a_\text{{max}}}}  \delta(\eta-\etaf)
\end{split}
\end{equation*}} \\ 
  \hline
 \\ Symmetric anti-bounce \vbox{\begin{equation}
      \begin{split}
           a(\eta) &= \frac{\amax}{2} \left[ 1 + \cos \left( H_0\left(\eta - \frac{\eta_{\text{f}} + \etai}{2}\right) \right) \right] \\
           \frac{H_0}{2} \left( \eta_\text{f} - \etai \right) &= \arccos \left( 2 \frac{\amin}{\amax} - 1 \right)\nonumber
      \end{split}
  \end{equation}} & Rectangular-well bounded by repulsive $\delta$-peaks \vbox{\begin{equation}
  \begin{split}
    V_\text{r}(\eta) &= - \frac{H_0^2}{4}\Theta(\eta - \etai) \Theta(\etaf - \eta)\\
    V_\text{s}(\eta)&= \frac{H_0}{2} \sqrt{\frac{\amax}{\amin} - 1} [\delta(\eta-\eta_{\text{i}}) + \delta(\eta-\eta_{\text{f}})]\nonumber
    \end{split}
\end{equation}}\\ 
\hline \hline
\end{tabular}
\caption{Hallmark scattering potential landscapes with regular and singular terms, and their cosmological partners. 
The dynamic range of the cosmological scenario, given in terms of $\amax$ and $\amin$, sets the strength of the scattering problem \cite{Schwabl2007}, which can be identified via the product $H_0 \times (\etaf - \etai)$ and corresponds to the area of the scattering landscape in case of the rectangular-barrier and the rectangular-well.
Additionally, the temporal range $\Delta t$ of the cosmological evolution determines the analogue Hubble parameter $H_0$ which depends also on experimental parameters as described in \cref{Appendix:HubbleinLab}. 
Therein, the inverses of the normalization scales $a_{\mathrm{min},\mathrm{max}}$ correspond to speeds of sound in the BEC.}
\label{tbl:Comp}
\end{table*}
\end{center}  

\section{Solutions of scattering problem}
\label{sec:AnalyticalSolutionOfElementaryScenarios}

In this section, we present solutions of the quantum mechanical scattering problem of the potential landscapes shown in \cref{fig:ScatteringPotentialsQuantumSimulator} with the corresponding expansion scenarios being depicted in \cref{fig:ExpansionScenarios}.
The analytic expressions underlying the landscapes are collected in \cref{tbl:Comp}, where the analogue Hubble parameter $H_0$ differs between each case and is set according to \cref{Appendix:HubbleinLab}. 
Supplementary calculations are performed in \cref{Appendix:BoundaryConditions}.
 
The resulting excitation power spectra are compared in  \cref{fig:ComparisonSpectra} for a flat spatial geometry, with the corresponding reflection amplitudes being depicted in \cref{fig:PolarPlotRk}. The impact of spatial curvature is exemplified with the power spectrum resulting from a linear expansion in \cref{fig:GeneralSolutionPlots}.

\subsection{Rectangular-barrier bounded by \texorpdfstring{$\delta$}{TEXT}-peaks}

Consider first a power-law expansion \eqref{eq:PowerLawScaleFactorII} with $q = 0$ (see blue, solid curve in \cref{fig:ExpansionScenarios}), which corresponds to a matter-dominated universe in $(2+1)$-spacetime dimensions. 
In the scattering analogy, this cosmological background is mapped to a rectangular-barrier potential landscape of height $H_0^2/4$ bounded by a pair of attractive and repulsive $\delta$-peaks of equal strength $\mathcal{H}/2 = H_0/2$ (see leftmost panel in \cref{fig:ScatteringPotentialsQuantumSimulator}). 

In this context, one can distinguish between decaying modes $k$ with $E_k < H_0^2/4$, that are outside the horizon and tunnel through the barrier, and oscillating modes $k$ with $E_k > V_0$, which are inside the horizon. The distinction between these modes is equivalent to the effective mass analysis employed in Ref.~\cite{Sanchez2022}. 

The scattering amplitudes for the decaying (or super-horizon) modes are
\begin{equation}
    \begin{aligned}
    \frac{r_k}{t_k} &= - \im \frac{ \delta_k}{2} \e^{-\im \omega_k (\eta_\text{f}-\eta_\text{i})} \sinh\left[\Lambda_k (\eta_\text{f}-\eta_\text{i})\right], \\[5pt]
    r_k &= - \frac{\delta_k \e^{-2\im \omega_k (\eta_{\text{f}}-\eta_{\text{i}})}}{\sigma_k - 2\im \coth\left[\Lambda_k (\eta_\text{f}-\eta_\text{i})\right]},
    \end{aligned}
    \label{eq:PotentialBarrier_R/T_Decaying}
\end{equation}
and consequently, the spectrum components \eqref{eq:SakharovOffset} and \eqref{eq:SakharovAmplitude} are
\begin{equation}
\begin{aligned}
N_k & = \frac{\abs{\delta_k}^2}{4} \sinh^2\left[\Lambda_k (\eta_\text{f}-\eta_\text{i})\right], \\[5pt]
\frac{\Delta N_k^0}{N_k} &= \abs{\delta_k}^{-1} \sqrt{ \sigma_k^2 + 4 \coth^2\left[\Lambda_k (\eta_\text{f}-\eta_\text{i})\right]},
\label{eq:PotentialBarrier_Occupation_Decaying} 
\end{aligned}
\end{equation}
where the auxiliary variables $\sigma_k$ and $\delta_k$ are defined in \cref{app:SingularBarrier}. 

The scattering amplitudes and the spectrum components corresponding to the oscillating (or sub-horizon) modes can be obtained from \cref{eq:PotentialBarrier_R/T_Decaying,eq:PotentialBarrier_Occupation_Decaying}, respectively, by performing the change of variables $\Lambda_k\rightarrow\im\mu_k$, which also affects the definition of the auxiliary variables $\sigma_k$ and $\delta_k$ accordingly, as it is explained in \cref{app:SingularBarrier}.

In contrast to the decaying (or sub-horizon) modes, certain oscillating (or super-horizon) modes are not reflected at all. 
Considering the reflection amplitude for the oscillating modes, one finds that multiple zero-crossings of $r_k$ occur at wave numbers $k$ where $\mu_k(\eta_{\text{f}}- \etai) = n \pi$ for all $n \in \mathbb{N}$. 
In that limit, the cotangent in \cref{eq:PotentialBarrier_ScatteringAmplitudes} is discontinuous, such that the phase, defined in \cref{eq:SakharovPhaseShift}, decreases abruptly by a magnitude of $\pi$ in the sense that 
\begin{equation}
\lim_{\mu_k(\eta_{\text{f}}- \etai) \searrow n \pi} \vartheta_k - \lim_{\mu_k(\eta_{\text{f}}- \etai) \nearrow n \pi} \vartheta_k = - \pi.
\label{eq:PotentialBarrier_PhaseJumpMagnitude} 
\end{equation}
This behavior is visualized in the left panel of \cref{fig:PolarPlotRk} as a spiral trajectory of $r_k$ in the complex plane, which intersects itself at the origin with a tangent slope angle of $\pi$. 
The corresponding phase jumps and zero-crossings are shown in \cref{fig:ComparisonSpectra} and can be understood as resonant-forward-scattering in the scattering analogy:
Incoming waves from the right, which are fully transmitted, scatter back at the left potential-step where $\eta=\eta_\text{i}$. Upon back-scattering, the mode experiences a phase jump by $\pi$, which is well known from quantum scattering theory at potential steps \cite{Schwabl2007}.
Then, the mode back-propagates and destructively interferes fully with the right-mover such that no net-reflection occurs.

Finally, one can investigate the influence of finite spatial curvature by considering $\omega_k = - h(k)$ as dispersion relation, with $h(k)$ being defined in \cref{eq:LaplacianEigenvalue}. This leads to the graphs shown in \cref{fig:GeneralSolutionPlots}.
Here, we set the curvature scale $\sqrt{\abs{\kappa}}$ relative to the (sound) horizon scale $\pi / \Delta \eta$. 
The curves associated to spatially spherical and flat cases have a strong overlap on super-curvature scales ($k > \sqrt{\kappa}$), whereas they significantly differ in shape and magnitude on sub-curvature scales ($k < \sqrt{\kappa}$). However, the infrared limit of both cases is equal as the dispersion relation is non-gapped for both cases. This limit is determined by the zero-energy-resonance of the analogue scattering landscape and depends entirely on the boundary-values of the scale-factor (cf. \cref{sec:ZeroEnergySolutions}).
For the hyperbolic case, there is an effective mass gap $\kappa^2/4$, such that the corresponding curve can be obtained by a shift of the ($\kappa = 0$)-curve towards the infrared, and particle production is weaker in total. Furthermore, due to the mass-gap, the zero-energy-state of the analogue scattering landscape is non-visible to the modes propagating on a spatially hyperbolic FLRW spacetime. 

\subsection{Double \texorpdfstring{$\delta$}{TEXT}-peak landscape}
A power-law expansion \eqref{eq:PowerLawScaleFactorII} with exponent ${q=1/2}$ corresponds to a radiation-dominated universe in $2+1$ dimensions (see green, dashed curve in \cref{fig:ExpansionScenarios}). Here, the regular terms of the scattering potential vanish as a result of conformal symmetry and only singular contributions at the boundaries remain. Thus, particle production occurs only due to discontinuous transition into or away from the radiation-dominated phase. In the scattering analogy, this corresponds to a potential landscape which consists of an attractive and a repulsive $\delta$-peak, each contributing with a characteristic frequency given by the conformal Hubble rate $\mathcal{H}_\mathrm{i,f}$ at the respective transitions (see second panel in \cref{fig:ScatteringPotentialsQuantumSimulator}).
The scattering amplitudes are given by
\begin{equation}
    \begin{aligned}
        \frac{r_k}{t_k} =& \frac{\im \e^{-\im \omega_k (\eta_\text{f} -\eta_\text{i} ) }}{2 \Upsilon_{k;\text{i}} \Upsilon_{k;\text{f}}} \Big \lbrace \sin[\omega_k (\eta_\text{f}-\eta_\text{i})]\\
        &+ \Upsilon_{k;\text{i}} \e^{-\im \omega_k (\eta_\text{f}-\eta_\text{i}) }  - \Upsilon_{k;\text{f}} \e^{ \im \omega_k (\eta_\text{f}-\eta_\text{i}) }  \Big \rbrace
    \end{aligned}
    \label{eq:DeltaPeaks_RT}
\end{equation}
and
\begin{equation}
    \begin{aligned}
    r_k &= \e^{- 2\im \omega_k (\eta_\text{f}-\eta_\text{i}) } \\
    &\times\frac{\Upsilon_{k;\text{f}} \e^{ \im \omega_k (\eta_\text{f}-\eta_\text{i})} - \Upsilon_{k;\text{i}} \e^{-\im \omega_k (\eta_\text{f}-\eta_\text{i}) } - \sin[\omega_k (\eta_\text{f}-\eta_\text{i})]}{\sin[\omega_k (\eta_\text{f}-\eta_\text{i})] - \Sigma_{k;\text{i},\text{f}} \e^{-\im \omega_k (\eta_\text{f}-\eta_\text{i})} },
    \end{aligned}
\end{equation} 
where $\Upsilon_{k;\text{i,f}} = 2\omega_k/\mathcal{H}_{\text{i,f}}$, and $\Sigma_{k; \text{i},\text{f}} = \omega_k \Delta \eta - 2\im \Upsilon_{k;\text{i}} \Upsilon_{k;\text{f}}$ with $\Delta \eta = \eta_\mathrm{f}-\eta_\mathrm{i}$. 
Here, the reflection amplitude does not have any zero-crossings because this would require $\sin[\omega_k (\eta_\text{f}-\eta_\text{i})] = 0$ and $\cos[\omega_k (\eta_\text{f}-\eta_\text{i})] = 0$ simultaneously. In the corresponding polar plot of $r_k$, shown in the central panel of \cref{fig:PolarPlotRk}, the absence of zero-crossings is manifested in the spiral not intersecting itself at the origin. As can be deduced from \cref{fig:GeneralSolutionPlots,fig:ComparisonSpectra}, the phase varies continuously through these points and no zero-crossings in $N_k$ and $\Delta N_k^0$ occur.

\subsection{Qualitative analysis of further power-law expansions}

\begin{itemize}
    \item \emph{Case $q>1/2$.} Here, one can distinguish between hard modes $k$, $E_k > V_0$, and soft modes, $E_k\ll V_0$. Hard modes are mainly transmitted with a small amplitude for reflection and they are inside the horizon, also during the expansion, $\eta_{\text{i}} \leq \eta \leq \eta_{\text{f}}$, and, consequently, oscillating. In contrast, modes with $E_k \ll V_0$ are exponentially decaying in the region $\eta_{\text{i}} \leq \eta \leq \eta_{\text{f}}$ and are outside the horizon. When thinking in terms of time evolution from small to large $\eta$, some modes can actually exit the horizon and change from oscillating to decaying. 
    \item \emph{Case  $q\leq 1/2$ .} Here, all modes are always oscillating and they oscillate faster with respect to $\eta$ in the region $\eta_{\text{i}} \leq \eta \leq \eta_{\text{f}}$ than outside. In this region, the notion of a cosmological horizon does not make sense anymore. 
    \item \emph{Ultraviolet limit.} Now, consider an arbitrary power-law scale factor in the limit $k \rightarrow \infty$ or $E_k \rightarrow \infty$. Here, the scattering potential becomes irrelevant and there is no reflection, i.e. no particle production is expected for large wave numbers. This can also be seen from the in-spiralling trajectories in \cref{fig:PolarPlotRk}. 
\end{itemize}

\subsection{Rectangular-well potential}
\begin{figure*}\subfloat{\includegraphics[width=0.33\textwidth,valign=B]{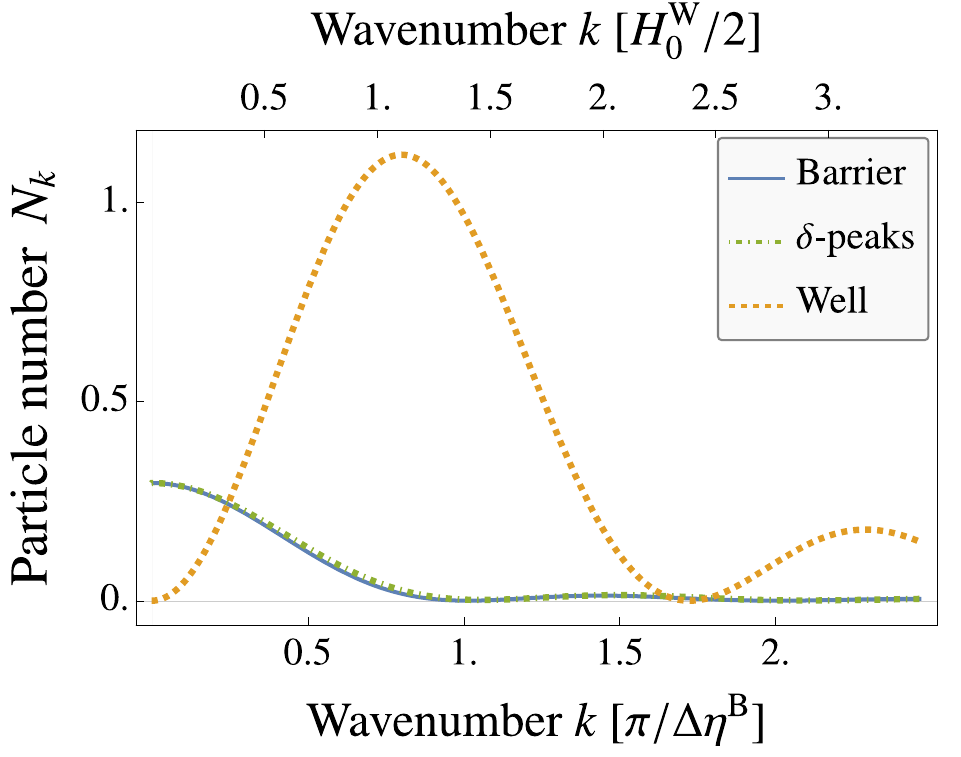}} \hfill
\subfloat{\includegraphics[width=0.33\textwidth]{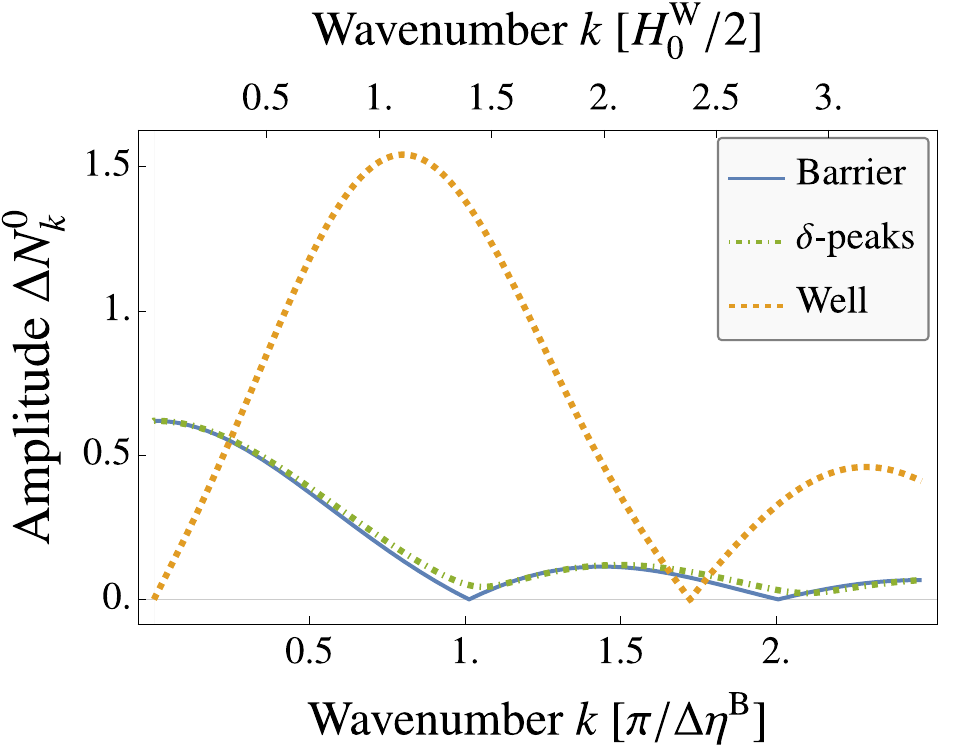}}%
\hfill
\subfloat{\includegraphics[width=0.33\textwidth]{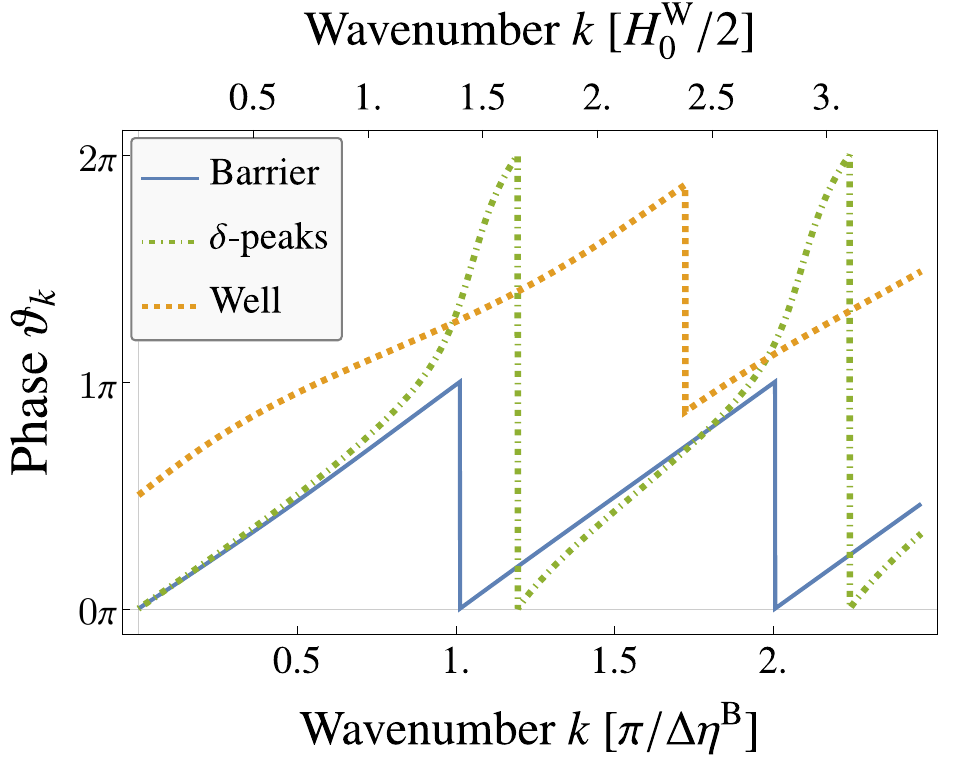}}
\caption{Particle number $N_k$, amplitude $\Delta N_k^0$ and phase $\vartheta_k$ (left to right image) of the excitation power spectrum corresponding to the expansion scenarios in \cref{fig:ExpansionScenarios} in case of a spatially flat spacetime ($\kappa = 0$).
$\Delta \eta^{\text{B}}$ is equal to the (sound) horizon $\etaf - \etai$ for the Barrier-scenario and $H_0^\text{W}$is the Hubble parameter for the Well-scenario. The only open parameter is the expansion strength which is set to $\amax/\amin = \sqrt{8}$. For clearer visibility, the phases $\vartheta_k$ were shifted by $\pi$.}
\label{fig:ComparisonSpectra}
\end{figure*}

\begin{figure*}
\raggedright
    \subfloat{
    \includegraphics[height=0.25\textwidth]{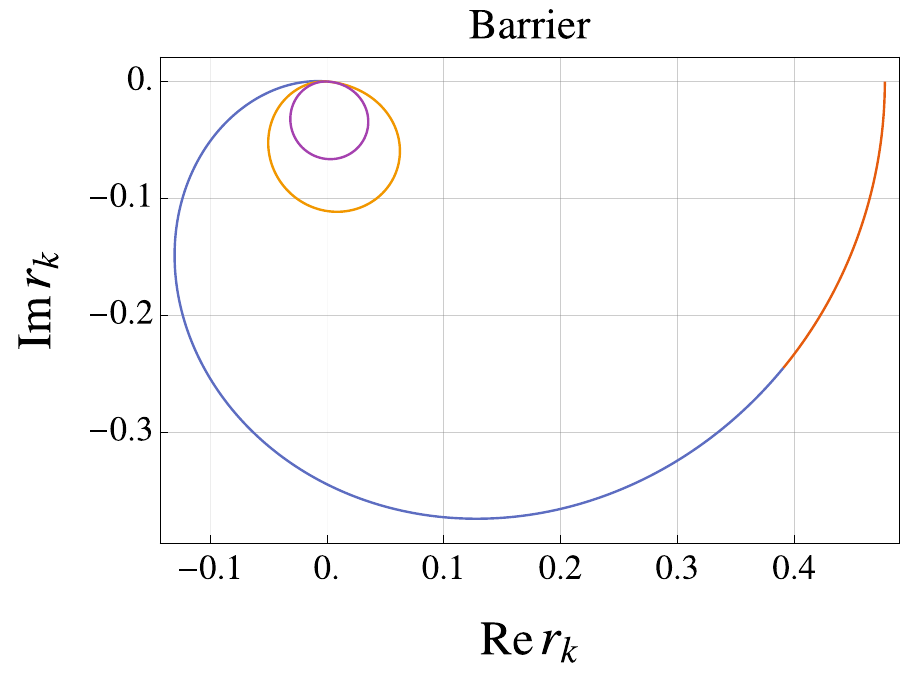}
    }
    \subfloat{ \hspace{0.5cm}
    \includegraphics[height=0.25\textwidth]{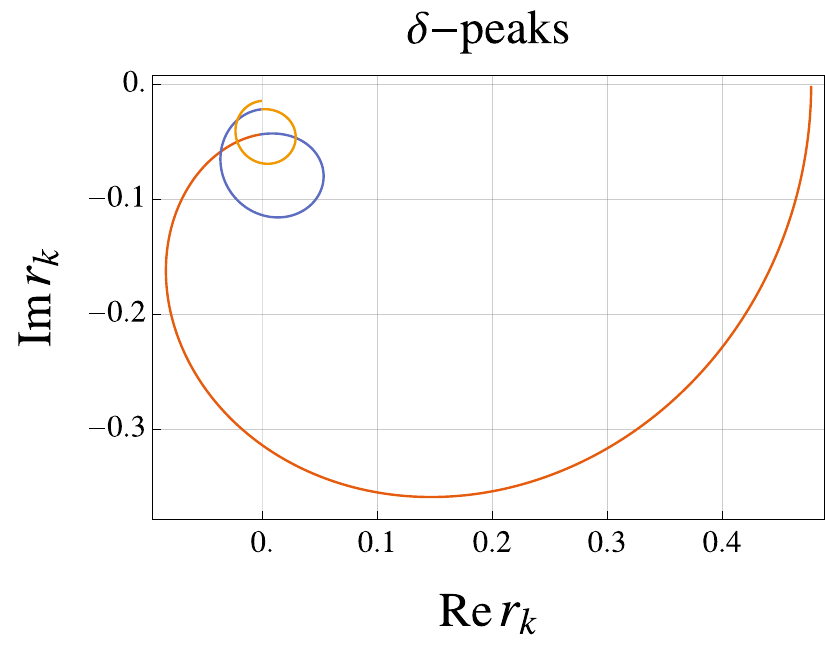}
    }
    \subfloat{ \hspace{0.5cm}
    \includegraphics[height=0.25\textwidth]{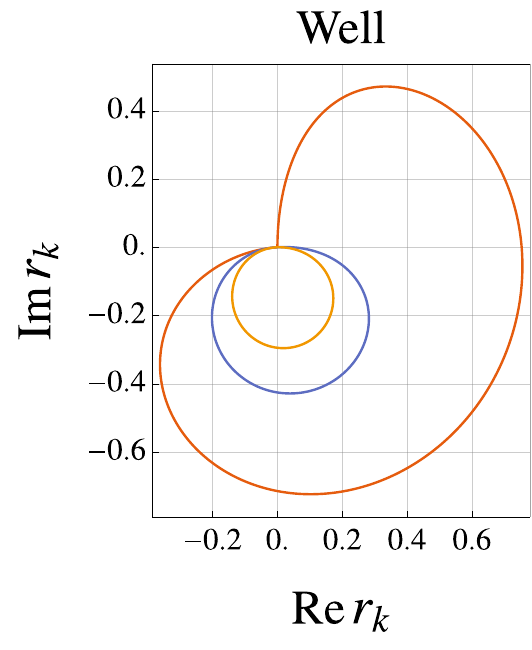}}
    \caption{Imaginary and real part of the reflection amplitude of the rectangular-barrier, $\delta$-peak landscape and rectangular-well (left to right image) for $\amax / \amin = \sqrt{8}$ and $\kappa = 0$. The wavenumber $k$ is the curve parameter and different colors indicate different reflection cycles. 
    Spiral trajectories intersecting themselves at the origin indicate phase jumps of magnitude $\pi$ which coincides with the tangent slopes at the origin. }
    \label{fig:PolarPlotRk}
\end{figure*}

\begin{figure*}
    \subfloat{\hspace{-5cm}
    \begin{minipage}[b]{\textwidth}
    \includegraphics[width=0.5\textwidth,valign=t]{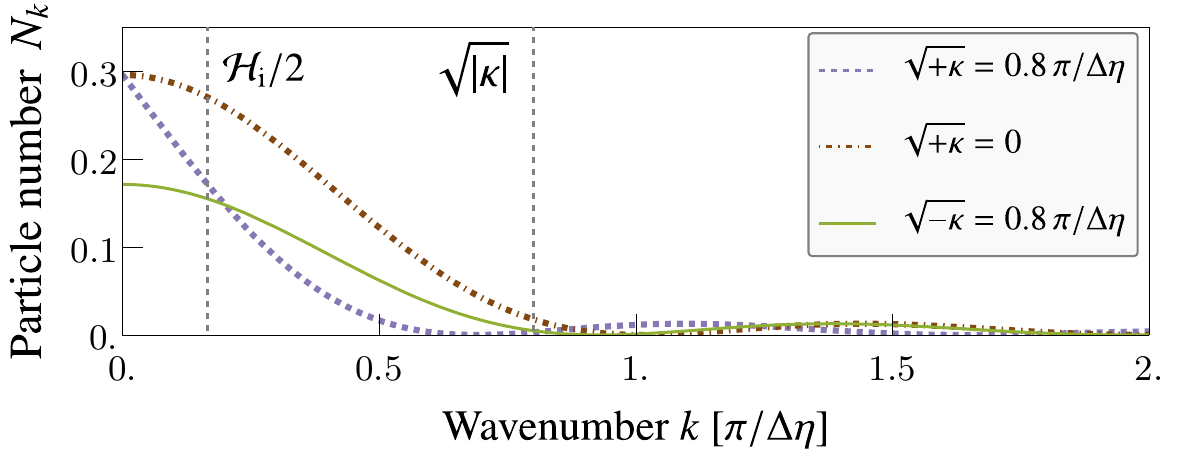}\vfill
    \includegraphics[width=0.5\textwidth]{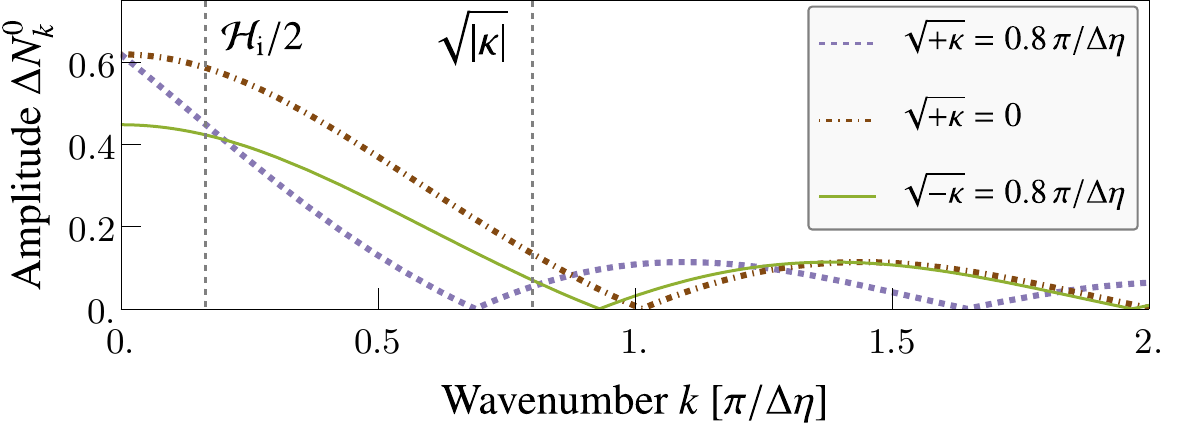}
    \end{minipage}
    }
    \subfloat{\hspace{-4.5cm}
    \includegraphics[width=0.4325\textwidth,valign=b]{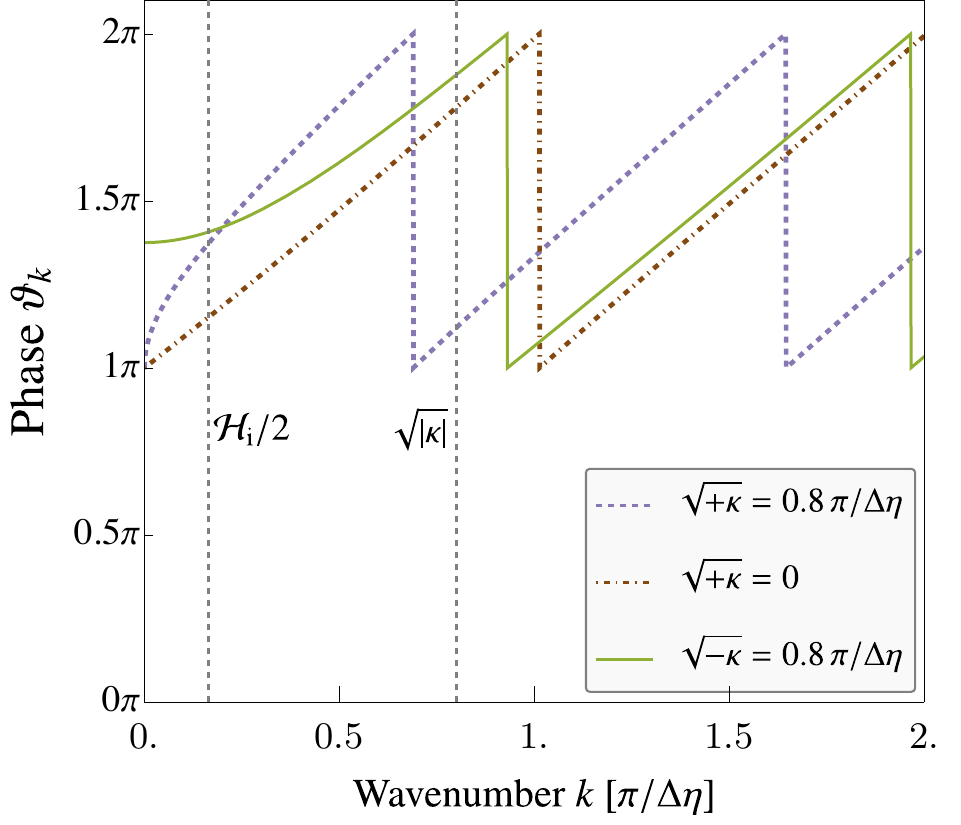}}
    \caption{Particle number $N_k$ (upper left panel), amplitude $\Delta N_k^0$ (lower left panel) and phase $\vartheta_k$ (right panel) of the rectangular-barrier landscape with $a_{\text{f}} / a_{\text{i}} = \sqrt{8}$ for different values of the spatial curvature $\kappa$. 
    Length scales are set relative to the comoving cosmological horizon $\Delta \eta = \etaf - \etai$ (which corresponds to the sound-horizon in the condensate). Acoustic peaks bounded by minima occur. The infrared regime of the spectrum is modified by an interplay between the $\delta$-peaks and curvature effects as indicated by the dashed gray lines. Due to the effective mass gap of the hyperbolic dispersion relation, particle production is weaker in total; also, the bound states at $k=0$ do not affect this case. As expected, for high momenta, curvature plays a only a small role.}
\label{fig:GeneralSolutionPlots}
\end{figure*}

The potential landscape of interest is visualized in the right panel of \cref{fig:ScatteringPotentialsQuantumSimulator} and consists of the regular part \eqref{eq:SquarePotentialWell_Definition} as well as the singular part \eqref{eq:WellSingular}. 
Here, both $\delta$-peaks are repulsive and of equal strength, i.e $\Hi \equiv \mathcal{H}(\eta_\text{i}) = - \mathcal{H}(\eta_{\text{f}})$ due to the symmetry of the analogue cosmological anti-bounce (orange, dotted curve in \cref{fig:ExpansionScenarios}).
In the opposite case of a symmetric bounce, both peaks would be attractive and of equal strength.  
The scattering amplitudes are 
\begin{equation}
\begin{aligned}
        \frac{r_k}{t_k} = - \frac{\im}{2} \e^{- \im \omega_k \text{$(\eta_{\text{f}}-\eta_{\text{i}})$}}&\bigg\{ \delta_k \sin[ \text{$(\eta_{\text{f}}-\eta_{\text{i}}) $} \text{$\mu_k $}]  \\
        &+ \frac{\Hi}{\omega_k} \cos[ \text{$(\eta_{\text{f}}-\eta_{\text{i}}) $} \text{$\mu_k $}]  \bigg\}.
\end{aligned}
    \label{eq:RkTkWell}
\end{equation}
and 
\begin{widetext}
\begin{equation}
    r_k = \e^{-2 \im \text{$\omega_k $} \text{$(\eta_{\text{f}}-\eta_{\text{i}})$} }  \frac{ \sin [ \text{$(\eta_{\text{f}}-\eta_{\text{i}}) $} \text{$\mu_k $}] \left(\Hi^2/4-\text{$\mu^2_k $}+\text{$\omega_k^2 $}\right)+ \Hi \text{$\mu_k $} \cos [ \text{$(\eta_{\text{f}}-\eta_{\text{i}}) $} \text{$\mu_k $}] }{\sin [ \text{$(\eta_{\text{f}}-\eta_{\text{i}}) $} \text{$\mu_k $}] \left[\text{$\mu_k $}^2+(\text{$\omega_k $}+\im \Hi^2/2)^2 \right]+2 \im \text{$\mu_k $} [\text{$\omega_k $}+\im (\text{$\Hi/2$})] \cos [ \text{$(\eta_{\text{f}}-\eta_{\text{i}}) $} \text{$\mu_k $}] }.
    \label{eq:RkWell}
\end{equation}
\end{widetext}
The structure of the rectangular-well landscape also admits resonant forward scattering modes, which are manifested as zero-crossings and phase jumps in the components of the particle spectrum depicted in \cref{fig:ComparisonSpectra,fig:GeneralSolutionPlots}. 
Thus, the curve of the corresponding reflection amplitude in the complex plane (shown in the right panel of \cref{fig:PolarPlotRk}) intersects itself at the origin - as for the rectangular-barrier scenario.

\subsection{Full transmission as desqueezing of the initial vacuum}
It is well known that cosmological particle production can be understood via squeezing and phase rotations of the Wigner ellipse in phase space \cite{Grishchuk1990,Albrecht1994,Wittemer2019,Martin2022,Brady2022}. 
In the present context, the emergence of fully transmitted modes in the analogue scattering landscape and the resulting zero-crossings in the excitation power spectrum can also be understood as a full desqueezing of the initial vacuum state. This type of desqueezing effect was already analyzed in a cosmological model involving an inflation-radiation-matter transition \cite{Grishchuk1990,Albrecht1994} where a significantly amount of desqueezing occurred during the matter-dominated epoch, thereby providing an alternative viewpoint on the emergence of Sakharov oscillations~\cite{Sakharov1966}.  

We will now show that a desqueezing effect occurs for all cosmological scenarios shown in \cref{fig:ExpansionScenarios}, where in particular, full desqueezing occurs when the analogue potential landscape (cref. \cref{fig:ScatteringPotentialsQuantumSimulator}) has a certain degree of symmetry, leading to a quantum recurrence of the initial vacuum state. Here, it is convenient to adopt the transfer-matrix-method (see for example \cite{Grosso2000,Markos2008} for an introduction), where the modefunction in region I is connected to the mode function in region III via 
\begin{equation}
    \left(\mqty{ a_k \mathrm{e}^{-\mathrm{i}\omega_k \eta_\mathrm{f}} \\ b_k \mathrm{e}^{\mathrm{i}\omega_k \eta_\mathrm{f}} }\right) = T    \left(\mqty{ c_k \mathrm{e}^{-\mathrm{i}\omega_k \eta_\mathrm{i}} \\ 0 }\right).
\end{equation}
Let us first consider a linear expansion which represents matter-domination in two spatial dimensions. The crucial observation is that the transfer-matrix of the analogue potential landscape can be factorized
\begin{equation}
\begin{aligned}
T &= T_\mathrm{step-up}\,  \times \,  M(-H_0/2) \, \times \, R(\eta_\mathrm{f}-\eta_\mathrm{i}) \\  &\qquad \times M(H_0/2) \, \times \, T_\mathrm{step-down},
\end{aligned}
\label{eq:FactorizedTransferMatrix}
\end{equation}
where 
\begin{equation}
    M(\mathcal{H})  = \frac{1}{\omega_k} \left(
\begin{array}{cc}
 \omega_k + \frac{\im}{2} \mathcal{H} & \frac{\im}{2} \, \mathcal{H} \\
- \frac{\im}{2} \, \mathcal{H} & \omega_k - \frac{\im}{2} \mathcal{H}\\
\end{array}
\right),
\label{eq:TransferDelta}
\end{equation}
describes the transfer along a repulsive ($+$) or attractive ($-$) delta-peak of strength $\mathcal{H}$ and 
\begin{equation}
    R(\eta) = \left( \mqty{ \mathrm{e}^{-\mathrm{i}\mu_k \eta} & 0 \\ 0 & \mathrm{e}^{\mathrm{i}\mu_k \eta}
    } \right)
\end{equation}
performs a phase-rotation with frequency ${\mu_k = \sqrt{\omega_k - H_0^2/4}}$. 
Furthermore, $T_\mathrm{step-up}$ and $T_\mathrm{step-down}$ transfer the mode function along an upwards (or downwards) potential step and are given by
\begin{equation}
\begin{aligned}
    T_\mathrm{step-up} &= \frac{1}{2\omega_k} 
    \left(\mqty{ \omega_k + \mu_k & \omega_k - \mu_k \\ \omega_k - \mu_k &\omega_k + \mu_k
    }\right), \\
    T_\mathrm{step-down} &= \frac{1}{2\mu_k} 
    \left(\mqty{ \mu_k + \omega_k & \mu_k - \omega_k \\ \mu_k - \omega_k & \mu_k + \omega_k 
    }\right).
\end{aligned}
\end{equation}
For fully transmitted modes, the squeezing by $T_\mathrm{step-up} \times M(-H_0/2)$ is entirely compensated by the desqueezing of $M(H_0/2) \times T_\mathrm{step-down}$ since a  full phase-rotation with $R(n \pi/\mu_k) = \mathds{1}_{2 \times 2}$ is performed between the two squeezing processes in \cref{eq:FactorizedTransferMatrix} and one has that 
$$T_\mathrm{step-up} \times T_\mathrm{step-down} = \mathds{1}_{2 \times 2} = M(-H_0/2) \times M(H_0/2). $$
The transfer matrices of the two remaining scenarios collected in \cref{tbl:Comp} and depicted in \cref{fig:ScatteringPotentialsQuantumSimulator} can be factorized similarly; however, in case of the rectangular-well, the presence of two equally repulsive $\delta$-peaks affects the condition for full transmission \eqref{eq:ZeroCrossingsWell} such that no full phase-rotation needs to be performed (in contrast to the rectangular-barrier). In case of the asymmetric $\delta$-peak-pair (central image of \cref{fig:ScatteringPotentialsQuantumSimulator}), the desqueezing effect does not lead to a quantum recurrence of the initial vacuum due to the asymmetry in attractiveness and repulsiveness of the $\delta$-peaks.

\section{Particle production in oscillating spacetimes as a scattering problem}
\label{sec:OscillatingSpacetimes}

Having examined the influence of non-differentiable junctures of scale factors on cosmological particle production, let us now construct a periodic cosmological scenario in which particle production occurs as a consequence of these junctures only.
The scattering analogue is the so-called Dirac comb \cite{Flügge1999,Griffiths1992} (also known as the Kronig-Penney model \cite{KronigPenney1931,Markos2008}), which is analytically solvable and serves as a minimal model to understand the energy spectrum of electrons in solids. Here, we will use it (consider \cref{fig:AlternatingDiracComb} for a visualization) to study the characteristic properties of particle spectra created by oscillating spacetimes  (displayed in \cref{fig:Comb_CyclicSpectrum,fig:Comb_SpectrumVsPhase}). 

\subsection{Construction of a Dirac comb}
The general solution to \cref{eq:ScatteringPotential} for a vanishing scattering potential is a radiation-dominated expansion (contraction) with 
\begin{equation}
    \begin{aligned}
        a_\text{e(c)}(t) &= a_\text{min} \left[1 \varpm \frac{3}{2} H_0 (t-t_\text{b}/2)\right]^{2/3}.
    \end{aligned}
\label{eq:combRadDomExpContr}
\end{equation}
Joining a contracting with an expanding period at $t_\text{b}/2$ via
\begin{equation}
    a_\text{rb}(t;t_\text{b}) =  a_\text{c}(t) \Theta(t_\text{b}/2 - t) + a_\text{e}(t) \Theta(t-t_\text{b}/2),
\end{equation}
yields a radiation-bounce that contributes with the singular term
\begin{equation}
\begin{aligned}
  &\frac{1}{2} \ddot{a}_\text{rb}(t;t_\text{b}) a_\text{rb}(t;t_\text{b}) \\
  &\supset \frac{1}{2} \left[\dot{a}_\text{e}(t_\text{b}/2)-\dot{a}_\text{c}(t_\text{b}/2)\right] a_\text{rb}(t_\text{b}/2) \delta(t-t_\text{b}/2) \\
    &= a_\text{min} H_0 \delta(\eta - \eta_\text{b}/2)
\end{aligned}
\label{eq:combPositiveSpike}
\end{equation}
to the scattering landscape, where
$\eta_\text{b}/2$ is defined to be the conformal time evaluated at $t_\text{b}/2$.

To construct a periodic lattice, we match $J$ symmetric radiation bounces in between the initial time $t_\text{i}$ and the final time $t_\text{f} = J t_\text{b} + t_\text{i}$. The times at which the bounces occur are
$t_\text{b}^{(j)} = (j + \tfrac{1}{2}) t_\text{b} + t_\text{i} $ and the matching times are $t_\text{m}^{(j)} = j t_\text{b} + t_\text{i}$ with $j \in \lbrace 1, \ldots, J-1 \rbrace$.
Around each matching time, the scale factor is
\begin{equation}
a_\text{rb}\left(t;t_\text{b}^{(j-1)}\right) \Theta\left(t_\text{m}^{(j)} - t\right) +  a_\text{rb}\left(t;t_\text{b}^{(j)}\right) \Theta\left(t - t_\text{m}^{(j)}\right) 
\end{equation}
with bounces at $t_\text{b}^{(j-1)}/2 + t_\text{i} $ and $t_\text{b}^{(j)}/2+ t_\text{i}$. Then, one has further contributions of the form
\begin{equation}
\begin{aligned}
&\frac{1}{2} \left[\dot{a}_\text{rb}\left(t_\text{m}^{(j)};t_\text{b}^{(j)}\right)
- \dot{a}_\text{rb}\left(t_\text{m};t_\text{b}^{(j-1)}\right)\right] \atwo\left(t_\text{m}^{(j)}\right) \delta\left(t-t_\text{m}^{(j)}\right) \\
&= \frac{1}{2} \left[\dot{a}_\text{c}\left(t_\text{m}^{(j)}\right)
- \dot{a}_\text{e}\left(t_\text{m}^{(j)}\right) \right] a_\text{e}\left(t_\text{m}^{(j)}\right) \delta\left(t-t_\text{m}^{(j)}\right) \\
&= -  H_0 \sqrt{\frac{a_\text{min}^3}{a_\text{max}}} \, \delta\left(\eta-\eta_\text{m}^{(j)}\right) 
\end{aligned}
\label{eq:combNegativeSpike}
\end{equation}
with the matching scale factor $a_\text{max} = a_\text{e}(t_\text{m}) = a_\text{c}(t_\text{m})$. 

It now becomes apparent that the $\delta$-peaks at the junctures of the radiation bounces are attractive since the conformal Hubble rate abruptly changes from negative to positive values there. These contributions can not be avoided since an abrupt transition from an expansion to a contraction (or vice versa) is necessary non-differentiable.
Thus, the constructed Dirac-comb is necessarily alternating in the bulk, which can be interpreted as a one-dimensional crystal. 

Finally, we have to account for $\delta$-peaks at initial and final times. From \cref{eq:IrregularTerms_General}, we find 
\begin{equation}
   - \frac{H_0}{2} \sqrt{\frac{a_\text{min}^3}{a_\text{max}}} \delta(\eta - \etai) - \frac{H_0}{2} \sqrt{\frac{a_\text{min}^3}{a_\text{max}}} \delta(\eta - \etaf) 
\end{equation}
as the boundary-contributions whose magnitude is necessarily smaller by a factor $1/2$ relative to the attractive bulk peaks, since the abrupt transition is one-sided at the outer boundaries, whereas it is two-sided in the bulk. 

\begin{figure*}
    \centering
    \includegraphics[scale=1]{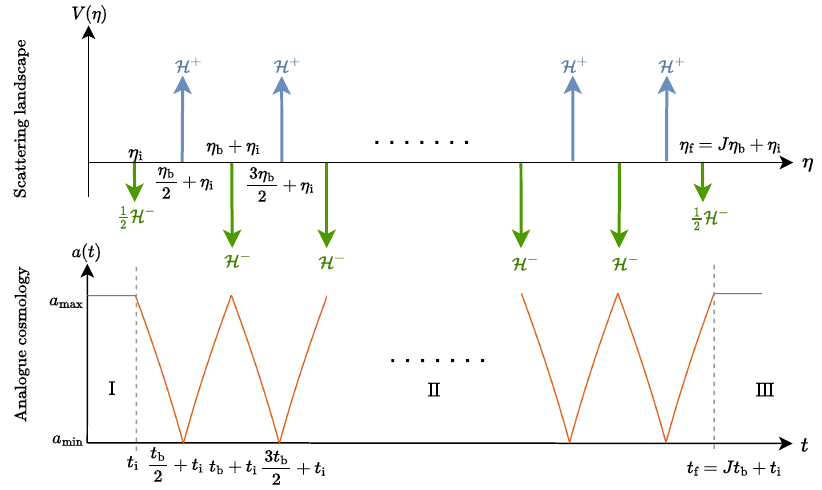}
    \caption{Alternating finite Dirac comb as the bulk of the scattering potential analogue to a radiation-dominated bouncing cosmology. Minima (Maxima) of the scale-factor in the form of cusps result in repulsive (attractive) $\delta$-peaks.
    The amplitudes of the comb  are related to the cosmological quantities via $\mathcal{H}^+ = H_0 a_\text{min}$ and $\mathcal{H}^- = \mathcal{H^+} \sqrt{a_\text{min}/ a_\text{max}} < \mathcal{H}^+$ }
    \label{fig:AlternatingDiracComb}
\end{figure*}

The cosmological evolution and the resulting lattice are depicted in \cref{fig:AlternatingDiracComb}. 

\subsection{Transfer matrix method}
We consider the scattering problem described in the previous section with the potential landscape depicted in \cref{fig:AlternatingDiracComb}, where $\mathcal{H}^+ = H_0 \, a_\text{min}$ and $\mathcal{H}^- = \mathcal{H^+} \sqrt{a_\text{min}/ a_\text{max}} $, and solve it analytically using the transfer matrix method. In this section, we sketch the structure of the calculation and give explicit expressions and further information in \cref{Appendix:DiracCombSupp}. 

\begin{figure}
    \includegraphics[scale=0.8]{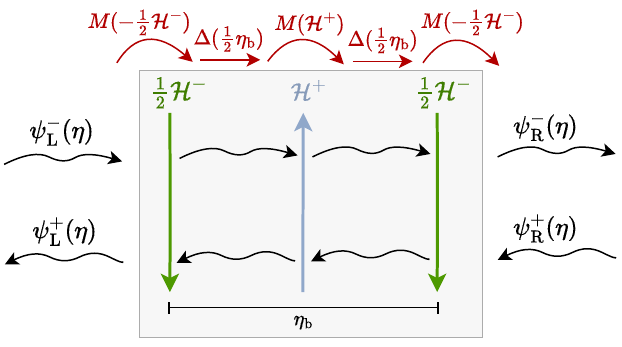}
    \caption{Transfer matrices along an elementary cell of an alternating Dirac-comb.}
    \label{fig:AlternatingDiracBlock}
\end{figure}

The transfer of the mode solution in region I to the mode solution in region III is described by the matrix equation 
\begin{equation}
\mqty( \psi_\text{III}^+(\eta_\text{f}) \\ \psi_\text{III}^-(\eta_\text{f})  ) = T^J(\mathcal{H}^-,\mathcal{H}^+,\eta_\text{b}) \mqty(\psi_\text{I}^+(\eta_\text{i})  \\ 0) , 
\end{equation}
where the superscripts $\pm$ indicate positive and negative frequency modes. The transfer along a single elementary cell of the lattice is 
\begin{equation}
    \begin{aligned}
       T = M\left(- \frac{\mathcal{H}^-}{2}\right) \, \Delta \left(\frac{\eta_\text{b}}{2}\right) \, M(\mathcal{H}^+) \, \Delta \left(\frac{\eta_\text{b}}{2}\right) M\left(- \frac{\mathcal{H}^-}{2}\right) ,
    \end{aligned}
    \label{eq:PeriodicTransferMatrix}
\end{equation}
where we recall that the transfer along an individual $\delta$-peak can be described by the matrix 
\begin{equation}
    M(\mathcal{H})  = \frac{1}{\omega_k} \left(
\begin{array}{cc}
 \omega_k + \frac{\im}{2} \mathcal{H} & \frac{\im}{2} \, \mathcal{H} \\
- \frac{\im}{2} \, \mathcal{H} & \omega_k - \frac{\im}{2} \mathcal{H}\\
\end{array}
\right),
\label{eq:MPlusMinus}
\end{equation}
that was already introduced in \cref{eq:TransferDelta} and the free-space transfer matrix by
\begin{equation}
    \Delta(\eta) = \mqty(\e^{-\im \omega_k \eta} & 0 \\ 0 & \e^{\im \omega_k \eta}),
    \label{eq:FreeSpaceTransferMatrix}
\end{equation}
which connects two $\delta$-peaks separated by a distance $\eta$. 
Here, we used the property $M(\mathcal{H}/2) \times M(\mathcal{H}/2) = M(\mathcal{H})$ in order to divide up the structure shown in \cref{fig:AlternatingDiracComb} into pure $N$-fold repetition of the elementary cell depicted in \cref{fig:AlternatingDiracBlock}.

In general, diagonal elements are associated with transmission, whereas the off-diagonal elements represent reflection. Since each degree of freedom of the diagonal or off-diagonal entries is associated to a direction in time,  its components satisfy $M^\pm_{11} = (M^\pm_{22})^*$ and $M^\pm_{12} = (M^\pm_{21})^*$, if the problem is time-reversal symmetric.
Furthermore, the unitarity of quantum mechanics constrains transfer matrices to be unimodular \cite{Grosso2000,Kalotas1991}, i.e.\ they are elements of the special linear group $\text{SL}(2,\mathbb{C})$.

Adopting the method described in Refs.~\cite{Kalotas1991,Markos2008}, we compute the matrix power of $T$ using the Cayley-Hamilton theorem. The characteristic equation for the unimodular operator $T$ is 
\begin{equation}
   T^2 - (\tr T)\, T + 1 = 0. 
   \label{eq:DiracCombCayleyHamilton}
\end{equation}
Iteration yields
\begin{equation}
    T^J = U_{J-1}\left(\frac{\tr T}{2}\right) T - U_{J-2} \left(\frac{\tr T}{2}\right),
    \label{eq:combMatrixPower}
\end{equation}
for $N \geq 2$, where $U_j(\tr T/2)$ are Chebyshev-polynomials of second kind and 
\begin{equation}
\begin{aligned}
\frac{\tr  T}{2} =& \cos(\omega_k \eta_\text{b}) + \bigg( \frac{\mathcal{H}^+}{2\omega_k} - \frac{\mathcal{H}^-}{2\omega_k}  \bigg) \sin(\omega_k \eta_\text{b}) \\
&- \frac{\mathcal{H^+}\mathcal{H^-}}{2\omega_k^2} \sin^2( \tfrac{1}{2} \omega_k \eta_\text{b}) \\
\equiv& \cos(q_k \eta_\text{b}).
\end{aligned}
\label{eq:AlternatingCombBandStructure}
\end{equation} 
The first two terms on the right-hand side of \cref{eq:AlternatingCombBandStructure} correspond to a chain of identical $\delta$-peaks of strength $(\mathcal{H}^+ - \mathcal{H}^-)$ separated by a distance $\eta_\text{b}$ (see for example~\cite{KronigPenney1931}). 
The third term is a quadratic modification stemming from an interaction between the two kinds of peaks, thereby resolving the sub-structure of the elementary cell.

\begin{figure}
\subfloat{\includegraphics[scale=0.74]{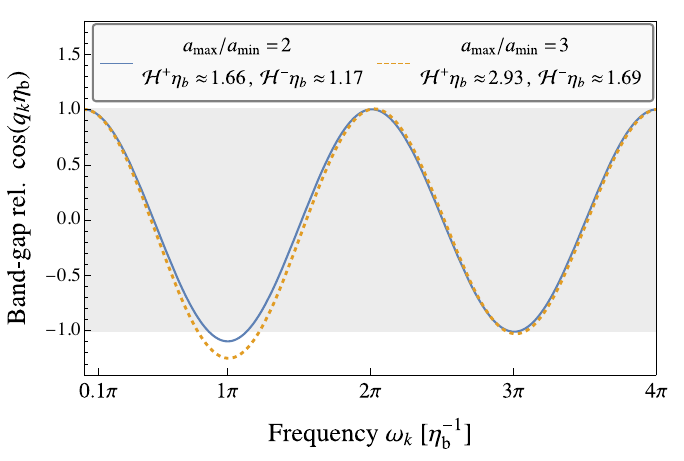}}
\caption{Trace of the elementary-cell transfer matrix with the associated transmission band highlighted in grey.
In the present context (consider \cref{Appendix:DiracCombSupp} for details), the graph can be interpreted as $\Re [1/t_k^{\text{(1)}}]$ where $t_k^{\text{(1)}}$ is the transmission amplitude of the elementary cell depicted in \cref{fig:AlternatingDiracBlock}.
Therefore the particle production peaks shown in \cref{fig:Comb_CyclicSpectrum,fig:Comb_SpectrumVsPhase} correspond to modes outside the transmission band, where the trace of the transfer matrix is extremal.}
\label{fig:Comb_TransmissionBand}
\end{figure}

One can now show that only modes which satisfy $\lvert \tr T \rvert \leq 2$ can propagate effectively through long systems which consist of identical copies of scattering potentials described by $T$ \cite{Markos2008}.
In the limit $J \to \infty$, these transmission-bands (cf. \cref{fig:Comb_TransmissionBand}) would correspond to the Bloch states, where the condition $\lvert \cos(q_k \eta_\mathrm{b}) \rvert  \leq 1 $ provides the energy-band structure and $q_k$ is the Bloch-momentum of the crystal.

In terms of this parametrization, the Chebyshev-polynomials from \cref{eq:combMatrixPower} read
\begin{equation}
    U_j(\cos(q_k \eta_\text{b})) = \frac{\sin[(j+1) q_k \eta_\text{b}]}{\sin  (q_k \eta_\text{b})},
    \label{eq:ChebyshevBlochVector}
\end{equation}
indicating exponential decay of modes into the potential for imaginary $q_k$, where 
$\lvert \tr T \rvert > 2$.

Finally, we can express the scattering amplitudes of the full chain in terms of the scattering amplitudes at an elementary cell $r_k^{\text{(1)}},t_k^{\text{(1)}}$. We have 
\begin{equation}
    \begin{aligned}
        \frac{r_k}{t_k} &= \frac{r_k^{(1)}}{t_k^{(1)}} \frac{\sin(J q_k \eta_\text{b})}{\sin (q_k \eta_\text{b})},  \\[3pt]
        t_k &= \frac{t_k^{(1)} \sin (q_k \eta_\text{b})}{\e^{-\im \omega_k \eta_\mathrm{b}} \sin (J q_k \eta_\text{b})  - t_k^{(1)} \sin( [J-1] q_k \eta_\text{b})}. 
    \end{aligned}
    \label{eq:MultipleCycle}
\end{equation} 

Using the scattering analogy of cosmological particle production, we can now also compute the particle spectrum of a periodically evolving cosmological spacetime in terms of the corresponding one-cycle expressions
\begin{equation}
    \begin{aligned}
        N_k &= N_k^{\text{(1)}} \frac{\sin^2(J q_k \eta_\text{b})}{\sin^2(q_k \eta_\text{b})},\\[3pt]
        \Delta N_k^0 &=  N_k^{1/2} \left[ 1 + N_k^{\text{(1)}}\frac{\sin^2(J q_k \eta_\text{b})}{\sin^2(q_k \eta_\text{b})} \right]^{1/2},
    \end{aligned}
    \label{eq:CombCyclicSpectra}
\end{equation}
and 
\begin{equation}
\vartheta_k = \arg \left[\frac{- r_k^{(1)} \sin(J q_k \eta_\mathrm{b}) \, \e^{-2\im \omega_k J \eta_\mathrm{b}}  }{\e^{-\im \omega_k \eta_\mathrm{b}} \sin(J q_k \eta_\mathrm{b}) - t_k^{(1)} \sin( (J-1) q_k \eta_\mathrm{b}) }\right].
 \label{eq:CombCyclicPhase}
\end{equation}

\begin{figure}
   \subfloat{
   \includegraphics[width=0.96\columnwidth]{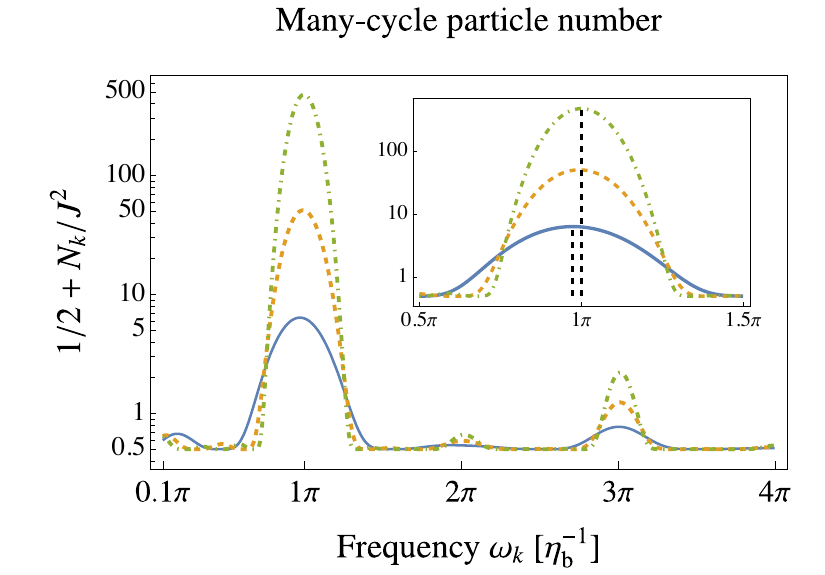}
   }
   \vfill
    \subfloat{
    \includegraphics[width=0.96\columnwidth]{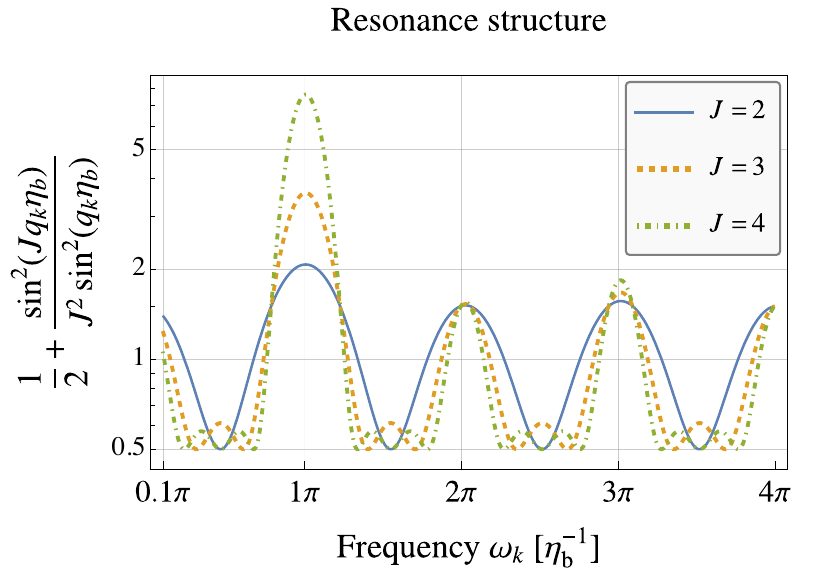}
     }  
    \vfill
        \subfloat{
    \includegraphics[width=0.96\columnwidth]{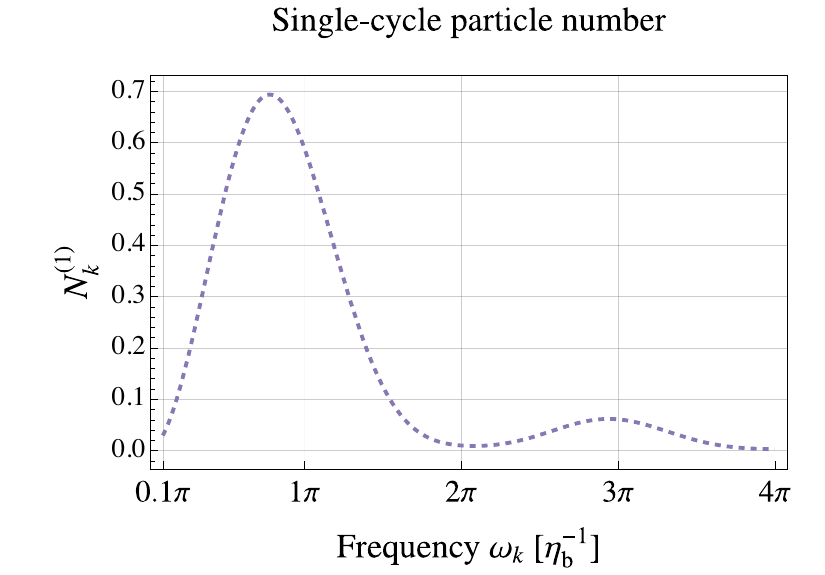}
    }
   \caption{Particle number $N_k$ (upper image) resulting from $J$ radiation bounces for bouncing ratio of $\amax/\amin = 3$. $N_k$ is compared to $U^2_{J-1}(\cos (q_k \eta_\mathrm{b}))$ (central image) which exhibits a typical resonance structure (cf. \cref{eq:ChebyshevBlochVector}) with peak locations independent of $J$. Since $N_k$ is the product of this Dirichlet-kernel with the single-bounce spectrum (lower image, $J=1$), the latter acts as an envelope for $N_k$. Thus, the peak locations shift with each cycle (see inset of top image) and converge to the corresponding peak locations of the Dirichlet-kernel with an increasing amount of cycles $J$.}
   \label{fig:Comb_CyclicSpectrum}
\end{figure}

\begin{figure}
    \centering
    \includegraphics[scale=0.73]{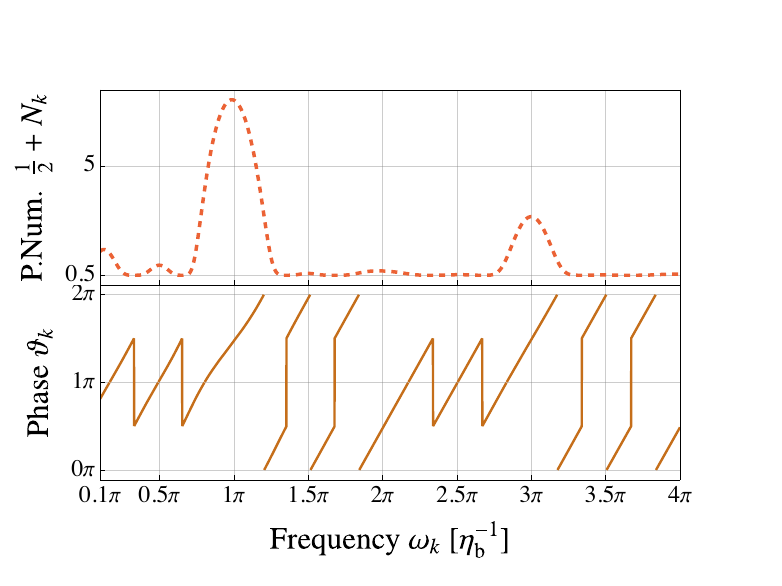}
    \caption{Structural relation between the phase of the power spectrum $\vartheta_k$ and the particle number $N_k$ produced by the oscillating cosmology analogue to the alternating Dirac comb shown in \cref{fig:AlternatingDiracComb} with $J=a_\mathrm{max}/a_\mathrm{min} = 3$. The dominant peaks in $N_k$ correspond to extrema in $\cos(q_k \eta_\text{b})$ around $\omega_k \eta_\text{b} \approx n \pi$ with $n \in \mathbb{N}\setminus \lbrace 0 \rbrace$. The more the mode lies outside the transmission band in \cref{fig:Comb_TransmissionBand}, the stronger the peak in $N_k$ is. 
    Also, the general trend that reflection becomes weaker with increasing mode energy can be withstood by secondary reflection resonances as can be seen by comparing the peaks at $\omega_k \eta_\text{b} \approx 3 \pi$ and $\omega_k \eta_\text{b} \approx 2 \pi$. Periods of continuous, close to linear evolution in the phase are connected by phase-jumps which corresponds to acoustic peaks $N_k$ being connected by zero-crossings. With \cref{eq:CombCyclicSpectra,eq:CombCyclicPhase} at hand, we can now also understand how many zero-crossings $N_k$ has at a given cycle number $J$: 
    The Chebyshev-polynomials $U^2_{J-1}(\cos q_k \eta_\text{b})$ that define the modulating factor in \cref{eq:CombCyclicSpectra} have $J-1$ roots in $\cos q_k \eta_b \in [-1,1]$ and monotonously increase/decrease outside given interval as a function of $\cos q_k \eta_\text{b}$.
    Now, between two adjacent dominant peaks, corresponding to extrema in $\cos q_k \eta_b$ (see \cref{fig:Comb_TransmissionBand}), the function $\cos q_k \eta_b$ runs through $[-1,1]$ once such that $(J-1)$ roots are taken by $N_k$ in addition to possible zero crossings stemming from the factor $N_k^{(1)}$ (see \cref{fig:Comb_CyclicSpectrum}). Equivalently, in the phase \eqref{eq:CombCyclicPhase}, whenever $r_k^{(1)}$ or $\sin(J q_k \eta_\mathrm{b})$ crosses zero, a jump with magnitude $\pi$ occurs. }
    \label{fig:Comb_SpectrumVsPhase}
\end{figure}

The factors underlying $N_k$ are displayed in \cref{fig:Comb_CyclicSpectrum}. Therein, the particle number after a single cycle $N_k^{(1)}$ acts as an envelope for the particle number after many cycles $N_k$ which exhibits resonant growth according to the modulating factor $\sin^2(J q_k \eta_\mathrm{b})/\sin^2(q_k \eta_b)$.  A characteristic pattern in the phase $\vartheta_k$ emerges and can be understood by comparison with the particle number as shown in \cref{fig:Comb_SpectrumVsPhase}. There, the zero-crossings and side maxima of $N_k$ for $J$ cycles are governed by the factor $\sin^2(J q_k \eta_\mathrm{b})$, and lead to an alternating pattern of growth and phase jump of magnitude $\pi$ in $\vartheta_k$. 

\section{Zero-energy resonances}
\label{sec:ZeroEnergySolutions}
\subsection{Emergence of irregular contributions}
The emergence of irregular contributions to the scattering landscape that were derived in \cref{fig:ScatteringPotentialsQuantumSimulator} is not a mere coincidence; in fact, these contributions affect the bound state structure of the potential such that there exists a zero-energy resonance. 

To realize this, it is instructive to investigate the stationary Schrödinger equation \eqref{eq:SchrodingerEq} for vanishing energy, $E_k=0$, in the case $m_\phi=\xi=0$, where the potential reads
\begin{equation}
V(\eta) = \frac{D-1}{2} \left[\frac{a''(\eta)}{a(\eta)} + \frac{D-3}{2}\left( \frac{a'(\eta)}{a(\eta)} \right)^2\right].
\label{eq:potentialMassless}
\end{equation}
This second order differential equation has two linearly independent solutions given by
\begin{equation}
    \varphi_1(\eta) = a^\frac{D-1}{2}(\eta), \qquad \varphi_2(\eta) = a^\frac{D-1}{2} (\eta)\tau(\eta), 
    \label{eq:ZeroEnergySolutions}
\end{equation}
where $\tau(\eta)$ is a solution to the differential equation $\mathrm{d}\tau/\mathrm{d}\eta = a^{1-D}(\eta)$. In the regions where $a(\eta)$ is constant, $\varphi_1(\eta)$ is also constant, while $\varphi_2(\eta)$ grows linearly with $\eta$. 
Accordingly, $\varphi_1(\eta)$ is the solution that satisfies the physical boundary conditions. 
If there are no cosmological singularities, $\varphi_1$ has no zero crossings or nodes, and is, therefore, the eigenfunction of eq.~\eqref{eq:SchrodingerEq} with lowest energy (in the quantum mechanical sense). 
In addition, this energy is vanishing, and it can be seen as a bound state precisely at the threshold to the scattering continuum. The fact that this zero-energy bound state solution always exists makes an interesting statement about the potential in eq.~\eqref{eq:potentialMassless}. It automatically realizes a potential with a bound state in resonance \cite{Senn1988,Boya2008}, which is conceptually very similar to Feshbach-resonances \cite{Feshbach1958,Pitaevskii2016} and was early envisioned by Wigner~\cite{Wigner1948}.

Constraining the analogue cosmological scenario to have no singularities restricts the analogue potential landscapes to those that have a single bound state at zero energy.
The peculiarity of this situation is highlighted by considering the case of purely attractive potentials, $V(\eta) < 0$, which generically have multiple bound states. According to \cref{eq:potentialMassless}, decelerating expansion scenarios are necessary to realize attractive landscapes in the most relevant case $D \leq 3$. 
Since such cosmological scenarios encounter a spacetime-singularity after a finite time, it thus becomes clear that attractive landscapes are incompatible with the demand of cosmological regularity. 
This can be explicitly seen in context of the rectangular-well scenario designed in \cref{sec:ScatteringPotentialsInTheQuantumSimulation}. 
There, the cosmological singularities were avoided by letting the dynamics abruptly cease, which resulted in a pair of repulsive $\delta$-peaks that shifted the lowest bound state of the scattering potential to zero energy (as we explicitly show in \cref{Appendix:BoundaryConditions}). 

\paragraph*{Generalization.}
More generally, one can consider the non-minimally coupled, massive case in which the zero energy Schrödinger equation assumes the form 

\begin{equation}
\varphi_1''(\eta) - \frac{V(\eta) - \xi D (D-1) \kappa}{1-\xi/\xi_\mathrm{c} } \varphi_1(\eta) = \frac{\xi_\mathrm{c} \, m_\phi^2}{\xi_\mathrm{c}-\xi} \varphi_1(\eta)^{\xi_\mathrm{c}+1}
\end{equation}
% \begin{equation}
%     V(\eta) = \left(1-  \frac{\xi}{\xi_\mathrm{conf}} \right) \frac{\varphi_1''(\eta)}{\varphi_1(\eta)} + \xi D (D-1) \kappa - \varphi_1(\eta)^{\xi_\mathrm{conf}} m^2
% \end{equation}
where $\xi_\mathrm{c} = (D-1)/(4D)$ is the conformal coupling constant and $\varphi_1(\eta)$ is the zero-energy solution to the minimally coupled, massless case given in \cref{eq:ZeroEnergySolutions}. 
Hence, in the massless case, $m_\phi=0$, one could still identify a modified scattering potential in which $\varphi_1(\eta)$ is a zero-energy solution. 

\subsection{Infrared limit of power spectrum}

Let us analyze in the following the existence of a homogeneous mode solution from a field theory perspective. 
In the minimally coupled massless case, the effective action \eqref{eq:EffAKG} for the homogeneous mode $\varphi$ can be written as 
\begin{equation}
    \Gamma_2[\varphi] = \frac{1}{2} \int \mathrm{d}t \, \mathrm{d}^D x \, a^D(t) (\partial_t \varphi(t))^2,
\end{equation}
resulting in the equation of motion
\begin{equation}
    \pdv[2]{}{\tau} \phi(\tau) = 0
\end{equation}
where we defined $\mathrm{d} t = a^D(t) \, \mathrm{d}\tau$ as in \cref{eq:ZeroEnergySolutions}. 
This behavior in the infrared limit resembles a well-known situation in cosmology where a mode equation of the same type as \cref{eq:SchrodingerEq} is equivalent to a conservation law if adiabatic super-horizon modes are considered \cite{Weinberg2003,Weinberg2008,MukhanovFeldmanBrandenberger1992,Lyth1984,Martin2005}. 
In the present case one can identify a conservation of
the conjugate momentum $\partial_\tau \phi = \delta \Gamma_2/\delta \dot{\varphi} = a^D(t) \dot{\varphi}(t)$.  

To establish a connection to the quantum-mechanical analogy, let us again introduce $\chi = a^{(D-1)/2} \varphi$. Here it is of convenient use that 
\begin{equation}
 a(\eta)^{D-1} \bra{0} [\chi'(\eta) - \tfrac{D-1}{2} \mathcal{H}(\eta) \chi(\eta)]^2 \ket{0}
\end{equation}
is a constant of motion, where $\ket{0}$ is the vacuum state. 

Evaluating this conservation law for the vacua in region I and region III of the $(2+1)$-dimensional setup introduced in \cref{subsec:DescriptionScatteringAnalogy}, one finds a non-trivial relation between the reflection amplitude at $k=0$ and the boundary values of the scale factor,
\begin{equation}
    \frac{a_\mathrm{i}}{a_\mathrm{f}} = \frac{1 + \abs{r_0}^2 - 2 \Re r_0}{1-\abs{r_0}^2}.
    \label{eq:r0Intermediate}
\end{equation}
We can further simplify this expression under use of Levinson's theorem \cite{Wellner1964,Boya2007,Boya2008}, whose one-dimensional version states that the zero-energy limit of the reflection amplitude is entirely determined by the number of bound states and by the phase difference between highly infrared and ultraviolet scatterers.
At this point, it is sufficient to use that it entails $r_0 \in \mathbb{R}$. Then, \cref{eq:r0Intermediate} simplifies to
\begin{equation}
r_0 = \frac{\af - \ai}{\af + \ai},
\label{eq:r0Final}
\end{equation}
which is in line with the reasoning employed in \cite{Senn1988} and the values displayed in \cref{fig:ComparisonSpectra}.
Hence, in the scattering analogy, the infrared limit of the power spectrum, is entirely fixed by the ratio $\ai/\af$ and does not depend on the cosmological history between these two values.
As a consequence, the infrared limits of the spectral components \cref{eq:SakharovOffset,eq:SakharovAmplitude,eq:SakharovPhaseShift} are 
\begin{equation}
    \begin{aligned}
        \lim_{k \to 0} N_k &= \frac{(\af - \ai)^2}{4\,  \af \, \ai}, \\
        \lim_{k \to 0} \Delta N_k^0 &= \frac{(\ai + \af)\lvert \af - \ai \rvert}{4\, \af \, \ai}, \\
        \lim_{k \to 0} \vartheta_k &= \begin{cases}
            \pi & \ai < \af ,\\
            0 & \ai > \af.
        \end{cases}
    \end{aligned}
\end{equation}

\section{Smooth transitions between stasis and dynamics in the quantum field simulation}
\label{subsec:SmoothSwitches}

Until now, we have considered that the derivative of the scale factor changes abruptly at $t_{\text{i}}$ and $t_{\text{f}}$. That is, in the QFS, the derivative of the scattering length departs from $0$ infinitely fast, which is an idealization that requires proper investigation. Hence, in this section, we study the effect of an experimentally realistic finite duration switch-on and -off of the derivative of the scale factor (and therefore, of the scattering length). The absence of discontinuities in the derivative will lead to a smoothing of the $\delta$-peaks in the scattering picture at $\eta_{\text{i}}$ and $\eta_{\text{f}}$. As it is known from the literature \cite{birrell_davies_1982,Visser1999,Glenz2009,Dominguez2023}, we expect the strength of particle production to decrease with an increasing amount of smoothing. 

\begin{figure}[t]
    \centering
    \includegraphics[width=0.38\textwidth]{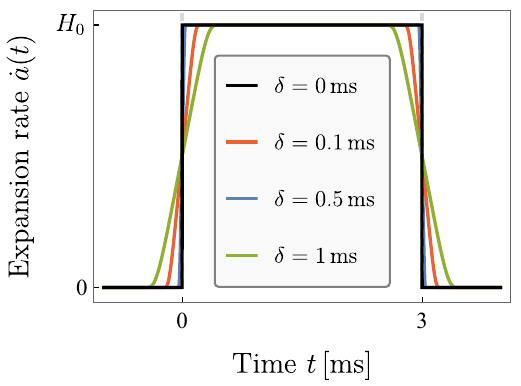}
    \caption{Time evolution of the derivative of the scale factor for different durations $\delta$ of the switch-on and -off. Note that the limiting case $\delta=0$ corresponds to the discontinuous behavior analyzed in previous sections.}
    \label{fig:SwitchOnAndOff}
\end{figure}

For implementing the smooth switch-on and -off, we will consider a $C^{\infty}$ step function $\Theta_{\sigma}(t)$ of width $\sigma$. 
The latter will be the duration of the switch-on, $\delta_{\text{on}}$, and the switch-off, $\delta_{\text{off}}$. 
In particular, we take a $C^{\infty}$ regularization which interpolates between $0$ and $1$ in the interval $(-\sigma/2,\sigma/2)$ by means of the function \mbox{$1/2+\tanh \left[\cot\left(\pi/2-\pi t/\sigma\right)\right]/2$} and is constant outside that interval. 
This precisely corresponds to the smoothing of the case $a(t) \sim t$. Furthermore, we will consider the particular situation in which $\delta_{\text{on}}=\delta_{\text{off}}=\delta$ for simplicity and symmetry reasons. 
Therefore, the switch-on process starts at $t_{\text{i}}-\delta/2$, until after a time $\delta$ the scale factor reaches the functional form $a(t) \sim t$ and its derivative becomes a constant. 
Then, at $t_{\text{f}}-\delta/2$, the switch-off starts, and it is at $t_{\text{f}}+\delta/2$ that the scale factor becomes constant again. 
This is illustrated in \cref{fig:SwitchOnAndOff} for different values of $\delta$. Note that the duration of the expansion is $(t_\text{f} - t_\text{i}) + \delta/2$, so that a process with a longer switch-off ends at the same final value of the scale factor.

\begin{figure*}[t]
    \centering
    \includegraphics[width=0.99\textwidth]{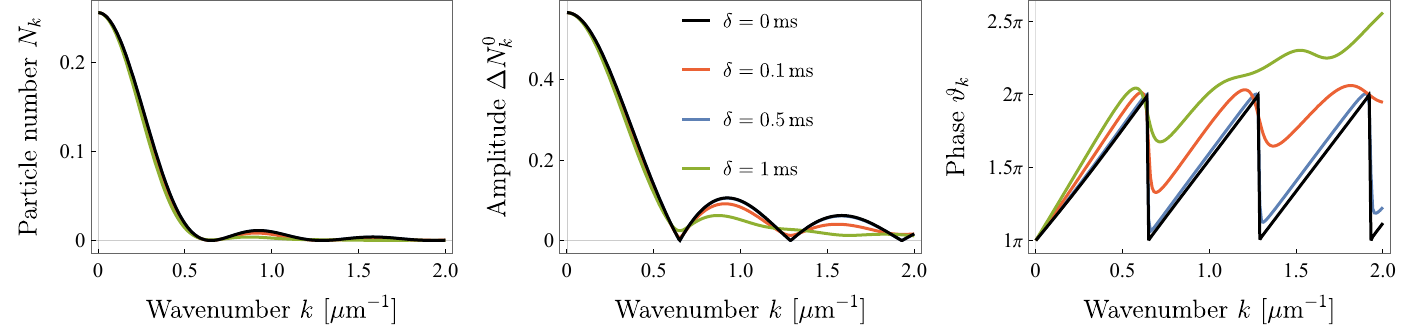}
    \caption{Effect of a finite duration $\delta$, symmetric switch-on and -off in $N_k$, $\Delta N_k^0$ and $\vartheta_k$. We considered an expansion of $\Delta t = 2 \, \text{ms}$ in which the scattering length is changed from $350 \, \text{a}_0$ to $50 \, \text{a}_0$, and a final speed of sound of $c_s=1.1 \mu\text{m}/ \text{ms}$ (consider \cref{Appendix:HubbleinLab} for characteristic scales in the experiment). }
    \label{fig:SpectraSwitch}
\end{figure*}

The spectrum of produced particles is obtained numerically for $\delta \neq 0$, and $N_k$, $\Delta N_k$ as well as $\vartheta_k$ are depicted in \cref{fig:SpectraSwitch}. We observe that the effect of the switch-on and -off is subtle on $N_k$, but it does have a larger impact on~$\Delta N_k$. The amplitude does not vanish anymore for $\delta \neq 0$, while interestingly the phase jump in~$\vartheta_k$, characteristic of the $q=0$ scale factor is smeared out for~$\delta$. Instead, a behavior corresponding to $q>0$ (accelerated expansion) can be observed for smooth switch-ons and -offs, i.e. oscillations in the phase instead of jumps. 
Of course, note that since the duration of the process slightly increases with $\delta$, there is an additional constant contributing to the slope of the phase $\vartheta_k$.

\section{Conclusions}
\label{Sec:Conclusions}

We have discussed here the very close parallels between three physics problems: cosmological particle production, quantum mechanical scattering and structure formation in Bose-Einstein condensates. 
For many examples, we found analytical solutions and solved others numerically. 
By building bridges between the different problems, we were able to transfer physics intuition and concrete results from one to another.

Probably the most exciting of the three concerns quantum fluctuations of a scalar field in the early universe. 
It is satisfying to see that for many cosmological histories, expanding and contracting, with singularities at early or late times, or periodic with a few cycles, one can find the spectrum of produced quantum excitations. 
A nice feature of the mapping to a quantum mechanical scattering problem is that it also works in the presence of cosmological singularities, like a big bang or big crunch, if there is an analytic continuation beyond it. 
This particular aspect warrants further study in future work. 

The conceptually simplest perspective is given by quantum mechanical scattering of a non-relativistic particle on a potential landscape. 
Being essentially a textbook problem, this allows to derive detailed insights, in both general and very concrete form.
By the very construction of the analogy, potential landscapes corresponding to the essential case of a minimally coupled, massless cosmological theory are such that they have a zero-energy or critical bound state.
This has by itself very interesting consequences for the transition amplitudes at large wavelenghts in the cosmology problem or the BEC.

The requirement to have such a critical bound state poses a restriction on the class of potentials that can be realized this way. Beyond that, however, there is much freedom, which means that many classical textbook problems, that mainly served for educational purposes so far, can now actually be realized in concrete laboratory experiments and provide an effective framework to intuitively understand emergent properties of (analogue) excitation spectra.

The quantum mechanical scattering perspective is also interesting for another, more conceptual reason. It may allow to address the inverse problem, where one aims to (re-)construct the scattering potential from the knowledge of scattering data \cite{GelfandLevitan1955,Marchenko1955,Chadan2013}, like transition and reflection amplitudes at different energies, as well as the bound state energies. This is another point that deserves further study in future work.

Finally, the third element in our trinity of problems, structure formation in an interacting Bose-Einstein condensate, has the big benefit of allowing a direct experimental realization. 
The most important ingredient here is the possibility to make the scattering length time-dependent, which in turn is possible through Feshbach resonances, where the s-wave scattering length is a function of the magnetic field strength. 
It is through this time dependence that different cosmological histories, or different quantum scattering potentials, can be realized. 
Through this non-equilibrium evolution one triggers the formation of spatial structure in a Bose-Einstein condensate that might initially have been in a homogeneous or harmonically trapped ground state, or in a thermal equilibrium state with small temperature. 
Different spatial wavelengths of this structure correspond to different energies in the quantum mechanical scattering problem. Seen as a quantum field simulator, it is therefore solving the one-dimensional stationary quantum scattering problem for all energies in parallel.

We have discussed here many examples for scattering potentials with their associated cosmological histories and realizations in the QFS. 
Some of them, like attractive or repulsive box potentials with Dirac peaks at the boundaries, could be solved analytically. 
It is even possible to realize potentials of purely distributional character (attractive and repulsive Dirac distribution peaks). In the laboratory system, these peaks arise from transitioning between a time-independent interaction strength to one with non-vanishing time-derivative. 
Of course, in a real experiment, such a transition might be modified on short time scales, for example by lag in the magnetic field, and we have consequently studied the effect of such modifications on the spectrum of produced excitations.

Dirac peaks can also be combined in a periodic sequence with a few periods. In terms of cosmology this would be a universe that expands and contracts periodically a few times. 
In the quantum mechanical scattering problem, this corresponds to a simple one-dimensional lattice structure with a few unit cells, and one can use the transfer matrix method to solve it.

In future work, we will extend our analysis to the ultraviolet regime in the QFS where the Bogoliubov modes acquire dispersion \cite{Bogolyubov:1947zz}, and thereby experience an emergent rainbow spacetime geometry \cite{Weinfurtner2006,Weinfurtner2009}. Furthermore we aim to include damping effects \cite{MicheliRobertson2022} which are expected to occur due to mode-interactions in the QFS. 

Taken together, we believe that with the present work we have taken a step forward to understand how modern quantum technology, using interacting Bose-Einstein condensates, can be used to perform simulations of two very interesting problems in the context of early time cosmology as well as quantum mechanics in one dimension.

\section*{Acknowledgements}
The authors thank Isabelle Bouchoule, Amaury Micheli, Markus Schröfl and Tim Stötzel for fruitful discussions. C.F.S acknowledges support by the Studienstiftung des Deutschen Volkes and the Deutsche Forschungsgemeinschaft (DFG) under Grant No 406116891 within the Research Training Group RTG 2522/1. Á. P.-L. acknowledges support through the MICIN (Ministerio de Ciencia, Innovación y Universidades, Spain) fellowship FPU20/05603 and the projects PID2020-118159GBC44, and PID2022-139841NB-I00, COST (European Cooperation in Science and Technology) Actions CA21106 and CA21136.
This work is supported by the Deutsche Forschungsgemeinschaft (DFG, German Research Foundation) under Germany’s Excellence Strategy, Grant No. EXC2181/1-390900948 (the Heidelberg STRUCTURES Excellence Cluster), and within the Collaborative Research Center SFB1225 (ISOQUANT, Project ID No. 273811115). 
N.L. acknowledges support by the Studienstiftung des Deutschen Volkes. 
 
\appendix 

\section{Characteristic scales and the analogue Hubble parameter}
\label{Appendix:HubbleinLab}

In the QFS, the cosmological scale factor $a(t)$ is simulated via a temporal variation of the s-wave scattering length $a_\mathrm{s}(t)$ via Feshbach resonances, which is for example accessible with the atomic species $^{39}\mathrm{K}$. 
In an effectively two-dimensional condensate, both quantities are related by
\begin{equation}
    a(t) = \frac{1}{\cs( a_\mathrm{s}(t))} = \left( \frac{m^3}{8 \pi \hbar^3 \omega_z  n_0^2 } \right)^{1/4} \frac{1}{\sqrt{\as(t)}},
    \label{eq:RelationScaleFactorScatteringLength}
\end{equation}
where $m$ is the atomic mass, $n_0$ is the density in the center of the trap and $\omega_z$ is the trapping frequency, which we assume to be given by $\omega_z = 2\pi \times 1500 \mathrm{Hz}$. 

To determine $n_0$, a speed of sound at some reference scattering length is used as an input parameter for the theory. It can be measured in the experiment through the propagation of density perturbations or Sakharov oscillations \cite{Viermann2022} and thereby serves as a conversion factor between cosmological scales (in natural units) and BEC-scales (in SI-units).
At $a_\mathrm{s, ref} = 50 \, \mathrm{a}_0$ (where $\mathrm{a}_0$ is the Bohr radius), a realistic value is $c_\mathrm{s,ref} = 1.1 \mu\mathrm{m}/\mathrm{ms}$.
With these values, the analogue Hubble parameter for the hallmark scenarios can be calculated:  
Consider first the power-law expansion \eqref{eq:PowerLawScaleFactorII}. At initial time $t_\text{i}$, \cref{eq:RelationScaleFactorScatteringLength} evaluates to
\begin{equation}
a(\ti) = [1 + (q+1) H_0 \ti]^{\frac{1}{q+1}}  = \frac{1}{\cs(a_\text{s}^{\text{max}})},
\label{eq:PolyRamps_HowToGetQIntermediate}
\end{equation}
where $a_\text{s}^{\text{max}} = \as(t_{\text{i}})$ is the scattering length at initial time.
Introducing the final value $a_\text{s}^{\text{min}} = a_\text{s}(\ti + \Delta t)$, we can combine \cref{eq:PolyRamps_HowToGetQIntermediate,eq:PowerLawScaleFactorII} to
\begin{equation}
1 + (q+1) H_0 \ti = \frac{(q+1) H_0 \Delta t}{\left(a_\text{s}^{\text{max}} / a_\text{s}^{\text{min}}  \right)^\frac{q+1}{2} - 1}.
\end{equation}
Inserting this expression into \cref{eq:PolyRamps_HowToGetQIntermediate}, one arrives at 
\begin{equation}
    H_0 = \frac{1}{\cs(a_\text{s}^{\text{max}})^{q+1}  \Delta t} \frac{\left(a_\text{s}^{\text{max}} / a_\text{s}^{\text{min}}\right)^\frac{q+1}{2} - 1}{(q+1)}.
\end{equation}

For the anti-bounce mimicking a rectangular well, the above procedure has to be adjusted due to the absence of an analytic expression for $a(t)$. Combining \cref{eq:SquareWellEtaF,eq:SymmetricExpansionContraction_Deltat}, we find here
\begin{align*}
    H_0 = \frac{\amax}{\Delta t} &\left[ \arccos \left( \frac{2\,  \amin}{\amax}-1 \right) 
    \right. \\
  &\left.+ 2  \sqrt{\frac{\amin}{\amax}- \left(\frac{\amin}{\amax}\right)^2 } \,\right]. 
\end{align*}
The laboratory parameters enter again through \cref{eq:RelationScaleFactorScatteringLength} with $\amax = 1/\cs(a_\text{s}^{\text{min}})$. 
Then, we finally find 
\begin{equation}
H_0 = \frac{\arccos \left( 2 \sqrt{\frac{a_\text{s}^{\text{min}}}{a_\text{s}^{\text{max}}}} - 1 \right) + 2  \sqrt{ \sqrt{\frac{a_\text{s}^{\text{min}}}{a_\text{s}^{\text{max}}}} - \frac{a_\text{s}^{\text{min}}}{a_\text{s}^{\text{max}}}}}{ \cs(a_\text{s}^{\text{min}}) \Delta t}. 
\end{equation}

Using these expressions for $H_0$ in combination with the conformal time formulae collected in \cref{tbl:Comp}, we find the numerical values shown in \cref{tbl:Scales}.

\begin{table}[ht]
    \centering
    \begin{tabular}{M{2cm} M{2cm}  M{2cm} }
     \hline \hline 
        Scenario & $H_0$ & $\Delta \eta$  \\ 
        \hline
        Barrier & $ 0.2 \, \mu\mathrm{m}^{-1} $  & $5.4 \, \mu \mathrm{m}$ \\
        $\delta$-peaks & $ 4.8 \, \mathrm{ms}^{1/2} \mu\mathrm{m}^{-3/2} $ & $5.0 \, \mu \mathrm{m}$ \\
        Well & $0.9\, \mu\mathrm{m}^{-1} $ & $4.4 \, \mu\mathrm{m}$ \\
        \hline \hline
    \end{tabular}
    \caption{Characteristic scales of the three hallmark scenarios for $a_\mathrm{s}^\mathrm{max} = 400 \, \mathrm{a}_0$, $a_\mathrm{s}^\mathrm{min} = 50 \, \mathrm{a}_0$,  $\Delta t = 3 \, \mathrm{ms}$.}
    \label{tbl:Scales}
\end{table}
There, the dimension of $H_0$ is different for the $\delta$-peak scenario because no global factor which absorbs the dimension of the analogue scale factor is employed in the parametrization \eqref{eq:PowerLawScaleFactor}.  
In contrast, using such a parameter in the Dirac-comb scenario (cf. \cref{eq:combRadDomExpContr}), we there find the parameters $H_0 = 0.8 \, \mathrm{ms}^{-1}$ and $\eta_\mathrm{b} = 3.6\,  \mu \mathrm{m}$ for $t_\mathrm{b} = 6\, \mathrm{ms}$, $a_\mathrm{s}^\mathrm{max} = 400 \, \mathrm{a}_0$, $a_\mathrm{s}^\mathrm{min} = 50 \, \mathrm{a}_0$.

\section{Supplementary calculations: Hallmark landscapes}
\subsection*{Scattering States}
\label{Appendix:BoundaryConditions}

\subsubsection*{Rectangular-barrier bounded by a repulsive and attractive $\delta$-peak }\label{app:SingularBarrier}
We want to find the reflection states of the potential landscape \cref{eq:ScatteringPotential_PowerLawExpansion_ConformalTimeQeq0}. To distinguish explicitly contributions from the barrier and the irregular $\delta$-peaks, we treat the barrier height $V_0 = H_0^2/4$ as a separate parameter and, thus, write
\begin{equation}
    \begin{split}
    V(\eta) =& V_0 \Theta(\eta-\eta_{\text{i}}) \Theta(\eta_{\text{f}} - \eta) \\
    &+ \frac{H_0}{2}[\delta(\eta-\eta_{\text{i}}) - \delta(\eta-\eta_{\text{f}})].
    \label{eq:PotentialBarrier_Definition}
    \end{split}
\end{equation}
Besides the expansion rate $H_0$, the scattering landscape is also determined by the number of e-folds. Explicitly one has for the present case that product between the width $(\eta_\text{f}-\etai)$ and the square root of the depth $\sqrt{V_0}$ of the barrier is directly constrained by the number of e-folds
\begin{equation}
\sqrt{V_0} \left(\eta_\text{f}-\etai\right) =  \frac{H_0}{2}  \int_{t_{\text{i}}}^{t_{\text{f}}} \frac{\dd t}{a(t)} = \frac{1}{2} \ln \left( \frac{\amax}{\amin} \right),
    \label{eq:PotentialBarrier_Width}
\end{equation}
which sets a dimensionless parameter that characterizes the strength of the scattering (or particle production) process.

\paragraph*{Decaying (or super-horizon) modes.}
\label{Appendix:PotentialBarrier_IrregularContributions_DecayingModes}

Let us discuss first the case $E_k<V_0$, i.e. decaying modes outside the horizon. In the scattering analogy, these modes correspond to tunneling waves. 
We make an Ansatz for the mode functions
\begin{equation}
   \psi_k(\eta) =  \begin{cases}
        c_k \e^{-\im \omega_k \eta} &\text{for}\quad \eta <\eta_\text{i}, \\[5pt] f_k \e^{-\Lambda_k \eta} + g_k \e^{\Lambda_k \eta} &\text{for}\quad \eta_\text{i} \leq \eta \leq \eta_{\text{f}}, \\[5pt]
        a_k \e^{-\im \omega_k \eta} + b_k \e^{\im \omega_k \eta} &\text{for}\quad   \eta_{\text{f}}<\eta,
        \end{cases} 
        \label{eq:PotentialBarrier_ModeSolutions_Decaying}
\end{equation}
with $\omega_k = \sqrt{E_k}$ and $\Lambda_k = \sqrt{V_0 - E_k}$.

Adjusting the boundary conditions for the mode derivative as described in the main text gives
 \begin{equation}
     \begin{aligned}
         c_k \e^{-\im \omega_k \eta_\text{i}} &= f_k \e^{-\Lambda_k \eta_\text{i}} + g_k \e^{\Lambda_k \eta_\text{i}}, \\[3pt]
 (H_0/2-\im \omega_k) c_k \e^{-\im \omega_k \eta_\text{i}} &= - \Lambda_k f_k \e^{-\Lambda_k \eta_\text{i}} + \Lambda_k g_k \e^{\Lambda_k \eta_\text{i}}.
     \end{aligned}
    \label{eq:PotentialBarrier_LeftBoundaryConditions_Decaying_Irregular}
 \end{equation}
Similarly, at $\eta = \eta_{\text{f}}$, 
\begin{equation}
    \begin{aligned}
        &f_k \e^{- \Lambda_k \eta_{\text{f}}} + g_k \e^{\Lambda_k \eta_{\text{f}}} = a_k \e^{-\im \omega_k \eta_{\text{f}}} + b_k \e^{\im \omega_k \eta_{\text{f}}}, \\[3pt]
        & -\Lambda_k f_k \e^{- \Lambda_k \eta_{\text{f}}} + \Lambda_k g_k \e^{\Lambda_k\eta_{\text{f}}} \\[5pt]
        &=  (H_0/2 - \im \omega_k ) a_k \e^{-\im \omega_k \eta_{\text{f}}} + (H_0/2 + \im \omega_k )  b_k \e^{\im \omega_k \eta_{\text{f}}} .
        \end{aligned} 
    \label{eq:PotentialBarrier_RightBoundaryConditions_Decaying_Irregular}
\end{equation}

Since the scattering amplitudes and, thus, also the spectrum only depend on ratios of the coefficients $a_k,b_k,c_k$, they are fully determined by this system of equations.
We find the amplitudes $a_k$ and $b_k$
\begin{equation}
 \begin{split}
    a_k =& \frac{\e^{\im \omega_k \eta_\text{f}}}{2} \Bigg[ \left(1+\frac{\Lambda_k + H_0/2}{\im \omega_k}\right) f_k \e^{-\Lambda_k \eta_\text{f}} \\
    & \quad \phantom{\frac{\e^{\im \omega_k\eta_\text{f}}}{2}}{+ \left(1-\frac{\Lambda_k - H_0/2}{\im \omega_k}\right) g_k \e^{\Lambda_k \eta_\text{f}} \Bigg]},\\
    b_k =& \frac{\e^{-\im \omega_k \eta_\text{f}}}{2} \Bigg[ \left(1-\frac{\Lambda_k + H_0/2}{\im \omega_k}\right) f_k \e^{-\Lambda_k \eta_\text{f}}\\
    & \quad \phantom{\frac{\e^{-\im \omega_k\eta_\text{f}}}{2}}{+ \left(1+\frac{\Lambda_k - H_0/2}{\im \omega_k}\right) g_k \e^{\Lambda_k \eta_\text{f}} \Bigg]},
\end{split} 
\end{equation}
 with
 \begin{equation}
 \begin{split}
     f_k &= \frac{\e^{-(\im \omega_k -\Lambda_k)\eta_\text{i}}}{2} \left(1-\frac{H_0/2 - \im \omega_k}{\Lambda_k}\right) c_k,\\
     g_k &= \frac{\e^{-(\im \omega_k +\Lambda_k)\eta_\text{i}}}{2} \left(1+\frac{H_0/2 -\im \omega_k}{\Lambda_k}\right) c_k.
 \end{split}
 \end{equation}

Therefore, 
 \begin{equation}
 \begin{split}
     a_k =& \e^{\im \omega_k(\eta_\text{f}-\eta_\text{i})} \Big \lbrace \cosh[\Lambda_k(\eta_\text{f}-\eta_\text{i})]\\
     &+ \frac{\im}{2} \sigma_k \sinh[\Lambda_k (\eta_\text{f}-\eta_\text{i})]\Big \rbrace c_k,\\
     b_k =&  -\frac{\im}{2} \delta_k \e^{-\im \omega_k(\eta_\text{f}+\eta_\text{i})}  \sinh[\Lambda_k (\eta_\text{f}-\eta_\text{i})] c_k
 \end{split}
 \end{equation}
with the auxiliary variables $\delta_k,\sigma_k$ defined as
\begin{equation}
    \begin{aligned}
    \sigma_k &= \frac{\Lambda_k^2 - \abs{\omega_k + \im H_0/2 }^2}{\omega_k \Lambda_k} = - 2 \frac{\omega_k}{\Lambda_k}, \\[5pt]
    \delta_k &=  \frac{\Lambda_k^2 + (\omega_k + \im H_0/2)^2}{\omega_k \Lambda_k} = \im \frac{H_0}{\Lambda_k}.
    \end{aligned}
    \label{eq:AuxiliaryVariables}
\end{equation}

\paragraph*{Oscillating (or sub-horizon) modes.}
\label{Appendix:PotentialBarrier_IrregularContributions_OscillatingModes}
Here, the Ansatz for the mode functions is the same as in \cref{eq:PotentialBarrier_ModeSolutions_Decaying} for the static regions I and III, but for region II now we have
\begin{equation}
    \psi_k (\eta) =\
        f_k \e^{-\im \mu_k \eta} + g_k \e^{\im \mu_k \eta} \quad\text{for}\quad \eta_{\text{i}} \leq \eta \leq \eta_{\text{f}},  
        \label{eq:PotentialBarrier_ModeSolutions_Oscillating}  
\end{equation}
with $\mu_k = \sqrt{E_k-V_0}.$ The results for the oscillating modes can be straightforwardly obtained through analytic continuation of the results for the decaying modes with $\Lambda_k= \im \mu_k$. 
However, we explicitly state them for completeness. At $\eta = \etai$, we have
 \begin{equation}
     \begin{aligned}
         c_k \e^{-\im \omega_k \eta_\text{i}} &= f_k \e^{-\im\mu_k \eta_\text{i}} + g_k \e^{\im\mu_k \eta_\text{i}}, \\[3pt]
 (H_0/2 - \im \omega_k) c_k \e^{-\im \omega_k \eta_\text{i}} &= - \im\mu_k f_k \e^{-\im\mu_k \eta_\text{i}} + \im\mu_k g_k \e^{\im\mu_k \eta_\text{i}},
     \end{aligned}
     \label{eq:PotentialBarrier_LeftBoundaryConditions_Oscillating_Irregular}
 \end{equation}
and at $\eta = \eta_{\text{f}}$
\begin{equation}
    \begin{aligned}
        &f_k \e^{- \im\mu_k \eta_{\text{f}}} + g_k \e^{\im\mu_k \eta_{\text{f}}} = a_k \e^{-\im \omega_k \eta_{\text{f}}} + b_k \e^{\im \omega_k \eta_{\text{f}}}, \\[3pt]
        & -\im\mu_k f_k \e^{- \im\mu_k \eta_{\text{f}}} + \im\mu_k g_k \e^{\im\mu_k\eta_{\text{f}}} \\[5pt]
        &=  (H_0/2 - \im \omega_k ) a_k \e^{-\im \omega_k \eta_{\text{f}}} + (H_0/2 + \im \omega_k )  b_k \e^{\im \omega_k \eta_{\text{f}}} .
        \end{aligned} 
        \label{eq:PotentialBarrier_RightBoundaryConditions_Oscillating_Irregular}
\end{equation}

For the amplitudes $a_k$ and $b_k$, we have
\begin{equation}
\begin{aligned}
    a_k =& \frac{\e^{\im \omega_k\eta_\text{f}}}{2} \Bigg[ \left(1+\frac{\im\mu_k + H_0/2}{\im \omega_k}\right) f_k \e^{-\im\mu_k \eta_\text{f}} \\ &\qquad \qquad + \left(1-\frac{\im\mu_k - H_0/2}{\im \omega_k}\right) g_k \e^{\im\mu_k \eta_\text{f}} \Bigg],\\
    b_k =& \frac{\e^{-\im \omega_k\eta_\text{f}}}{2} \Bigg[ \left(1-\frac{\im\mu_k + H_0/2}{\im \omega_k}\right) f_k \e^{-\im\mu_k \eta_\text{f}} \\ &\qquad \qquad + \left(1+\frac{\im\mu_k - H_0/2}{\im \omega_k}\right) g_k \e^{\im\mu_k \eta_\text{f}} \Bigg],
\end{aligned}
\end{equation}
with
\begin{equation}
 \begin{split}
     f_k &= \frac{\e^{-i(\omega_k -\mu_k)\eta_\text{i}}}{2} \left(1-\frac{H_0/2 - \im \omega_k}{\im\mu_k}\right) c_k,\\
     g_k &= \frac{\e^{-i(\omega_k +\mu_k)\eta_\text{i}}}{2} \left(1+\frac{H_0/2 -\im \omega_k}{\im\mu_k}\right) c_k.
 \end{split}
 \end{equation}

Therefore, 
\begin{equation}
 \begin{split}
     a_k =& \e^{\im \omega_k(\eta_\text{f}-\eta_\text{i})} \Big \lbrace \cos[\mu_k(\eta_\text{f}-\eta_\text{i})]\\
     &- \frac{1}{2} \sigma_k \sin[\mu_k (\eta_\text{f}-\eta_\text{i})]\Big \rbrace c_k,\\
     b_k =& \frac{1}{2} \delta_k \e^{-\im \omega_k(\eta_\text{f}+\eta_\text{i})}  \sin[\mu_k (\eta_\text{f}-\eta_\text{i})] c_k
 \end{split}
 \end{equation}
with the same auxiliary variables $\sigma_k,\delta_k$ as for the decaying case taking into account the change of variables $\Lambda_k \rightarrow \im \mu_k$
\begin{equation}
    \begin{aligned}
   \sigma_k &= \im \frac{\abs{\omega_k + \im H_0/2}^2 + \mu_k^2}{\omega_k \mu_k} = 2 \frac{\omega_k}{\mu_k}, \\[5pt]
    \delta_k &=  -\im\frac{(\omega_k + \im H_0/2)^2 - \mu_k^2}{\omega_k \mu_k} = \im \frac{H_0}{\mu_k}. 
    \end{aligned}
    \label{eq:GeneralizedAuxiliaryVariables}
\end{equation}
Note that now $\sigma_k$ is purely imaginary.

The scattering amplitudes for the oscillating (or sub-horizon) modes are
\begin{equation}
    \begin{aligned}
    \frac{r_k}{t_k} &=  \frac{ \delta_k}{2}  \e^{-\im \omega_k (\eta_\text{f}-\eta_\text{i})} \sin[\mu_k(\eta_\text{f}-\eta_\text{i})], \\[5pt]
    r_k &= - \frac{\delta_k \e^{-2\im \omega_k (\eta_\text{f}-\eta_\text{i}) }}{\sigma_k - 2  \cot[\mu_k (\eta_\text{f}-\eta_\text{i})]}.
    \end{aligned}
    \label{eq:PotentialBarrier_ScatteringAmplitudes}
\end{equation}
Then, the components of the spectrum are 
\begin{align}
    N_k &= \frac{\abs{\delta_k}^2}{4}  \sin^2[\mu_k (\eta_\text{f}-\eta_\text{i})], \\[5pt]
    \frac{\Delta N_k^0}{N_k} &= \abs{\delta_k}^{-1} \sqrt{ \abs{\sigma_k}^2 + 4 \cot^2[\mu_k (\eta_\text{f}-\eta_\text{i})].}  \label{eq:PotentialBarrier_Occupation_Oscillating}
\end{align}
Note that, in contrast to the super-horizon (or tunneling) regime, there exist zero-crossings of $N_k$ and $\Delta N_k^0$ when 
\begin{equation}
    \mu_k (\etaf - \etai) = \sqrt{E_k - V_0} = n \pi
    \label{eq:ZeroCrossingsBarrier}
\end{equation}
for $n \in \mathbb{N}$, i.e. incoming modes are fully transmitted.

\subsubsection*{Double $\delta$-peak-landscape}
\label{Appendix:DeltaPeaks}

In this section, we consider the potential landscape \eqref{eq:ScatteringPotential_PowerLawExpansion_ConformalTimeQneq0} for $q=1/2$ leading to an attractive and repulsive $\delta$-peak separated by a distance $\etaf - \etai$, i.e.
\begin{equation}
V(\eta) = \frac{\Hi}{2} \delta(\eta - \etai) - \frac{\Hf}{2} \delta(\eta-\etaf),
\label{eq:DeltaPeaks_Potential}
\end{equation}
where the magnitude of the peaks is
\begin{equation}
    \frac{\Hi}{2} = \frac{H_0}{2\sqrt{a_\text{i}}}, \quad \frac{\Hf}{2} = \frac{H_0}{2\sqrt{a_\text{f}}}.
    \label{eq:DeltaPeaksTransitionRates}
\end{equation}

It will turn out useful later that the separation ${\etaf - \etai}$ of the $\delta$-peaks can be expressed solely through these magnitudes, i.e. 
\begin{equation}
\etaf - \etai = \int_{t_{\text{i}}}^{t_{\text{f}}} \frac{\dd t}{a(t)}=  \frac{2}{H_0} (\sqrt{a_{\text{f}}} - \sqrt{a_{\text{i}}}) = 2 \left(\Hf^{-1} - \Hi^{-1}\right) . 
    \label{eq:DeltaPeaks_Width}
\end{equation}

The Ansatz for the mode functions is the same as the one for the oscillatory modes of the power-law expansion with $q=0$ given in \cref{eq:PotentialBarrier_ModeSolutions_Oscillating}.

The boundary conditions at $\eta = \eta_\text{i}$ are 
 \begin{equation}
     \begin{aligned}
         c_k \e^{-\im \omega_k \eta_\text{i}} &= f_k \e^{-\im\omega_k \eta_\text{i}} + g_k \e^{\im\omega_k \eta_\text{i}}, \\[3pt]
 (\Hi/2-\im \omega_k) c_k \e^{-\im \omega_k \eta_\text{i}} &= - \im \omega_k f_k \e^{-\im \omega_k \eta_\text{i}} + \im \omega_k g_k \e^{\im \omega_k \eta_\text{i}}.
     \end{aligned}
     \label{eq:PotentialBarrier_LeftBoundaryConditions_Deltas_Irregular}
 \end{equation}
and at $\eta = \eta_{\text{f}}$
\begin{equation}
    \begin{aligned}
        &f_k \e^{- \im \omega_k \eta_{\text{f}}} + g_k \e^{\im \omega_k \eta_{\text{f}}} = a_k \e^{-\im \omega_k \eta_{\text{f}}} + b_k \e^{\im \omega_k \eta_{\text{f}}}, \\[3pt]
        & -\im \omega_k f_k \e^{- \im \omega_k \eta_{\text{f}}} + \im \omega_k g_k \e^{\im \omega_k\eta_{\text{f}}} \\[5pt]
        &=  (\Hf/2 - \im \omega_k ) a_k \e^{-\im \omega_k \eta_{\text{f}}} + (\Hf/2 + \im \omega_k )  b_k \e^{\im \omega_k \eta_{\text{f}}} .
        \end{aligned} 
        \label{eq:PotentialBarrier_RightBoundaryConditions_Deltas_Irregular}
\end{equation}
For the final amplitudes, we have
\begin{equation}
\begin{split}
    a_k &= \frac{1}{4} \Bigg[\left(2-\im\frac{\Hf/2}{\omega_k}\right) \left(2+\im\frac{\Hi/2}{\omega_k}\right) \\ & \qquad \qquad  - \frac{\Hf/2}{\omega_k}  \frac{\Hi/2}{\omega_k} \e^{2\im \omega_k (\eta_\text{f}-\eta_\text{i})}\Bigg] c_k,\\
    b_k &= \frac{\im}{4} \Bigg[\left(2 + \im \frac{\Hi/2}{\omega_k}\right) \frac{\Hf/2}{\omega_k}  \e^{-2\im \omega_k \eta_\text{f}} \\ & \qquad \qquad - \left(2 + \im \frac{\Hf/2}{\omega_k}\right)  \frac{\Hi/2}{\omega_k} \e^{-2\im \omega_k \eta_\text{i}} \Bigg] c_k.
\end{split}
\end{equation}

Introducing the re-scaled frequency variables
\begin{equation}
\begin{split}
    \Upsilon_{k;\text{i}} = \frac{2 \omega_k}{\mathcal{H}_{\text{i}}}, \quad \Upsilon_{k;\text{f}} = \frac{2 \omega_k}{\mathcal{H}_{\text{f}}}, \\
    \Sigma_{k; \text{i},\text{f}} = \omega_k (\eta_\text{f}-\eta_\text{i})- 2\im \Upsilon_{k;\text{i}} \Upsilon_{k;\text{f}},
\end{split}
\end{equation} 
we find the mean occupation
\begin{equation}
\begin{split}
 N_k =& \frac{1}{4 \Upsilon^2_{k,\text{i}} \Upsilon^2_{k,\text{f}}} \bigg[ \Upsilon_{k;\text{i}}^2 + \Upsilon_{k;\text{f}}^2 + \sin^2[\omega_k (\eta_\text{f}-\eta_\text{i})] \\
         &- 2 \Upsilon_{k;\text{i}} \Upsilon_{k;\text{f}} \cos[2 \omega_k (\eta_\text{f}-\eta_\text{i}) ]\\
         &+ (\Upsilon_{k;\text{i}} - \Upsilon_{k;\text{f}}) \sin[2 \omega_k (\eta_\text{f}-\eta_\text{i}) ]  \bigg],
\end{split}
\end{equation}
which can be used to express the amplitude of oscillations as
\begin{equation}
    \begin{split}
        \Delta N_k^0 =& \frac{\sqrt{N_k}}{\abs{t_k}}\\
        =&  \frac{\sqrt{N_k}}{2 \Upsilon_{k;\text{i}} \Upsilon_{k;\text{f}}} \abs{\Sigma_{k;\text{i},\text{f}} \, \e^{-\im \omega_k (\eta_\text{f}-\eta_\text{i}) } - \sin[\omega_k (\eta_\text{f}-\eta_\text{i})] },
    \end{split}
\end{equation}
where we used that 
\begin{equation}
    t_k = -\im \e^{-\im \omega_k (\eta_\text{f}-\eta_\text{i})} \frac{2 \Upsilon_{k;\text{i}} \Upsilon_{k;\text{f}}}{\Sigma_{k;\text{i},\text{f}}  \e^{-\im \omega_k (\eta_\text{f}-\eta_\text{i}) } - \sin[\omega_k (\eta_\text{f}-\eta_\text{i})]}.
\end{equation} 
Here, there are no zero-crossings in $N_k$ and $\Delta N_k^0$ as this would require both $\cos[\omega_k (\eta_\text{f}-\eta_\text{i})]$ and $\sin[\omega_k (\eta_\text{f}-\eta_\text{i})]$ to vanish.

    \begin{figure*}
    \vfill
    \subfloat{\hspace{-5cm}
    \begin{minipage}[b]{\textwidth}
    \includegraphics[width=0.5\textwidth,valign=t]{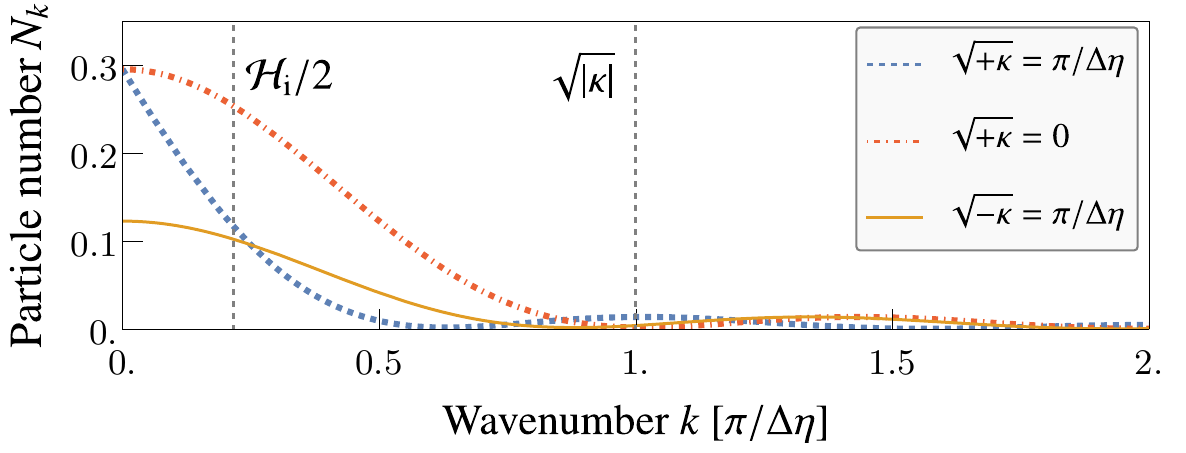}\vfill
    \includegraphics[width=0.5\textwidth]{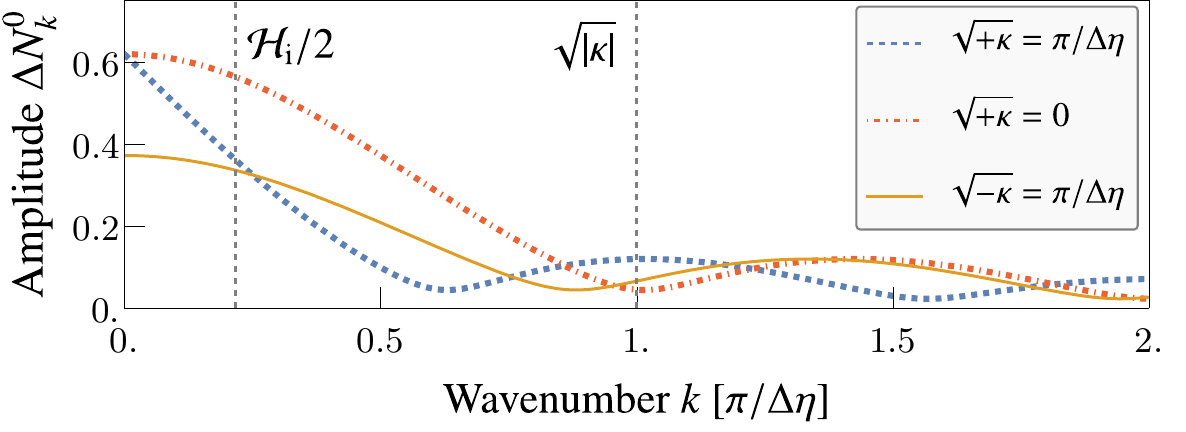}
    \end{minipage}
    }
    \subfloat{\hspace{-4.5cm}
    \includegraphics[width=0.4325\textwidth,valign=b]{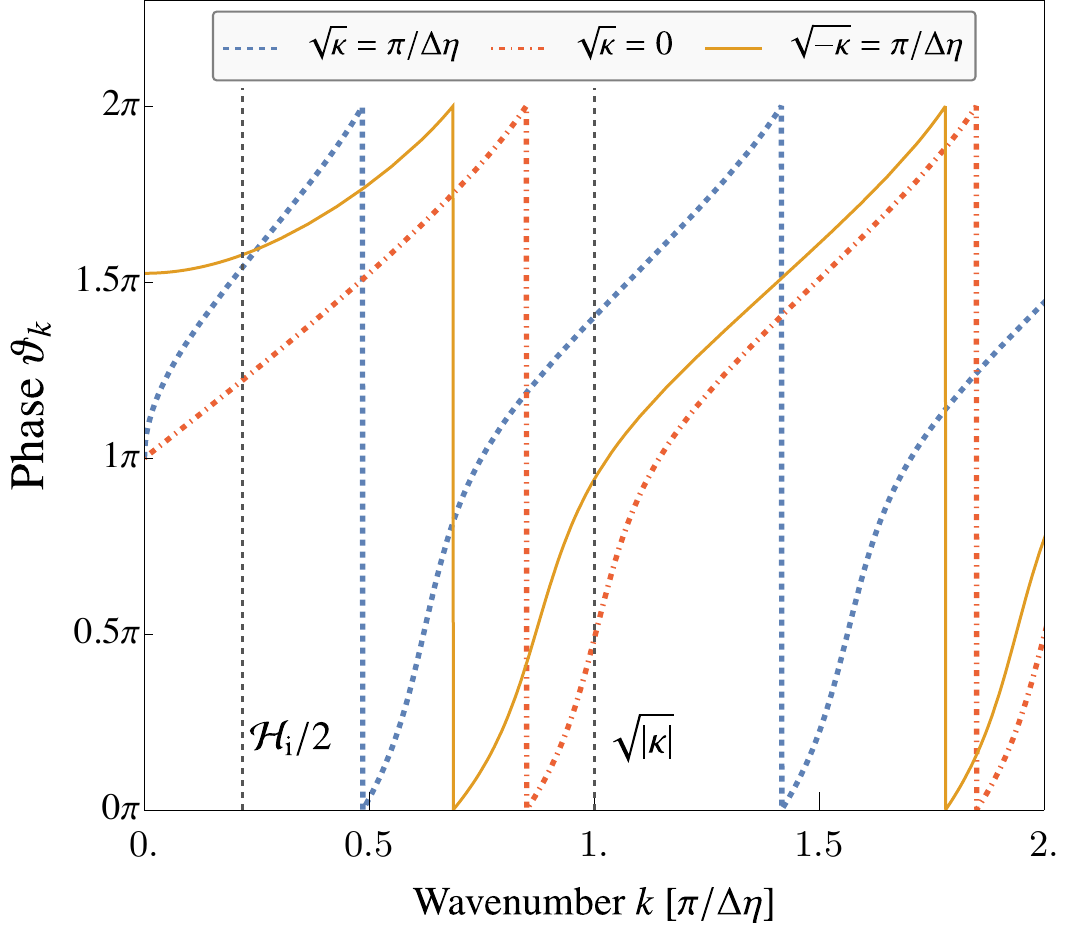}}\vfill
    \caption{Particle number $N_k$ (upper left panel), amplitude $\Delta N_k^0$ (lower left panel) and phase $\vartheta_k$ (right panel) of the $\delta$-peak landscape with $a_{\text{f}} / a_{\text{i}} = \sqrt{8}$
    for different values of the spatial curvature $\kappa$.}
\end{figure*}

\subsubsection*{Rectangular-well bounded by repulsive $\delta$-peaks}
In this section, we consider the potential landscape 
\begin{equation}
\begin{split}
    V(\eta) =& - \frac{H_0^2}{4} \Theta(\eta - \etai) \Theta(\etaf - \eta)\\
    &+ \frac{\Hi}{2} \delta(\eta-\etai) - \frac{\Hf}{2} \delta(\eta-\etaf),
\end{split}
\end{equation}
with $V_0= H_0^2/4$ and
\begin{equation}
\Hi = H_0 \sqrt{\frac{\amax}{\amin}-1} = - \Hf.
\label{eq:WellIrreg}
\end{equation}

\label{Appendix:SquareWell}
The boundary conditions at $\eta = \etai$ are 
\begin{equation}
    \begin{aligned}
        c_k \e^{-\im \omega_k \eta_\text{i}} &= f_k \e^{-\im\mu_k \eta_\text{i}} + g_k \e^{\im\mu_k \eta_\text{i}}, \\[5pt]
        (\Hi/2 -\im \omega_k )c_k \e^{-\im\omega_k \eta_\text{i}}  &= - \im \mu_k f_k \e^{-\im\mu_k \eta_\text{i}} + \im \mu_k g_k \e^{\im\mu_k \eta_\text{i}}
        \end{aligned}
\end{equation}
and at $\eta = \eta_{\text{f}}$
\begin{equation}
    \begin{aligned}
        &f_k \e^{-\im \mu_k \eta_{\text{f}}} + g_k \e^{\im \mu_k \eta_{\text{f}}} = a_k \e^{-\im \omega_k \eta_{\text{f}}} + b_k \e^{\im \omega_k \eta_{\text{f}}}, \\[5pt]
        &- (\im \mu_k + \Hf/2) f_k \e^{-\im \mu_k \eta_{\text{f}}} +(\im \mu_k -\Hf/2 ) g_k \e^{\im \mu_k \eta_{\text{f}}}  \\[5pt]
        &=  -\im \omega_k a_k \e^{-\im \omega_k \eta_{\text{f}}} + \im \omega_k b_k \e^{\im \omega_k \eta_{\text{f}}} .
        \end{aligned} 
\end{equation}
Then, we find for the amplitudes
\begin{equation}
\begin{aligned}
    a_k &=  - \frac{\im}{2}  \frac{\e^{\im \omega_k (\eta_{\text{f}} - \etai) }}{\omega_k \mu_k} \bigg \lbrace \mu_k^2 + \left(\omega_k + \im \frac{\Hi}{2} \right)^2 \sin[\mu_k (\eta_{\text{f}} - \etai)] \\  &\qquad \qquad \qquad  + \im \mu_k \left(\omega_k + \im \frac{\Hi}{2} \right) \cos[\mu_k (\eta_{\text{f}} - \etai)] \bigg \rbrace c_k ,\\[5pt]
    b_k &= - \frac{\im}{2} \e^{- \im \omega_k (\eta_{\text{f}} - \etai)} \bigg \lbrace \delta_k \sin[\mu_k (\eta_{\text{f}} - \etai)] \\[5pt]
    &\qquad \qquad \qquad \qquad + \frac{\Hi}{\omega_k} \cos[\mu_k (\eta_{\text{f}} - \etai)] \bigg \rbrace c_k.
\end{aligned}
\end{equation}
with $\delta_k$ defined in \cref{eq:AuxiliaryVariableWell}.

As for the rectangular-barrier, we can make the Ansatz \cref{eq:PotentialBarrier_ModeSolutions_Oscillating} with the frequency now acquiring a positive shift $\mu_k = \sqrt{\omega_k^2 + V_0}$ due to the attractive well. 
Anticipating a structural similarity of solutions to the case of the potential barrier, we define
\begin{equation}
    \delta_k = \frac{\abs{\omega_k + \im \Hi/2 }^2 - \mu_k^2}{\mu_k \omega_k}.
    \label{eq:AuxiliaryVariableWell}
\end{equation}
From \cref{eq:RkTkWell}, we find that $N_k$ and thus also $\Delta N_k^0$ have zero-crossings when the condition 
\begin{equation}
    \tan[\mu_k (\eta_{\text{f}} - \etai)] = - \frac{\Hi}{\omega_k \delta_k} = - \frac{2 \mu_k}{\sqrt{V_0}}\frac{\sqrt{\frac{\amax}{\amin} - 1}}{\frac{\amax}{\amin} - 2}
    \label{eq:ZeroCrossingsWell}
\end{equation}
is fulfilled. This transcendental equation has no closed form solution, but we can still consider some interesting limiting cases. 

For example, at $\omega_k=0$ this condition evaluates to
\begin{equation}
    \tan \left[ \sqrt{V_0}
    (\eta_{\text{f}} - \etai) \right] = -2 \frac{\sqrt{\frac{\amax}{\amin} - 1}}{\frac{\amax}{\amin} - 2},
\end{equation}
which is fulfilled through \cref{eq:SquareWellEtaF} and can be shown through direct insertion.

\begin{figure*}
    \subfloat{\hspace{-5cm}
\begin{minipage}[b]{\textwidth}
\includegraphics[width=0.5\textwidth,valign=t]{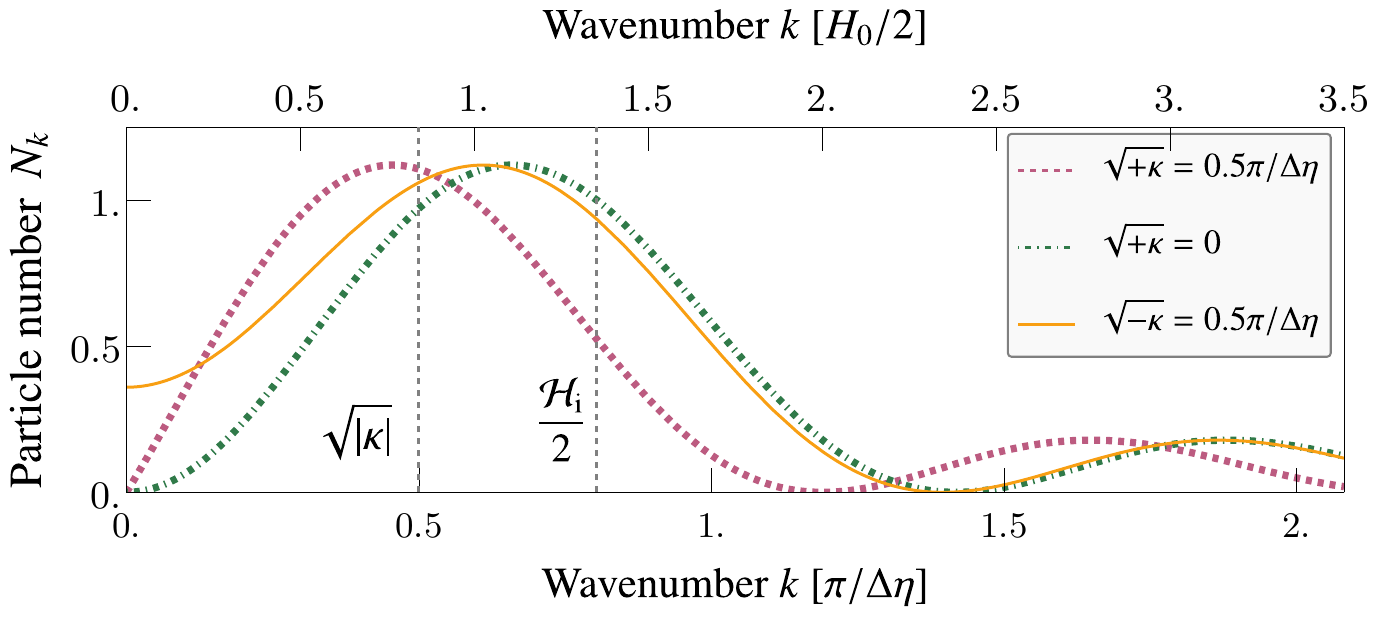}\vfill
\includegraphics[width=0.5\textwidth]{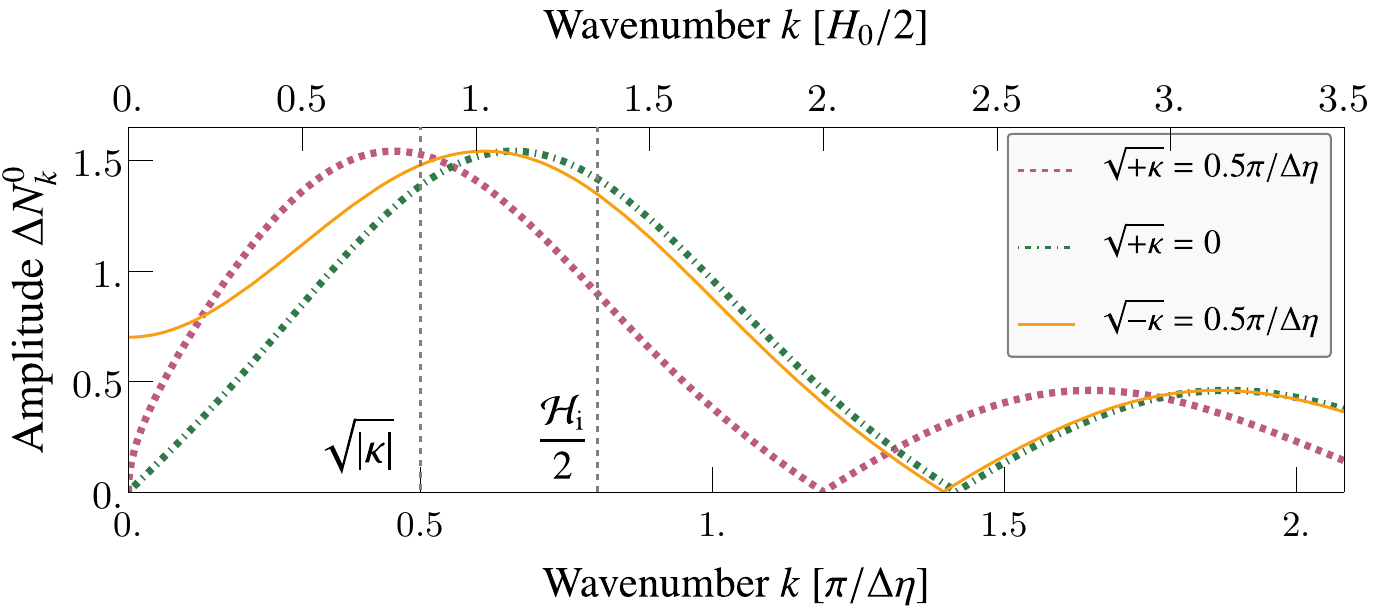}
\end{minipage}
}
\subfloat{\hspace{-4.5cm}
\includegraphics[width=0.4425\textwidth,valign=b]{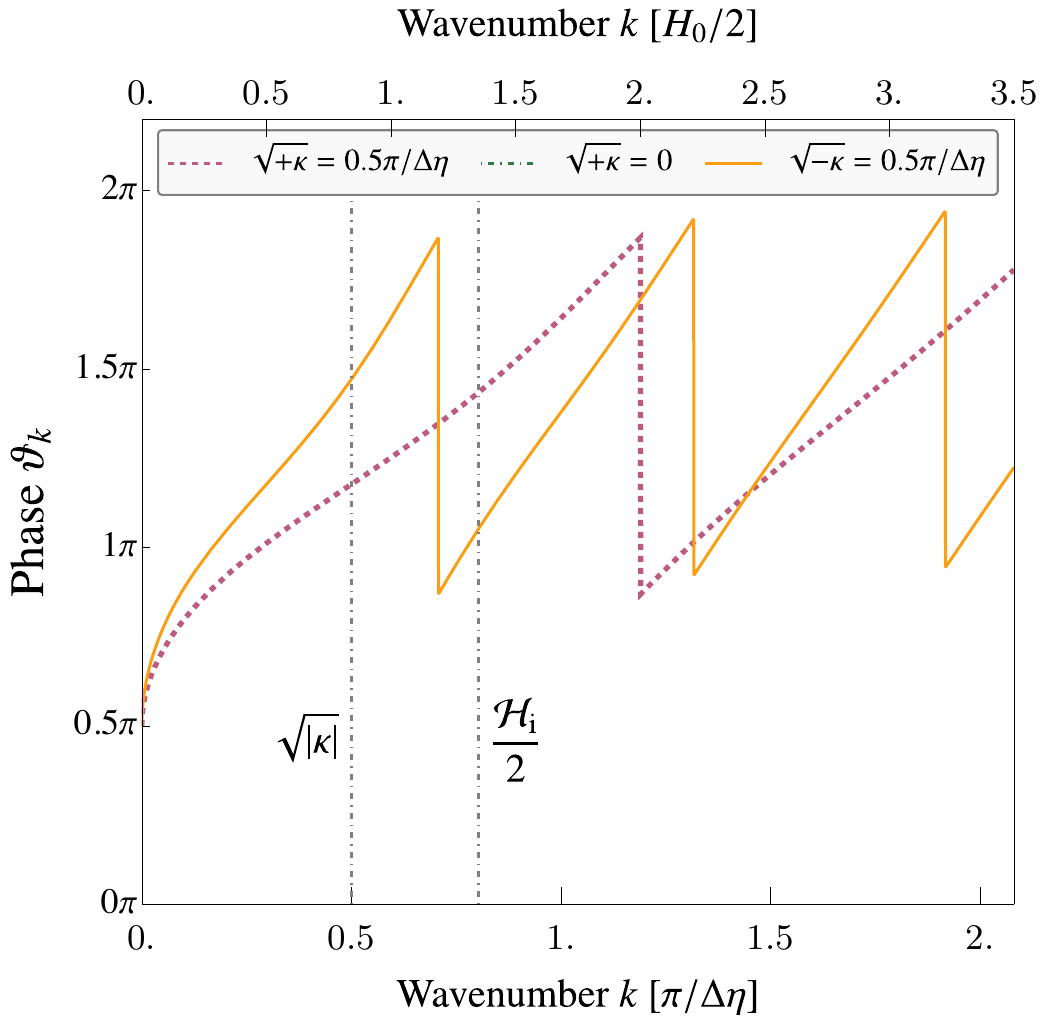}}\vfill
\caption{Particle number $N_k$ (upper left panel), amplitude $\Delta N_k^0$ (lower left panel) and phase $\vartheta_k$ (right panel)  of the rectangular-well landscape with $a_{\text{f}} / a_{\text{i}} = 1$ for different values of curvature $\kappa$. In the last panel, the phase was shifted by $\pi$ for clearer visibility.}
\end{figure*}

\subsection*{Infrared limit and bound state structure}
\label{Appendix:BoundStates}

\subsubsection*{Rectangular-well bounded by two repulsive $\delta$-peaks}
\label{subsec:BoundStatesWell}
Let us solve the Schrödinger equation for the potential landscape \eqref{eq:SquarePotentialWell_Definition} with negative energy eigenvalues $E_k<0$. 
We have
\begin{equation}
    \begin{aligned}
    \psi_k''(\eta) = - E_k \psi_k(\eta) \quad  &\text{ for }\eta < \eta_{\text{i}} \text{ and } \eta > \eta_{\text{f}},\\
    \psi_k''(\eta) = - (E_k + V_0) \psi_k(\eta) \quad &\text{ for } \eta_{\text{i}} \leq \eta \leq \eta_{\text{f}}.
    \end{aligned}
\end{equation}
As we showed, the presence of the $\delta$-peaks will only enter via modified matching conditions.
A suitable Ansatz that has regular asymptotic behavior is
\begin{equation}
     \psi_k(\eta) =
    \begin{cases}
   a\, \e^{\varpi_k \eta} &\text{ for } \eta < \eta_{\text{i}}, \\ b \sin(\varkappa_k \eta + \varphi) &\text{ for } \eta_{\text{i}}  \leq \eta \leq \eta_{\text{f}}, \\ c\, \e^{-\varpi_k \eta} &\text{ for } \eta > \eta_{\text{f}},
    \end{cases}
    \label{eq:BoundStatesAnsatz}
\end{equation}
where $ \varpi_k = \sqrt{\abs{E_k}}$ and $\varkappa_k = \sqrt{V_0 - \abs{E_k}}$. Taking into account the derivative jumps induced by the $\delta$-peaks, one finds at $\eta=\eta_{\text{i}}$ the matching conditions 
\begin{equation}
    \begin{aligned}
        a \e^{\varpi_k\etai} &= b \sin(\varkappa_k\etai + \varphi), \\ 
        a(\varpi_k+ \Hi/2) \e^{\varpi_k\etai} &= \varkappa_kb \cos(\varkappa_k\etai + \varphi),
    \end{aligned}
    \label{eq:WellBoundStatesLeftMatching}
\end{equation}
and at $\eta=\eta_{\text{f}}$
\begin{equation}
    \begin{aligned}
        b \sin(\varkappa_k\eta_{\text{f}} + \varphi) &= c\e^{-k \eta_{\text{f}}}, \\
        \varkappa_kb \cos(\varkappa_k\eta_{\text{f}} + \varphi) - b \Hf/2  \sin(\varkappa_k\eta_{\text{f}} + \varphi) &= - \varpi_kc\e^{- k \eta_{\text{f}}}. 
    \end{aligned}
    \label{eq:WellBoundStatesRightMatching}
\end{equation}
Dividing the second equation by the first, respectively, we find for the phase-shift $\varphi = \arccot[(\varpi_k+ \Hi/2)/\varkappa_k]- \varkappa_k \etai$ and upon insertion of $\varphi$ the final equation
\begin{equation}
    - \frac{\varpi_k}{\varkappa_k} = \frac{(\varpi_k+ \Hi/2 ) \cos[\varkappa\Delta \eta] - \varkappa_k\sin[\varkappa_k\Delta \eta]}{\varkappa_k \cos[\varkappa_k \Delta \eta] + (\varpi_k+ \Hi/2)  \sin[\varkappa_k \Delta \eta] }  - \frac{\Hf}{2\varkappa_k}
    \label{eq:BoundStateWell}
\end{equation}
with $\Delta \eta = \etaf-\etai$.

Using that $\Hi = - \Hf$ in the present case, it can be graphically shown that the transcendental equation \eqref{eq:BoundStateWell} has only one solution for $\varkappa_k \Delta \eta \in [0,\pi)$, which is the domain in which the analogue cosmological scenario has now a singularity (cf. \cref{eq:SquareWellEtaF}).
Furthermore, using \cref{eq:SquareWellEtaF} and \cref{eq:WellIrreg}, it is straightforward to show that this solution lies at $E_k = 0$ where $\varkappa_k = H_0/2$.

\subsubsection*{Rectangular-barrier bounded by a repulsive and attractive $\delta$-peak}
\label{subsec:BoundStatesBarrier}
Let us again consider the landscape \cref{eq:PotentialBarrier_Definition}, but explicitly distinguish the magnitude of the $\delta$-peaks $\Hi/2$ and $\Hf/2$ for illustrative reasons that become clearer later on. The Schrödinger equation for bound states with energy $E_k < 0$ is here
\begin{equation}
    \begin{aligned}
    \psi_k''(\eta) = - E_k \psi_k(\eta)  &\quad \text{ for } \eta < \eta_{\text{i}} \text{ and } \eta > \eta_{\text{f}},\\
    \psi_k''(\eta) = - (E_k - V_0) \psi_k(\eta) &\quad \text{ for } \eta_{\text{i}} \leq \eta \leq \eta_{\text{f}}.
    \end{aligned}
\end{equation}
A suitable Ansatz is
\begin{equation}
     \psi_k(\eta) =
    \begin{cases}
   a\, \e^{\varpi_k \eta} &\text{ for } \eta < \eta_{\text{i}}, \\ b \sinh(\varkappa_k \eta + \varphi) &\text{ for } \eta_{\text{i}}  \leq \eta \leq \eta_{\text{f}}, \\ c\, \e^{-\varpi_k \eta} &\text{ for } \eta > \eta_{\text{f}},
    \end{cases}
    \label{eq:BoundStateWavefunctionBarrier}
\end{equation}
with $\varpi_k = \sqrt{\abs{E_k}}$ and $\varkappa = \sqrt{V_0 - E_k}$. 
Then, the matching conditions are 
\begin{equation}
    \begin{aligned}
    a \e^{\varpi_k\etai} &= b \sinh(\varkappa_k\etai + \varphi), \\ 
    a(\varpi_k+ \Hi/2) \e^{\varpi_k\etai} &= \varkappa_kb \cosh(\varkappa_k\etai + \varphi),
    \end{aligned}
    \label{eq:BarrierBoundStatesLeftMatching}
\end{equation}
and
\begin{equation}
    \begin{aligned}
    b \sinh(\varkappa_k\eta_{\text{f}} + \varphi) &= c\e^{-k \eta_{\text{f}}}, \\
    \varkappa_kb \cosh(\varkappa_k\eta_{\text{f}} + \varphi) - b \frac{\Hf}{2}  \sinh(\varkappa_k\eta_{\text{f}} + \varphi) &= - \varpi_kc\e^{- k \eta_{\text{f}}}. 
    \end{aligned}
    \label{eq:BarrierBoundStatesRightMatching}
\end{equation}
Combining these equations, we find
\begin{equation}
    - \frac{\varpi_k}{\varkappa_k} =\frac{ \left(\varpi_k + \Hi/2 \right) \cosh[ \varkappa_k \Delta \eta] + \varkappa_k \sinh[ \varkappa_k \Delta \eta]}{\varkappa_k \cosh[\varkappa_k \Delta \eta] + \left(\varpi_k + \Hi/2 \right) \sinh[ \varkappa_k \Delta \eta]} - \frac{\Hf}{2\varkappa_k}
    \label{eq:BoundStatesBarrier}
\end{equation}
under use of ${\varphi = \text{arcoth}[(\varpi + \Hi/2)/\varkappa_k] - \varkappa_k \etai}$. 

In the present case, we have $\Hf = \Hi > 0$, such that the right-hand side can become negative due to the attractive $\delta$-peak contributing with $\Hf$. In particular at $E_k=0$, where $\varkappa_k = H_0/2$, both sides of \cref{eq:BoundStatesBarrier} identically vanish as can be deduced with \cref{eq:PotentialBarrier_Width}. Now, since the left-hand side is always negative for $\abs{E_k} \in (0,V_0] $, whereas the right-hand side is always positive there, the transient case $E_k = 0$ constitutes the only solution to the bound state equation \eqref{eq:BoundStatesBarrier}.

\subsubsection*{Double $\delta$-peak landscape }
Since we left the transition rates $\Hi$ and $\Hf$ as open parameters, \cref{eq:BoundStatesBarrier} also gives a condition for bound states for the present case with $V_0 \to 0$ and \cref{eq:DeltaPeaksTransitionRates} for the transition rates. In that case, we have 
\begin{equation}
    - \varkappa_k = \frac{(\varkappa_k + \tfrac{1}{2} \Hi) \cosh (\varkappa \Delta \eta) + \varkappa_k \sinh (\varkappa \Delta \eta)}{\cosh (\varkappa \Delta \eta) + (1 + \tfrac{\Hi}{2\varkappa_k}) \sinh( \varkappa_k \Delta \eta)} - \frac{\Hf}{2}. 
    \label{eq:BoundStateDeltas}
\end{equation}
We again find this equation to be fulfilled at the limit $\varkappa_k = \sqrt{\abs{E_k}} \to 0$, under use \cref{eq:DeltaPeaks_Width}. Moreover, it is also equivalent to the condition derived in \cite{Senn1988}. 
Furthermore, according to \cite{Senn1988}, there is only one bound state in the present scattering landscape. This can be seen from the derivative of \cref{eq:BoundStateDeltas}: The left-hand side decreases with increasing $\abs{E_k}$, while the right-hand side is monotonously increasing in $\abs{E_k}$; such that the only possible bound state lies at $E_k = 0$. 
 
\section{Supplementary expressions: Dirac comb}
\label{Appendix:DiracCombSupp}
The general properties and applications of the transfer matrix formalism are well summarized in \cite{Grosso2000,Markos2008}. Since our conventions differ slightly from these works, let us state properties of particular relevance for the analysis employed in the main text and furthermore provide supplementary expressions.

The components of the transfer matrix are computed from the matching conditions of the wave-function. Since we deal with an infinitesimally extended $\delta$-potential, there are only two matching conditions. At a repulsive or attractive $\delta$-peak of magnitude $\pm \mathcal{H}^\pm$ located at $\eta = \eta_*$ they read
\begin{equation}
    A_< \e^{- \im \omega_k \eta_*} + B_< \e^{\im \omega_k \eta_*} = A_> \e^{- \im \omega_k \eta_*} + B_> \e^{\im \omega_k \eta_*},
    \end{equation}
and
\begin{equation}
\begin{aligned}
    &A_> \e^{- \im \omega_k \eta_*} - B_> \e^{\im \omega_k \eta_*} \\
    =& A_< \e^{- \im \omega_k \eta_*} - B_< \e^{\im \omega_k \eta_*}
    \pm \im \frac{\mathcal{H^\pm}}{\omega_k}   (A_< \e^{- \im \omega_k \eta_*} + B_< \e^{\im \omega_k \eta_*}).
\end{aligned}
\end{equation}
Expressing these equations in matrix form defines the transfer matrix $M$ via
\begin{equation}
    \mqty(A_> \, \e^{- \im \omega_k \eta_*} \\ B_> \, \e^{\im \omega_k \eta_*}) = M(\pm \mathcal{H}^\pm) \mqty(A_< \, \e^{- \im \omega_k \eta_*} \\ B_< \, \e^{\im \omega_k \eta_*}),
\end{equation}
resulting in the expressions for \cref{eq:MPlusMinus}. 
Here, we adopt the convention that the transfer matrix relates wavefunctions (cf. \cite{Markos2008}), rather than amplitudes (cf. \cite{Grosso2000}).

Let us now extend the situation to a general scattering problem with an incoming left-mover. The problem can be formualted as the transfer of the state
\begin{equation}
    \mqty(\psi_\mathrm{L}^-  \\ 0) =  \mqty(A_\mathrm{L}^- \e^{-\im \omega_k \eta_\mathrm{i}}  \\ 0)
\end{equation}
on the left of the potential to 
\begin{equation}
    \mqty(\psi_\mathrm{R}^+ \\ \psi_\mathrm{R}^-) = \mqty(A_\mathrm{R} \e^{-\im \omega_k \eta_\mathrm{f}} \\ B_\mathrm{R} \e^{\im \omega_k \eta_\mathrm{f}})
\end{equation}
on the right. 
In this situation, the components of the transfer matrix are can be identified with scattering amplitudes $r_k = B_\mathrm{R}/A_\mathrm{R}$ and $t_k = A_\mathrm{L}/A_\mathrm{R}$ via 
\begin{equation}
    M = \mqty( 1/t_k \, \e^{-\im \omega_k \Delta \eta} & (r_k/t_k)^* \, \e^{-\im \omega_k (\Delta \eta - 2 \etaf) }  \\ r_k/t_k \,  \e^{\im \omega_k (\Delta \eta - 2 \etaf) }  & (1/t_k)^*\e^{\im \omega_k \Delta \eta})
\label{eq:TransferMatrix_vs_SMatrix}
\end{equation}
with $\Delta \eta = \etaf - \etai$.

In the scattering analogy of cosmological particle production, the components of the particle spectrum defined in \cref{eq:SakharovOffset,eq:SakharovAmplitude,eq:SakharovPhaseShift} then follow from 
\begin{equation}
\begin{aligned}
        N_k = \abs{M_{21}}^2,\,
        \Delta N_k^0 = \abs{M_{21}\, M_{11}},\,
        \vartheta_k = \arg(-M_{21}/M_{11}),
\end{aligned}
\label{eq:SpectrumFromTransfer}
\end{equation}
where a transfer from an initial vacuum state to a final vacuum state is described.
Finally, the complete expression for the transfer matrix of the elementary cell depicted in \cref{fig:AlternatingDiracBlock} is
\begin{align}
        T_{11} &= \frac{\im \,  \e^{-2 \im \omega \eta_\text{b}}}{32 \omega_k^3} \bigg[ (\e^{2 \im \omega \eta_\text{b}} - 1) \mathcal{H}^- - 4 \im \omega_k \bigg] \nonumber \\
        &\quad \times \bigg[(\mathcal{H}^- + 4 \im \omega_k)(\mathcal{H}^+ - 2 \im \omega_k) \nonumber \\
        &\quad \qquad - \e^{2 \im \omega \eta_\text{b}} \mathcal{H}^- (\mathcal{H}^+ + \im \omega_k) \bigg], \label{eq:CellTransferMatrix11} \\
        T_{12} &= \frac{\im}{16 \omega_k^3} \bigg[ \mathcal{H}^+ [(\mathcal{H}^-)^2 + 8 \omega_k^2 ] \nonumber \\
        &\qquad \qquad - \mathcal{H}^- (\mathcal{H}^+ \mathcal{H}^- + 8 \omega_k^2) \cos(2 \omega_k \eta_\text{b}) \nonumber \\ 
        &\qquad \qquad + 2 \mathcal{H}^- (\mathcal{H}^- - 2 \mathcal{H}^+) \omega_k \sin(2 \omega_k \eta_\text{b}) \bigg], \label{eq:CellTransferMatrix12} \\
T_{21} &= T_{12}^* \\
T_{22} &= T_{11}^*. 
\end{align}

\section{Cosmological scattering potentials in (D+1) spacetime dimensions}
\label{sec:ScatteringAnalogy}

\subsection*{Power-law expansions}
\label{subsec:PowerLaw}

\begin{figure}
\centering 
\subfloat{\includegraphics[width=\columnwidth]{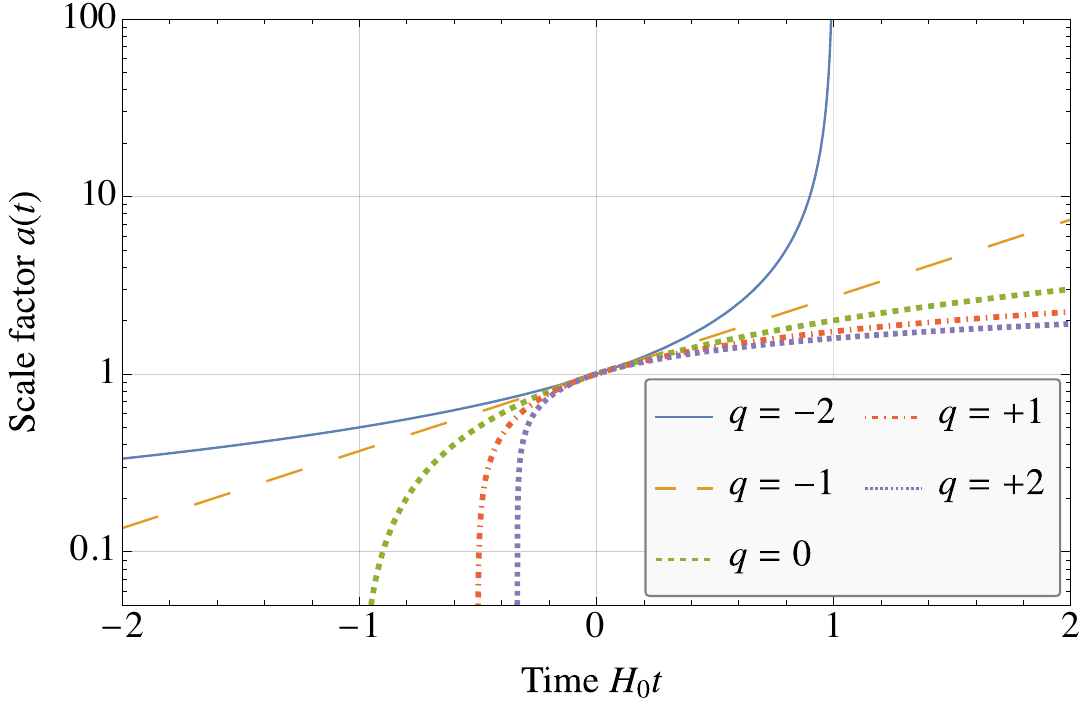}} \vfill \subfloat{\includegraphics[width=\columnwidth]{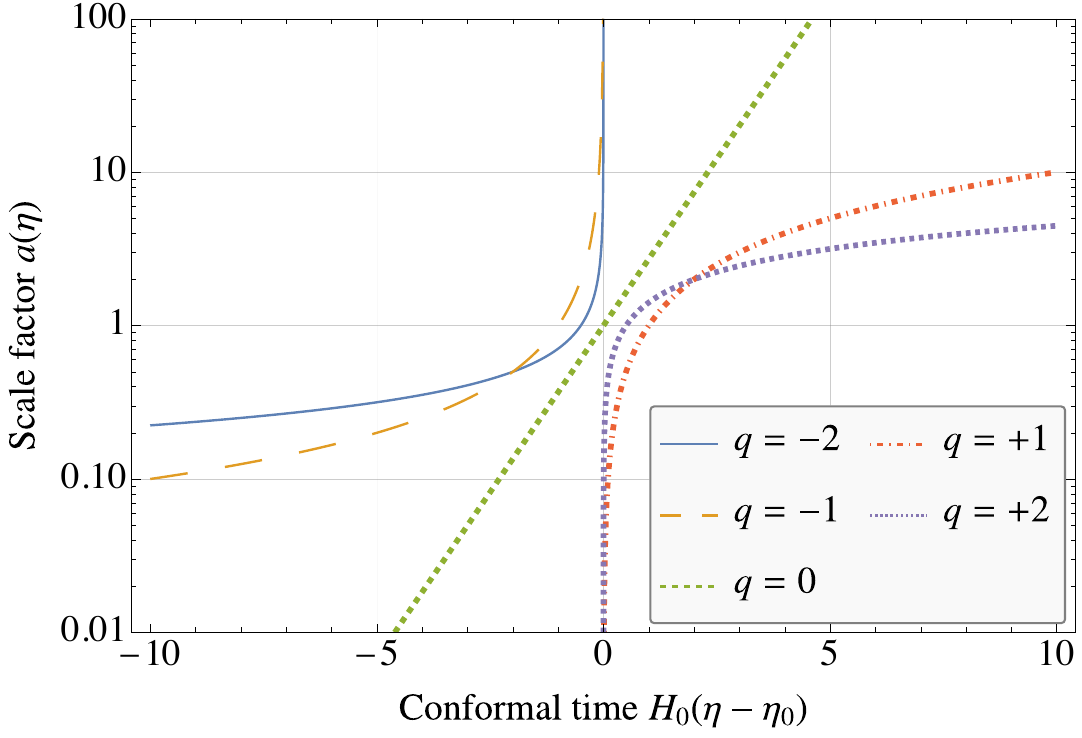}}
\caption{\label{fig:scaleFactorIllustrationTime} Power-law class of scale factors as a function of time $t$ (see eq.\ \eqref{eq:PowerLawScaleFactor}, upper panel), and as a function of conformal time $\eta$ (see eq.\ \eqref{eq:scaleFactorPowerLawConformal}, lower panel).}
\end{figure}

Let us return to the power-law scale factors introduced in \cref{sec:ScatteringPotentialsInTheQuantumSimulation} and analyze it in general cosmological context. We have
\begin{equation}
    a(t) = \left[1+ (q+1)H_0 t \right]^\frac{1}{q+1},
    \label{eq:PowerLawScaleFactor}
\end{equation}
with constant deceleration parameter $q$. For $q>0$, the expansion or contraction is decelerated, while $q<0$ corresponds to the accelerated case. Note that for $q \to -1$, the scale factor becomes of exponential form. 
Furthermore, we restrict $t$ such that  $-1< (q+1) H_0 t<\infty$ (see Fig.\ \ref{fig:scaleFactorIllustrationTime} for an illustration). For $q>-1$, this ensures that the ``big bang'' singularity at $H_0t=-1/(q+1)$ is avoided, where  the scale factor vanishes. For $q<-1$, one has instead a ``big rip'' singularity at positive $H_0 t=1/|q-1|$, where the scale factor diverges.

The Hubble rate, $ H(t) = \dot a(t) / a(t)$, for the power-law scale factors is
\begin{equation}
    H(t) = \frac{H_0}{a^{q+1}(t)},
\end{equation}
and, accordingly, $H_0\equiv H(0)$ parametrizes the Hubble rate at time $t=0$. 
A positive (negative) value of $H_0$ corresponds to an expanding (contracting) universe, while $H_0=0$ corresponds to a static universe. 

The relation between cosmic and conformal time reads
\begin{equation}
    \eta-\eta_0 = \frac{1}{H_0 q} \left[1+(q+1) H_0 t \right]^{q/(q+1)} ,
\label{eq:PowerLawConformalTime}
\end{equation}
where the integration constants have been fixed such that $t=0$ corresponds to $\eta=\eta_0+1/(qH_0)$.
The limit $q \to 0$ of \cref{eq:PowerLawConformalTime} can be taken with $\etai \to -1/(qH_0)$ resulting in a logarithmic form of $\eta(t)$. 
Then, the scale factor can also be written as a power-law in conformal time $\eta$,
\begin{equation}
    a(\eta) =  
\begin{cases}
   \left[q H_0 (\eta-\eta_0)\right]^{1/q} & q \neq 0 \\
     \exp[H_0 (\eta-\eta_0)] & q = 0.
\end{cases}
\label{eq:scaleFactorPowerLawConformal}
\end{equation}
Note that $\eta=\eta_0$ corresponds to the big bang or big rip singularities for $q>0$ and $q<0$, respectively. Consequently, the allowed regime for the scale factor is $\eta>\eta_0$ in the former case and $\eta<\eta_0$ in the latter. 

The scattering potentials corresponding to power-law scale factors \eqref{eq:scaleFactorPowerLawConformal} are (for $m_\phi=\xi=0$),
\begin{equation}
\begin{aligned}
V_{\text{pl}}(\eta) =\begin{cases}
   \frac{D-1}{2q} \left[  \frac{D-1}{2q} - 1 \right] \frac{1}{(\eta-\eta_0)^2} & \text{for} \quad q \neq 0, \\
    \left(\frac{D-1}{2}\right)^2 H_0^2  & \text{for} \quad q=0.
\end{cases} 
\end{aligned}
\label{eq:meffPowerLaw}
\end{equation}
Having the solution for the case of general $D$ and $q$, we can check whether there are equivalences between different cases on the level of the mode functions. "Indeed, the scattering potential \eqref{eq:meffPowerLaw} for de Sitter space in $D=3$ spatial dimensions, where $q=-1$, and the scattering potential in $D=2$ spatial dimensions for $q=-1/2$ are equal."
However, the curvatures differ substantially in the two cases. For the exponential expansion in $D=3$ one finds
\begin{equation}
    \mathcal{R}(\eta) = 6 H_0^2 \left[2 + \kappa (\eta - \eta_0)^2 \right],
\end{equation}
whereas for the quadratic expansion in $D=2$ one has
\begin{equation}
    \mathcal{R}(\eta) = \frac{H_0^4}{8} (\eta - \eta_0)^2 \left[8 + \kappa (\eta - \eta_0)^2 \right].
\end{equation}

\begin{figure}
    \centering
    \includegraphics[width=\columnwidth]{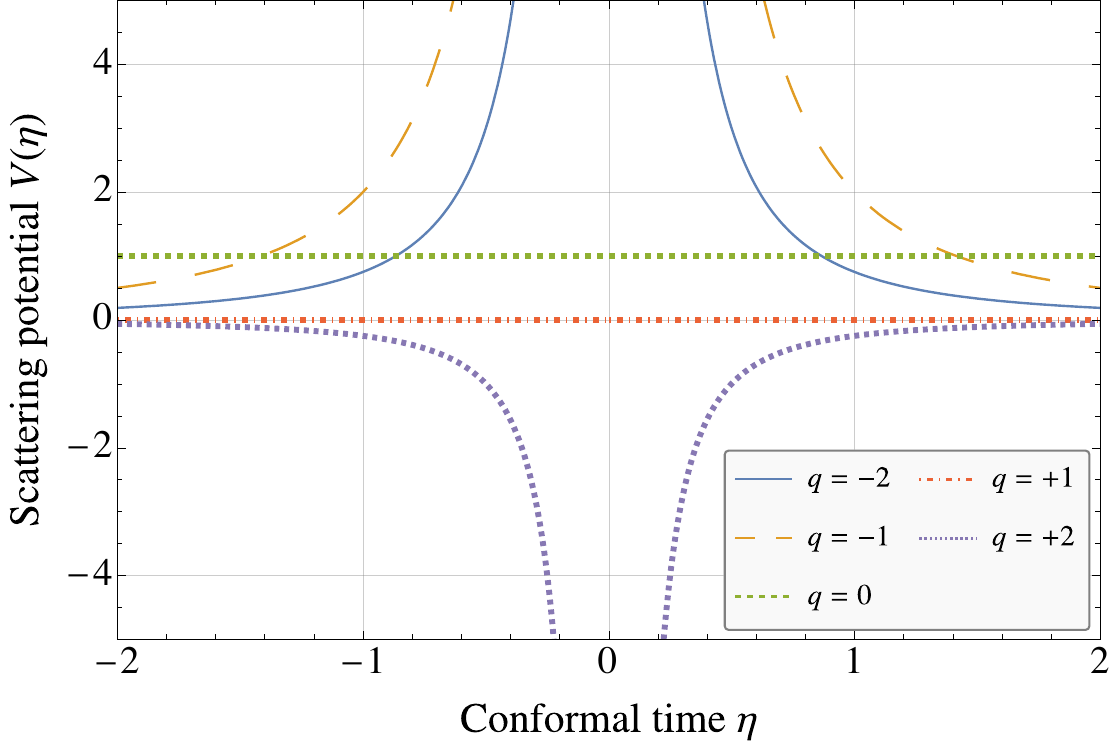}
    \caption{Scattering potential landscape of the power-law class as a function of conformal time for the case of $D+1=3+1$ spacetime dimensions. For the case $q=0$, $V(\eta)$ is displayed in units of $H_0^2$.}
\label{fig:GeneralPowerLawScatteringPotential}
\end{figure}

The mode equation \eqref{eq:SchrodingerEq} admits exact analytic solutions in case of the power-law potential landscape \eqref{eq:meffPowerLaw} (see Ref.~\cite{Wands1998,FinelliBrandenberger2002}). 
The case $q \neq 0$ yields Bessel's differential equation leading to the mode function
\begin{equation}
\begin{aligned}
        \psi_k^{\pm}(\eta) &= \sqrt{\pm\eta} \bigg[ c_1(k) H_\alpha^{(1)} \left(\pm\sqrt{-h(k)} \eta\right) \\
        &\hspace{2cm} + c_2(k) H_\alpha^{(2)} \left(\pm\sqrt{-h(k)} \eta\right) \bigg], 
\end{aligned}
\label{eq:PowerLawBesselSolutions}
\end{equation}
with
\begin{equation}
    \alpha = \frac{1}{2} \left(\frac{D-1}{q} - 1\right),
    \label{eq:BesselSolution}
\end{equation} 
where $+$ and $-$ signs correspond to the cases $q>0$ and $q<0$, respectively. 
% We work with Hankel functions $H_\alpha^{(1),(2)}$, which are of convenient use in scattering phenomena due to their asymptotic properties.

The case $q=0$ yields the mode function solutions
\begin{equation}
\begin{aligned}
    \psi_k(\eta) &= c_1(k) \e^{H_0 m \eta} + c_2(k) \e^{- H_0 m \eta},
    \label{eq:PowerLawTrigSolutions}
\end{aligned}
\end{equation}
with 
\begin{equation}
    m = \left[ \left(\frac{D-1}{2}\right)^2 + \frac{h(k)}{H_0^2} \right]^{1/2},
    \label{eq:mLinExp}
\end{equation}
where the superhorizon modes with $-h(k) > \left(\tfrac{D-1}{2}\right)^2 H_0^2$ decay and correspond to tunneling modes of the scattering problem, whereas the subhorizon modes oscillate.

The conformal time bases in \cref{eq:PowerLawBesselSolutions,eq:PowerLawTrigSolutions} can be transformed to the cosmic time bases chosen in our previous work \cite{Sanchez2022}, where specific power-law expansions in $D=2$ spatial dimensions were investigated.

\subsection*{Bouncing and anti-bouncing spacetimes}
\label{Appendix:(Anti-)BouncingScatteringStates}

\begin{figure}
    \centering
    \includegraphics[width=\columnwidth]{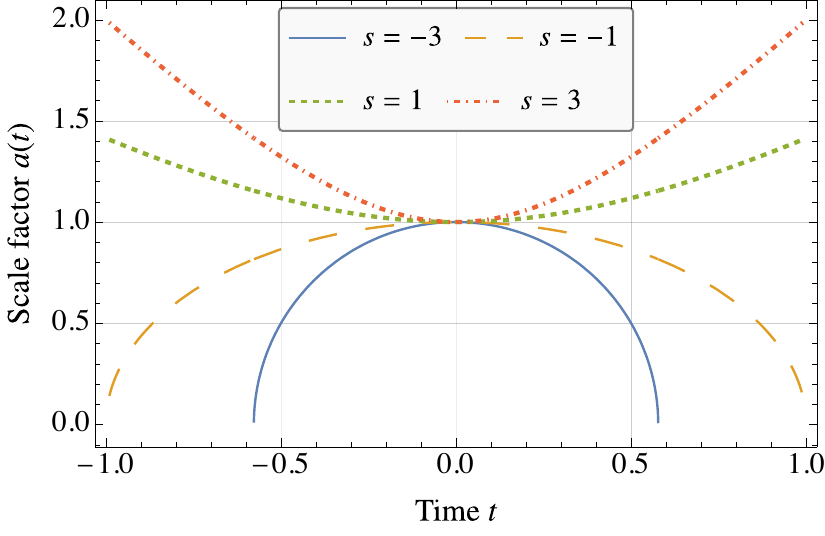}
    \caption{(Anti-)Bouncing class of scale factors as a function of cosmic time $t$. for the case $D+1=3+1$}
    \label{fig:GeneralBounceScaleFactors}
\end{figure}

Another interesting class of scale factors are of the form
\begin{equation}
    a(t) = \sqrt{1 + s t^2}.
    \label{eq:(Anti-)BouncingScaleFactors}
\end{equation}
On one hand, the case $s>0$ corresponds to a bounce with $a(t)\to \sqrt{s}|t|$ at asymptotically %or large negative or positive 
times $|t|\to \infty$ (several examples are shown in \cref{fig:GeneralBounceScaleFactors}). The turn-over point between contraction and expansion is $t=0$. This model was investigated in \cite{Shaikh2022} as a quintom bounce scenario. 

On the other hand, the case $s<0$ and cosmic time restricted to $\abs{t} < 1/\sqrt{-s} $ (to avoid cosmological singularities), the scale factor in eq.\ \eqref{eq:(Anti-)BouncingScaleFactors} describes an anti-bounce of a universe that is expanding after a big bang until the turnover time $t=0$ and, then, contracts again towards a big crunch. 

The Hubble rate for the bouncing and anti-bouncing scale factors is
\begin{equation}
    H(t) =  \frac{s t}{1 + s t^2 },
    \label{eq:(Anti-)BouncingHubbleRate}
\end{equation}
whereas the deceleration parameter \eqref{eq:DecelerationParameter} 
\begin{equation}
    q(t) = \frac{1}{s t^2},
\end{equation}
indicating that the bouncing scenario ($s>0$) is always accelerating, while the anti-bouncing ($s<0$) is always decelerating.

Here, the relation between conformal and cosmic time is given by
\begin{equation}
    \eta - \eta_0 = 
    \begin{cases} 
     \text{arcsinh}( \sqrt{s} t)/\sqrt{s} & \text{for  bounce},   \\
     \arcsin( \sqrt{-s} t)/\sqrt{-s} & \text{for anti-bounce},
    \end{cases}
\end{equation}
where we fixed the (anti-)bouncing time $\eta(t=0) = \eta_0$.
With this, we find a simple form of the scale factor as a function of conformal time,
\begin{equation}
    a(\eta) = \begin{cases}
        \cosh[ \sqrt{s} (\eta- \eta_0)] & \text{ for bounce }, \\
        \cos[ \sqrt{-s} (\eta- \eta_0)] & \text{ for anti-bounce.}
    \end{cases}
\end{equation}

\begin{figure*}
    \centering
    \includegraphics[width=0.925\columnwidth]{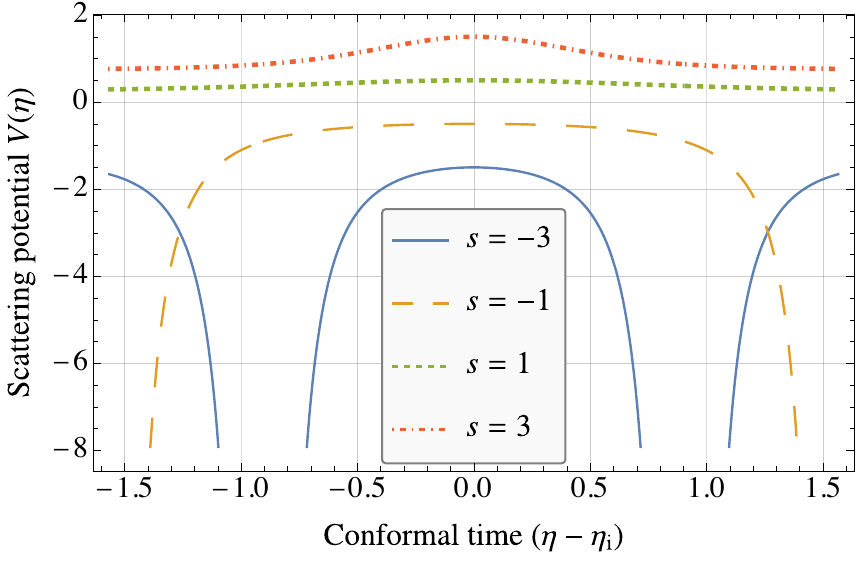}%
    \hspace{2em}
    \includegraphics[width=0.9\columnwidth,valign=b]{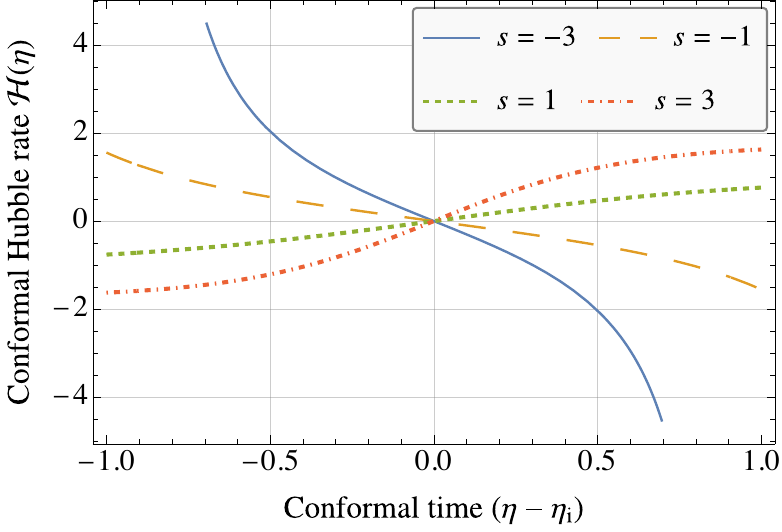}
    \caption{Scattering potential (left) and conformal Hubble rate (right) of the (anti-)bouncing class as a function of conformal time for the case of $D=2$ spatial dimensions.For the anti-bouncing class, the Hubble rate falls from positive infinity towards negative infinity, whereas it is bounded for the bouncing class such that the argument of the Legendre functions in \cref{eq:BounceGeneralSolution} lies within $(-1,1)$.}
    \label{fig:GeneralBounceScatteringPotential}
\end{figure*}

\label{subsec:AntiBouncingGeneral}
The corresponding scattering potential for the bouncing and anti-bouncing cosmologies shown in fig.~\ref{fig:GeneralBounceScatteringPotential} is given by
\begin{equation}
    V(\eta) =  \frac{D-1}{2} \left[ s + \frac{D-3}{2} \mathcal{H}^2(\eta) \right],
    \label{eq:AntiBounceScatteringPotential}
\end{equation}
with the conformal Hubble rate 
\begin{equation}
\mathcal{H}(\eta) = \begin{cases}
    \sqrt{s}  \tanh\left[\sqrt{s} (\eta- \eta_0)\right] & \text{for bounce}, \\
   - \sqrt{-s}  \tan\left[\sqrt{-s} (\eta- \eta_0)\right] & \text{for anti-bounce},
\end{cases}
\label{eq:ConfHubbleRate(Anti-)Bounce}
\end{equation}
as shown in \cref{fig:GeneralBounceScatteringPotential}.
Interestingly, analytic solutions of the mode equation also exist for the (anti-)bouncing case.
Here, \cref{eq:SchrodingerEq} can be written as an associated Legendre differential equation with the general solution for the bouncing case
\begin{equation}
    \psi_k^\text{b}(\eta) = c_1(k) \, P_l^m \left( \frac{\mathcal{H}(\eta)}{\sqrt{s}} \right) + c_2(k) \, Q_l^m \left( \frac{\mathcal{H}(\eta)}{\sqrt{s}}  \right),
    \label{eq:BounceGeneralSolution}
\end{equation}
with
\begin{equation}
\begin{aligned}
      l = \frac{D-3}{2} \quad \text{and} \quad
      m = \left[ \frac{(D-1)^2}{4} + \frac{h(k)}{s}  \right]^{1/2},
\end{aligned}
\label{eq:Bouncelm}
\end{equation}
and for the anti-bouncing case one finds 
\begin{equation}
\begin{aligned}
    \psi_k^{\text{ab}}(\eta) &= c_1(k) \, P_l^m \left( \im \frac{\mathcal{H}(\eta)}{\sqrt{-s} } \right) + c_2(k) \, Q_l^m \left( \im \frac{\mathcal{H}(\eta)}{\sqrt{-s} } \right),
\end{aligned}
\label{eq:AntiBounceGeneralSolution}
\end{equation}
with
\begin{equation}
\begin{aligned}
      l = \frac{D-3}{2} \quad \text{and} \quad
      m = \left[ \frac{(D-1)^2}{4} - \frac{h(k)}{s}  \right]^{1/2}.  
\end{aligned}
\label{eq:AntiBouncelm}
\end{equation}
Finally, note that for $D=3$, the potential
\cref{eq:AntiBounceScatteringPotential} becomes constant, i.e.\ we have $V_{\text{b/ab}} = s$, with $s<0$ for the anti-bounce and $s>0$ for the bounce. 
On the level of the mode functions, this reveals an interesting duality between the bouncing scenario \eqref{eq:(Anti-)BouncingScaleFactors} in $D=3$ dimensions and the linear expansion (cf. \eqref{eq:PowerLawScaleFactor} with $q=0$). However, to admit scattering processes, the potential landscape (and thus also the analogue cosmology) needs to have transitions between different epochs.
These appear naturally in an analogue quantum field simulation, as we showed in \cref{sec:ScatteringPotentialsInTheQuantumSimulation}.

\begin{widetext}
\subsection*{Scattering amplitudes of bouncing and anti-bouncing spacetimes}
Let us consider the (anti)-bouncing scale-factors \eqref{eq:(Anti-)BouncingScaleFactors} as the dynamical evolution in region II and compute the scattering amplitudes in region III.  
We make an Ansatz for the mode functions for a bouncing potential \eqref{eq:AntiBounceScatteringPotential}
\begin{equation}
   \psi_k(\eta) =  \begin{cases}
        c_k \e^{-\im \omega_k \eta} & \eta <\eta_\text{i}, \\[5pt] f_k \, P_l^m \left( \frac{\mathcal{H}(\eta)}{\sqrt{s}} \right) + g_k \, Q_l^m \left( \frac{\mathcal{H}(\eta)}{\sqrt{s}}  \right) &\eta \in [\eta_{\text{i}}, \eta_{\text{f}}], \\[5pt]
        a_k \e^{-\im \omega_k \eta} + b_k \e^{\im \omega_k \eta} & \eta > \eta_{\text{f}} ,
        \end{cases} 
        \label{eq:PotentialBouncing_ModeSolutions_Bouncing}
\end{equation}
with $\omega_k = \sqrt{E_k}$.
The boundary conditions at $\eta = \etai$ are 
\begin{equation}
    \begin{aligned}
        c_k \e^{-\im \omega_k \eta_\text{i}} &=  f_k \, P_l^m \left(\frac{\mathcal{H}_\text{i}}{\sqrt{s}}\right) + g_k \, Q_l^m  \left(\frac{\mathcal{H}_\text{i}}{\sqrt{s}}\right), \\[5pt]
        -\im \omega_k c_k \e^{-\im\omega_k \eta_\text{i}} &= f_k\Bigg[(l+1) \Hi P_{l}^m \left( \frac{\mathcal{H}_\text{i}}{\sqrt{s}} \right) - \sqrt{s} (l-m+1)\, P_{l+1}^m \left( \frac{\mathcal{H}_\text{i}}{\sqrt{s}} \right)\Bigg] \\
        &\quad \qquad +g_k\Bigg[(l+1) \Hi Q_{l}^m \left( \frac{\mathcal{H}_\text{i}}{\sqrt{s}} \right) - \sqrt{s} (l-m+1)\, Q_{l+1}^m \left( \frac{\mathcal{H}_\text{i}}{\sqrt{s}} \right)\Bigg], \\
    \end{aligned}
\end{equation}
and at $\eta = \eta_{\text{f}}$
\begin{equation}
    \begin{aligned}
            &f_k \, P_l^m \left( \frac{\mathcal{H}_\text{f}}{\sqrt{s}} \right) + g_k \, Q_l^m \left( \frac{\mathcal{H}_\text{f}}{\sqrt{s}}  \right) = a_k \e^{-\im \omega_k \eta_{\text{f}}} + b_k \e^{\im \omega_k \eta_{\text{f}}}, \\[5pt]
        &f_k\Bigg[(l+1) \Hf P_{l}^m \left( \frac{\mathcal{H}_\text{f}}{\sqrt{s}} \right) - \sqrt{s} (l-m+1)\, P_{l+1}^m \left( \frac{\mathcal{H}_\text{f}}{\sqrt{s}} \right)\Bigg] +g_k\Bigg[(l+1) \Hf Q_{l}^m \left( \frac{\mathcal{H}_\text{f}}{\sqrt{s}} \right) - \sqrt{s} (l-m+1)\, Q_{l+1}^m \left( \frac{\mathcal{H}_\text{f}}{\sqrt{s}} \right)\Bigg]\\
        &=  -\im \omega_k a_k \e^{-\im \omega_k \eta_{\text{f}}} + \im \omega_k b_k \e^{\im \omega_k \eta_{\text{f}}} .
        \end{aligned} 
\end{equation}
Then, we find for the amplitudes
\begin{equation}
\begin{split}
    a_k &= \frac{c_k e^{\im \omega_k (\etaf-\etai)}}{2\im \omega_k \sqrt{s}} \frac{\Gamma(l-m+1)}{\Gamma(l+m+1)} \\
    &\times \Bigg\{ \left[ \left(\im \omega_k - (l+1) \Hf\right) P_{l}^m \left(\frac{\mathcal{\Hf}}{\sqrt{s}}\right) + \sqrt{s}(l-m+1) P_{l+1}^m \left(\frac{\mathcal{\Hf}}{\sqrt{s}}\right)  \right]\\
    &\times \left[ \left(\im \omega_k + (l+1) \Hi\right) P_{l}^m \left(\frac{\mathcal{\Hi}}{\sqrt{s}}\right) - \sqrt{s}(l-m+1) P_{l+1}^m \left(\frac{\mathcal{\Hi}}{\sqrt{s}}\right)  \right]\\
    &-\left[ \left(\im \omega_k - (l+1) \Hf\right) Q_{l}^m \left(\frac{\mathcal{\Hf}}{\sqrt{s}}\right) + \sqrt{s}(l-m+1) Q_{l+1}^m \left(\frac{\mathcal{\Hf}}{\sqrt{s}}\right)  \right]\\
    &\times \left[ \left(\im \omega_k + (l+1) \Hi\right) Q_{l}^m \left(\frac{\mathcal{\Hi}}{\sqrt{s}}\right) - \sqrt{s}(l-m+1) Q_{l+1}^m \left(\frac{\mathcal{\Hi}}{\sqrt{s}}\right)  \right]\Bigg\},\\
    b_k=& \frac{c_k e^{-\im \omega_k (\etaf+\etai)}}{2\im \omega_k \sqrt{s}} \frac{\Gamma(l-m+1)}{\Gamma(l+m+1)} \\
    &\times\Bigg\{ \left[ \left(\im \omega_k + (l+1) \Hf\right) P_{l}^m \left(\frac{\mathcal{\Hf}}{\sqrt{s}}\right) - \sqrt{s}(l-m+1) P_{l+1}^m \left(\frac{\mathcal{\Hf}}{\sqrt{s}}\right)  \right]\\
    &\times \left[ \left(\im \omega_k + (l+1) \Hi\right) P_{l}^m \left(\frac{\mathcal{\Hi}}{\sqrt{s}}\right) - \sqrt{s}(l-m+1) P_{l+1}^m \left(\frac{\mathcal{\Hi}}{\sqrt{s}}\right)  \right]\\
    &-\left[ \left(\im \omega_k + (l+1) \Hf\right) Q_{l}^m \left(\frac{\mathcal{\Hf}}{\sqrt{s}}\right) - \sqrt{s}(l-m+1) Q_{l+1}^m \left(\frac{\mathcal{\Hf}}{\sqrt{s}}\right)  \right]\\
    &\times \left[ \left(\im \omega_k + (l+1) \Hi\right) Q_{l}^m \left(\frac{\mathcal{\Hi}}{\sqrt{s}}\right) - \sqrt{s}(l-m+1) Q_{l+1}^m \left(\frac{\mathcal{\Hi}}{\sqrt{s}}\right)  \right]\Bigg\}.
\end{split}
\end{equation}
We make an similar Ansatz for the mode functions for an anti-bouncing potential  \eqref{eq:AntiBounceScatteringPotential}
\begin{equation}
   \psi_k(\eta) =  \begin{cases}
        c_k \e^{-\im \omega_k \eta} & \eta <\eta_\text{i}, \\[5pt] f_k \, P_l^m \left( \im \frac{\mathcal{H}(\eta)}{\sqrt{-s} } \right) + g_k \, Q_l^m \left( \im \frac{\mathcal{H}(\eta)}{\sqrt{-s} }  \right) & \eta \in [\eta_{\text{i}}, \eta_{\text{f}}],\\[5pt]
        a_k \e^{-\im \omega_k \eta} + b_k \e^{\im \omega_k \eta} &\eta_{\text{f}}<\eta,
        \end{cases}
    \label{eq:PotentialAntiBouncing_ModeSolutions_Bouncing}
\end{equation}
with $\omega_k$ and $\mathcal{H}(\eta)$ as above.
The boundary conditions at $\eta = \etai$ are 
\begin{equation}
    \begin{aligned}
         c_k \e^{-\im \omega_k \eta_\text{i}} &= f_k \, P_l^m \left(\frac{\im \Hi}{\sqrt{-s} } \right) + g_k \, Q_l^m  \left( \frac{\im \Hi}{\sqrt{-s} } \right), \\[5pt]
        \left(\frac{\Hi}{2} - \im \omega_k \right) c_k \e^{-\im\omega_k \eta_\text{i}} &= f_k\Bigg[(l+1) \Hi P_{l}^m \left( \frac{\im \Hi}{\sqrt{\sqrt{-s}} } \right) + \im \sqrt{-s} (l-m+1)\, P_{l+1}^m \left( \frac{\im \Hi}{\sqrt{-s}} \right)\Bigg] \\
        &\hspace{1cm} +g_k\Bigg[(l+1) \Hi Q_{l}^m \left(\frac{\im \Hi}{\sqrt{-s} } \right) + \im \sqrt{-s} (l-m+1)\, Q_{l+1}^m \left(  \frac{\im \Hi}{\sqrt{-s} } \right) \Bigg].\\
        \end{aligned}
\end{equation}
As the scattering potential for the anti-bouncing case has singularities, the boundary conditions at $\eta = \eta_{\text{f}}$
\begin{equation}
    \begin{aligned}
        &f_k \, P_l^m \left( \im\frac{\mathcal{H}_\text{f}}{\sqrt{-s}} \right) + g_k \, Q_l^m \left( \im\frac{\mathcal{H}_\text{f}}{\sqrt{-s}}  \right) = a_k \e^{-\im \omega_k \eta_{\text{f}}} + b_k \e^{\im \omega_k \eta_{\text{f}}}, \\[5pt]
        &f_k\Bigg[(l+1) \Hf P_{l}^m \left( \im\frac{\mathcal{H}_\text{f}}{\sqrt{-s}} \right) + \im \sqrt{-s} (l-m+1)\, P_{l+1}^m \left(\im\frac{\mathcal{H}_\text{f}}{\sqrt{-s}}\right)\Bigg]\\
        &\qquad +g_k\Bigg[(l+1) \Hf Q_{l}^m \left( \im\frac{\mathcal{H}_\text{f}}{\sqrt{-s}} \right) +\im \sqrt{-s} (l-m+1)\, Q_{l+1}^m \left( \im\frac{\mathcal{H}_\text{f}}{\sqrt{-s}} \right)\Bigg] \\
        &=  \left(\frac{\Hf}{2} - \im \omega_k \right) a_k \e^{-\im \omega_k \eta_{\text{f}}} + \left(\frac{\Hf}{2} + \im \omega_k \right)  b_k \e^{\im \omega_k \eta_{\text{f}}} .
        \end{aligned} 
\end{equation}
Then, we find for the amplitudes
\begin{equation}
\begin{split}
    a_k=&  - \frac{c_k e^{\im \omega_k (\etaf-\etai)}}{2\omega_k \sqrt{-s}} \frac{\Gamma(l-m+1)}{\Gamma(l+m+1)} \\
    &\times\Bigg\{ \left[ \left(\left(l+\frac{1}{2}\right) \Hf - \im \omega_k \right) P_{l}^m \left(\im\frac{\mathcal{H}_\text{f}}{\sqrt{-s}} \right) + \im \sqrt{-s}(l-m+1) P_{l+1}^m \left(\im\frac{\mathcal{H}_\text{f}}{\sqrt{-s}}\right)  \right]\\
    &\times \left[ \left(\left(l+\frac{1}{2}\right) \Hi + \im \omega_k \right) Q_{l}^m \left(\im\frac{\mathcal{H}_\text{i}}{\sqrt{-s}} \right) + \im \sqrt{-s}(l-m+1) Q_{l+1}^m \left(\im\frac{\mathcal{H}_\text{i}}{\sqrt{-s}} \right)  \right]\\
    &-\left[ \left(\left(l+\frac{1}{2}\right) \Hf - \im \omega_k \right) Q_{l}^m \left(\im\frac{\mathcal{H}_\text{f}}{\sqrt{-s}} \right) + \im \sqrt{-s}(l-m+1) Q_{l+1}^m \left(\im\frac{\mathcal{H}_\text{f}}{\sqrt{-s}} \right)  \right]\\
    &\times \left[ \left(\left(l+\frac{1}{2}\right) \Hi + \im \omega_k \right) P_{l}^m \left(\im\frac{\mathcal{H}_\text{i}}{\sqrt{-s}} \right) + \im \sqrt{-s}(l-m+1) P_{l+1}^m \left(\im\frac{\mathcal{H}_\text{i}}{\sqrt{-s}} \right)  \right]\Bigg\},\\
    b_k=& \frac{c_k e^{- \im \omega_k (\etaf+\etai)}}{2\omega_k \sqrt{-s}} \frac{\Gamma(l-m+1)}{\Gamma(l+m+1)} \\
    &\times\Bigg\{ \left[ \left(\left(l+\frac{1}{2}\right) \Hf + \im \omega_k \right) P_{l}^m \left(\im\frac{\mathcal{H}_\text{f}}{\sqrt{-s}} \right) + \im \sqrt{-s}(l-m+1) P_{l+1}^m \left(\im\frac{\mathcal{H}_\text{f}}{\sqrt{-s}}\right)  \right]\\
    &\times \left[ \left(\left(l+\frac{1}{2}\right) \Hi + \im \omega_k \right) Q_{l}^m \left(\im\frac{\mathcal{H}_\text{i}}{\sqrt{-s}} \right) + \im \sqrt{-s}(l-m+1) Q_{l+1}^m \left(\im\frac{\mathcal{H}_\text{i}}{\sqrt{-s}} \right)  \right]\\
    &-\left[ \left(\left(l+\frac{1}{2}\right) \Hf + \im \omega_k \right) Q_{l}^m \left(\im\frac{\mathcal{H}_\text{f}}{\sqrt{-s}} \right) + \im \sqrt{-s}(l-m+1) Q_{l+1}^m \left(\im\frac{\mathcal{H}_\text{f}}{\sqrt{-s}} \right)  \right]\\
    &\times \left[ \left(\left(l+\frac{1}{2}\right) \Hi + \im \omega_k \right) P_{l}^m \left(\im\frac{\mathcal{H}_\text{i}}{\sqrt{-s}} \right) + \im \sqrt{-s}(l-m+1) P_{l+1}^m \left(\im\frac{\mathcal{H}_\text{i}}{\sqrt{-s}} \right)  \right]\Bigg\}.
\end{split}
\end{equation}
\end{widetext}

\clearpage 
\bibliography{main}

\begin{thebibliography}{115}%
\makeatletter
\providecommand \@ifxundefined [1]{%
 \@ifx{#1\undefined}
}%
\providecommand \@ifnum [1]{%
 \ifnum #1\expandafter \@firstoftwo
 \else \expandafter \@secondoftwo
 \fi
}%
\providecommand \@ifx [1]{%
 \ifx #1\expandafter \@firstoftwo
 \else \expandafter \@secondoftwo
 \fi
}%
\providecommand \natexlab [1]{#1}%
\providecommand \enquote  [1]{``#1''}%
\providecommand \bibnamefont  [1]{#1}%
\providecommand \bibfnamefont [1]{#1}%
\providecommand \citenamefont [1]{#1}%
\providecommand \href@noop [0]{\@secondoftwo}%
\providecommand \href [0]{\begingroup \@sanitize@url \@href}%
\providecommand \@href[1]{\@@startlink{#1}\@@href}%
\providecommand \@@href[1]{\endgroup#1\@@endlink}%
\providecommand \@sanitize@url [0]{\catcode `\\12\catcode `\$12\catcode
  `\&12\catcode `\#12\catcode `\^12\catcode `\_12\catcode `\%12\relax}%
\providecommand \@@startlink[1]{}%
\providecommand \@@endlink[0]{}%
\providecommand \url  [0]{\begingroup\@sanitize@url \@url }%
\providecommand \@url [1]{\endgroup\@href {#1}{\urlprefix }}%
\providecommand \urlprefix  [0]{URL }%
\providecommand \Eprint [0]{\href }%
\providecommand \doibase [0]{https://doi.org/}%
\providecommand \selectlanguage [0]{\@gobble}%
\providecommand \bibinfo  [0]{\@secondoftwo}%
\providecommand \bibfield  [0]{\@secondoftwo}%
\providecommand \translation [1]{[#1]}%
\providecommand \BibitemOpen [0]{}%
\providecommand \bibitemStop [0]{}%
\providecommand \bibitemNoStop [0]{.\EOS\space}%
\providecommand \EOS [0]{\spacefactor3000\relax}%
\providecommand \BibitemShut  [1]{\csname bibitem#1\endcsname}%
\let\auto@bib@innerbib\@empty
%</preamble>
\bibitem [{\citenamefont {Mukhanov}\ and\ \citenamefont
  {Winitzki}(2007)}]{mukhanov_winitzki_2007}%
  \BibitemOpen
  \bibfield  {author} {\bibinfo {author} {\bibfnamefont {V.}~\bibnamefont
  {Mukhanov}}\ and\ \bibinfo {author} {\bibfnamefont {S.}~\bibnamefont
  {Winitzki}},\ }\href {https://doi.org/10.1017/CBO9780511809149} {\emph
  {\bibinfo {title} {Introduction to Quantum Effects in Gravity}}}\ (\bibinfo
  {publisher} {Cambridge University Press},\ \bibinfo {year}
  {2007})\BibitemShut {NoStop}%
\bibitem [{\citenamefont {Birrell}\ and\ \citenamefont
  {Davies}(1982)}]{birrell_davies_1982}%
  \BibitemOpen
  \bibfield  {author} {\bibinfo {author} {\bibfnamefont {N.~D.}\ \bibnamefont
  {Birrell}}\ and\ \bibinfo {author} {\bibfnamefont {P.~C.~W.}\ \bibnamefont
  {Davies}},\ }\href {https://doi.org/10.1017/CBO9780511622632} {\emph
  {\bibinfo {title} {Quantum Fields in Curved Space}}},\ Cambridge Monographs
  on Mathematical Physics\ (\bibinfo  {publisher} {Cambridge University
  Press},\ \bibinfo {year} {1982})\BibitemShut {NoStop}%
\bibitem [{\citenamefont {Fulling}(1989)}]{Fulling1989}%
  \BibitemOpen
  \bibfield  {author} {\bibinfo {author} {\bibfnamefont {S.~A.}\ \bibnamefont
  {Fulling}},\ }\href {https://doi.org/10.1017/cbo9781139172073} {\emph
  {\bibinfo {title} {Aspects of Quantum Field Theory in Curved Space-Time}}}\
  (\bibinfo  {publisher} {Cambridge University Press},\ \bibinfo {year}
  {1989})\BibitemShut {NoStop}%
\bibitem [{\citenamefont {"Wald}(1995)}]{Wald1995}%
  \BibitemOpen
  \bibfield  {author} {\bibinfo {author} {\bibfnamefont {R.~M.}\ \bibnamefont
  {"Wald}},\ }\href@noop {} {\emph {\bibinfo {title} {{Quantum Field Theory in
  Curved Space-Time and Black Hole Thermodynamics}}}},\ "Chicago Lectures in
  Physics"\ (\bibinfo  {publisher} {"University of Chicago Press"},\ \bibinfo
  {address} {"Chicago, IL"},\ \bibinfo {year} {"1995"})\BibitemShut {NoStop}%
\bibitem [{\citenamefont {Parker}(1969)}]{Parker1969}%
  \BibitemOpen
  \bibfield  {author} {\bibinfo {author} {\bibfnamefont {L.}~\bibnamefont
  {Parker}},\ }\href {https://doi.org/10.1103/PhysRev.183.1057} {\bibfield
  {journal} {\bibinfo  {journal} {Phys. Rev.}\ }\textbf {\bibinfo {volume}
  {183}},\ \bibinfo {pages} {1057} (\bibinfo {year} {1969})}\BibitemShut
  {NoStop}%
\bibitem [{\citenamefont {Hawking}(1975)}]{Hawking1975}%
  \BibitemOpen
  \bibfield  {author} {\bibinfo {author} {\bibfnamefont {S.~W.}\ \bibnamefont
  {Hawking}},\ }\href {https://doi.org/10.1007/BF02345020} {\bibfield
  {journal} {\bibinfo  {journal} {Comm. Math. Phys.}\ }\textbf {\bibinfo
  {volume} {43}},\ \bibinfo {pages} {199} (\bibinfo {year} {1975})}\BibitemShut
  {NoStop}%
\bibitem [{\citenamefont {Landau}\ and\ \citenamefont
  {Lifshitz}(2013)}]{LandauLifshitzQuantum}%
  \BibitemOpen
  \bibfield  {author} {\bibinfo {author} {\bibfnamefont {L.~D.}\ \bibnamefont
  {Landau}}\ and\ \bibinfo {author} {\bibfnamefont {E.~M.}\ \bibnamefont
  {Lifshitz}},\ }\href@noop {} {\emph {\bibinfo {title} {Quantum mechanics:
  non-relativistic theory}}},\ Vol.~\bibinfo {volume} {3}\ (\bibinfo
  {publisher} {Elsevier},\ \bibinfo {year} {2013})\BibitemShut {NoStop}%
\bibitem [{\citenamefont {Sakurai}\ and\ \citenamefont
  {Napolitano}(2020)}]{Sakurai_Napolitano_2020}%
  \BibitemOpen
  \bibfield  {author} {\bibinfo {author} {\bibfnamefont {J.~J.}\ \bibnamefont
  {Sakurai}}\ and\ \bibinfo {author} {\bibfnamefont {J.}~\bibnamefont
  {Napolitano}},\ }\href@noop {} {\emph {\bibinfo {title} {Modern Quantum
  Mechanics}}},\ \bibinfo {edition} {3rd}\ ed.\ (\bibinfo  {publisher}
  {Cambridge University Press},\ \bibinfo {year} {2020})\BibitemShut {NoStop}%
\bibitem [{\citenamefont {Griffiths}\ and\ \citenamefont
  {Schroeter}(2018)}]{griffiths_schroeter_2018}%
  \BibitemOpen
  \bibfield  {author} {\bibinfo {author} {\bibfnamefont {D.~J.}\ \bibnamefont
  {Griffiths}}\ and\ \bibinfo {author} {\bibfnamefont {D.~F.}\ \bibnamefont
  {Schroeter}},\ }\href {https://doi.org/10.1017/9781316995433} {\emph
  {\bibinfo {title} {Introduction to Quantum Mechanics}}},\ \bibinfo {edition}
  {3rd}\ ed.\ (\bibinfo  {publisher} {Cambridge University Press},\ \bibinfo
  {year} {2018})\BibitemShut {NoStop}%
\bibitem [{\citenamefont {Schwabl}(2007)}]{Schwabl2007}%
  \BibitemOpen
  \bibfield  {author} {\bibinfo {author} {\bibfnamefont {F.}~\bibnamefont
  {Schwabl}},\ }\href {https://doi.org/10.1007/978-3-540-73675-2_3} {\emph
  {\bibinfo {title} {Quantenmechanik (QM I): Eine Einf{\"u}hrung}}}\ (\bibinfo
  {publisher} {Springer Berlin Heidelberg},\ \bibinfo {address} {Berlin,
  Heidelberg},\ \bibinfo {year} {2007})\ pp.\ \bibinfo {pages}
  {47--97}\BibitemShut {NoStop}%
\bibitem [{\citenamefont {Fl{\"u}gge}(1999)}]{Flügge1999}%
  \BibitemOpen
  \bibfield  {author} {\bibinfo {author} {\bibfnamefont {S.}~\bibnamefont
  {Fl{\"u}gge}},\ }\bibinfo {title} {One-body problems without spin},\ in\
  \href {https://doi.org/10.1007/978-3-642-61995-3_2} {\emph {\bibinfo
  {booktitle} {Practical Quantum Mechanics}}}\ (\bibinfo  {publisher} {Springer
  Berlin Heidelberg},\ \bibinfo {address} {Berlin, Heidelberg},\ \bibinfo
  {year} {1999})\ pp.\ \bibinfo {pages} {25--331}\BibitemShut {NoStop}%
\bibitem [{\citenamefont {Boya}(2008)}]{Boya2008}%
  \BibitemOpen
  \bibfield  {author} {\bibinfo {author} {\bibfnamefont {L.}~\bibnamefont
  {Boya}},\ }\href {https://doi.org/10.1393/ncr/i2008-10030-4} {\bibfield
  {journal} {\bibinfo  {journal} {La Rivista del Nuovo Cimento}\ }\textbf
  {\bibinfo {volume} {31}},\ \bibinfo {pages} {75–139} (\bibinfo {year}
  {2008})}\BibitemShut {NoStop}%
\bibitem [{\citenamefont {Calzetta}\ and\ \citenamefont
  {Hu}(2008)}]{CalzettaHu2008}%
  \BibitemOpen
  \bibfield  {author} {\bibinfo {author} {\bibfnamefont {E.~A.}\ \bibnamefont
  {Calzetta}}\ and\ \bibinfo {author} {\bibfnamefont {B.-L.~B.}\ \bibnamefont
  {Hu}},\ }\href {https://doi.org/10.1017/CBO9780511535123} {\emph {\bibinfo
  {title} {{Nonequilibrium Quantum Field Theory}}}},\ Cambridge Monographs on
  Mathematical Physics\ (\bibinfo  {publisher} {Cambridge University Press},\
  \bibinfo {address} {Cambridge},\ \bibinfo {year} {2008})\BibitemShut
  {NoStop}%
\bibitem [{\citenamefont {Viermann}\ \emph {et~al.}(2022)\citenamefont
  {Viermann}, \citenamefont {Sparn}, \citenamefont {Liebster}, \citenamefont
  {Hans}, \citenamefont {Kath}, \citenamefont {Parra-L{\'o}pez}, \citenamefont
  {Tolosa-Sime{\'o}n}, \citenamefont {S{\'a}nchez-Kuntz}, \citenamefont {Haas},
  \citenamefont {Strobel}, \citenamefont {Floerchinger},\ and\ \citenamefont
  {Oberthaler}}]{Viermann2022}%
  \BibitemOpen
  \bibfield  {author} {\bibinfo {author} {\bibfnamefont {C.}~\bibnamefont
  {Viermann}}, \bibinfo {author} {\bibfnamefont {M.}~\bibnamefont {Sparn}},
  \bibinfo {author} {\bibfnamefont {N.}~\bibnamefont {Liebster}}, \bibinfo
  {author} {\bibfnamefont {M.}~\bibnamefont {Hans}}, \bibinfo {author}
  {\bibfnamefont {E.}~\bibnamefont {Kath}}, \bibinfo {author} {\bibfnamefont
  {{\'A}.}~\bibnamefont {Parra-L{\'o}pez}}, \bibinfo {author} {\bibfnamefont
  {M.}~\bibnamefont {Tolosa-Sime{\'o}n}}, \bibinfo {author} {\bibfnamefont
  {N.}~\bibnamefont {S{\'a}nchez-Kuntz}}, \bibinfo {author} {\bibfnamefont
  {T.}~\bibnamefont {Haas}}, \bibinfo {author} {\bibfnamefont {H.}~\bibnamefont
  {Strobel}}, \bibinfo {author} {\bibfnamefont {S.}~\bibnamefont
  {Floerchinger}},\ and\ \bibinfo {author} {\bibfnamefont {M.~K.}\ \bibnamefont
  {Oberthaler}},\ }\href {https://doi.org/10.1038/s41586-022-05313-9}
  {\bibfield  {journal} {\bibinfo  {journal} {Nature}\ }\textbf {\bibinfo
  {volume} {611}},\ \bibinfo {pages} {260} (\bibinfo {year}
  {2022})}\BibitemShut {NoStop}%
\bibitem [{\citenamefont {Tolosa-Sime\'on}\ \emph {et~al.}(2022)\citenamefont
  {Tolosa-Sime\'on}, \citenamefont {Parra-L\'opez}, \citenamefont
  {S\'anchez-Kuntz}, \citenamefont {Haas}, \citenamefont {Viermann},
  \citenamefont {Sparn}, \citenamefont {Liebster}, \citenamefont {Hans},
  \citenamefont {Kath}, \citenamefont {Strobel}, \citenamefont {Oberthaler},\
  and\ \citenamefont {Floerchinger}}]{Tolosa2022}%
  \BibitemOpen
  \bibfield  {author} {\bibinfo {author} {\bibfnamefont {M.}~\bibnamefont
  {Tolosa-Sime\'on}}, \bibinfo {author} {\bibfnamefont {A.}~\bibnamefont
  {Parra-L\'opez}}, \bibinfo {author} {\bibfnamefont {N.}~\bibnamefont
  {S\'anchez-Kuntz}}, \bibinfo {author} {\bibfnamefont {T.}~\bibnamefont
  {Haas}}, \bibinfo {author} {\bibfnamefont {C.}~\bibnamefont {Viermann}},
  \bibinfo {author} {\bibfnamefont {M.}~\bibnamefont {Sparn}}, \bibinfo
  {author} {\bibfnamefont {N.}~\bibnamefont {Liebster}}, \bibinfo {author}
  {\bibfnamefont {M.}~\bibnamefont {Hans}}, \bibinfo {author} {\bibfnamefont
  {E.}~\bibnamefont {Kath}}, \bibinfo {author} {\bibfnamefont {H.}~\bibnamefont
  {Strobel}}, \bibinfo {author} {\bibfnamefont {M.~K.}\ \bibnamefont
  {Oberthaler}},\ and\ \bibinfo {author} {\bibfnamefont {S.}~\bibnamefont
  {Floerchinger}},\ }\href {https://doi.org/10.1103/PhysRevA.106.033313}
  {\bibfield  {journal} {\bibinfo  {journal} {Phys. Rev. A}\ }\textbf {\bibinfo
  {volume} {106}},\ \bibinfo {pages} {033313} (\bibinfo {year}
  {2022})}\BibitemShut {NoStop}%
\bibitem [{\citenamefont {S\'anchez-Kuntz}\ \emph {et~al.}(2022)\citenamefont
  {S\'anchez-Kuntz}, \citenamefont {Parra-L\'opez}, \citenamefont
  {Tolosa-Sime\'on}, \citenamefont {Haas},\ and\ \citenamefont
  {Floerchinger}}]{Sanchez2022}%
  \BibitemOpen
  \bibfield  {author} {\bibinfo {author} {\bibfnamefont {N.}~\bibnamefont
  {S\'anchez-Kuntz}}, \bibinfo {author} {\bibfnamefont {A.}~\bibnamefont
  {Parra-L\'opez}}, \bibinfo {author} {\bibfnamefont {M.}~\bibnamefont
  {Tolosa-Sime\'on}}, \bibinfo {author} {\bibfnamefont {T.}~\bibnamefont
  {Haas}},\ and\ \bibinfo {author} {\bibfnamefont {S.}~\bibnamefont
  {Floerchinger}},\ }\href {https://doi.org/10.1103/PhysRevD.105.105020}
  {\bibfield  {journal} {\bibinfo  {journal} {Phys. Rev. D}\ }\textbf {\bibinfo
  {volume} {105}},\ \bibinfo {pages} {105020} (\bibinfo {year}
  {2022})}\BibitemShut {NoStop}%
\bibitem [{\citenamefont {Unruh}(1981)}]{Unruh1981}%
  \BibitemOpen
  \bibfield  {author} {\bibinfo {author} {\bibfnamefont {W.~G.}\ \bibnamefont
  {Unruh}},\ }\href {https://doi.org/10.1103/PhysRevLett.46.1351} {\bibfield
  {journal} {\bibinfo  {journal} {Phys. Rev. Lett.}\ }\textbf {\bibinfo
  {volume} {46}},\ \bibinfo {pages} {1351} (\bibinfo {year}
  {1981})}\BibitemShut {NoStop}%
\bibitem [{\citenamefont {Unruh}(1995)}]{Unruh1995}%
  \BibitemOpen
  \bibfield  {author} {\bibinfo {author} {\bibfnamefont {W.~G.}\ \bibnamefont
  {Unruh}},\ }\href {https://doi.org/10.1103/PhysRevD.51.2827} {\bibfield
  {journal} {\bibinfo  {journal} {Phys. Rev. D}\ }\textbf {\bibinfo {volume}
  {51}},\ \bibinfo {pages} {2827} (\bibinfo {year} {1995})}\BibitemShut
  {NoStop}%
\bibitem [{\citenamefont {William G.~Unruh}(2007)}]{UnruhSchuetzhold2007}%
  \BibitemOpen
  \bibinfo {editor} {\bibfnamefont {R.~S.}\ \bibnamefont {William G.~Unruh}},\
  ed.,\ \href {https://doi.org/10.1007/3-540-70859-6} {\emph {\bibinfo {title}
  {Quantum Analogues: From Phase Transitions to Black Holes and Cosmology}}}\
  (\bibinfo  {publisher} {Springer Berlin Heidelberg},\ \bibinfo {year}
  {2007})\BibitemShut {NoStop}%
\bibitem [{\citenamefont {Volovik}(2009)}]{Volovik2009}%
  \BibitemOpen
  \bibfield  {author} {\bibinfo {author} {\bibfnamefont {G.~E.}\ \bibnamefont
  {Volovik}},\ }\href
  {https://doi.org/10.1093/acprof:oso/9780199564842.001.0001} {\emph {\bibinfo
  {title} {The {U}niverse in a {H}elium {D}roplet}}}\ (\bibinfo  {publisher}
  {Oxford University Press},\ \bibinfo {address} {Oxford},\ \bibinfo {year}
  {2009})\BibitemShut {NoStop}%
\bibitem [{\citenamefont {Feynman}(1982)}]{Feynman1982}%
  \BibitemOpen
  \bibfield  {author} {\bibinfo {author} {\bibfnamefont {R.~P.}\ \bibnamefont
  {Feynman}},\ }\href {https://doi.org/10.1007/bf02650179} {\bibfield
  {journal} {\bibinfo  {journal} {International Journal of Theoretical
  Physics}\ }\textbf {\bibinfo {volume} {21}},\ \bibinfo {pages} {467–488}
  (\bibinfo {year} {1982})}\BibitemShut {NoStop}%
\bibitem [{\citenamefont {Jaksch}\ \emph {et~al.}(1998)\citenamefont {Jaksch},
  \citenamefont {Bruder}, \citenamefont {Cirac}, \citenamefont {Gardiner},\
  and\ \citenamefont {Zoller}}]{Jaksch1998}%
  \BibitemOpen
  \bibfield  {author} {\bibinfo {author} {\bibfnamefont {D.}~\bibnamefont
  {Jaksch}}, \bibinfo {author} {\bibfnamefont {C.}~\bibnamefont {Bruder}},
  \bibinfo {author} {\bibfnamefont {J.~I.}\ \bibnamefont {Cirac}}, \bibinfo
  {author} {\bibfnamefont {C.~W.}\ \bibnamefont {Gardiner}},\ and\ \bibinfo
  {author} {\bibfnamefont {P.}~\bibnamefont {Zoller}},\ }\href
  {https://doi.org/10.1103/physrevlett.81.3108} {\bibfield  {journal} {\bibinfo
   {journal} {Physical Review Letters}\ }\textbf {\bibinfo {volume} {81}},\
  \bibinfo {pages} {3108–3111} (\bibinfo {year} {1998})}\BibitemShut
  {NoStop}%
\bibitem [{\citenamefont {Bloch}\ \emph {et~al.}(2012)\citenamefont {Bloch},
  \citenamefont {Dalibard},\ and\ \citenamefont {Nascimbène}}]{Bloch2012}%
  \BibitemOpen
  \bibfield  {author} {\bibinfo {author} {\bibfnamefont {I.}~\bibnamefont
  {Bloch}}, \bibinfo {author} {\bibfnamefont {J.}~\bibnamefont {Dalibard}},\
  and\ \bibinfo {author} {\bibfnamefont {S.}~\bibnamefont {Nascimbène}},\
  }\href {https://doi.org/10.1038/nphys2259} {\bibfield  {journal} {\bibinfo
  {journal} {Nature Physics}\ }\textbf {\bibinfo {volume} {8}},\ \bibinfo
  {pages} {267–276} (\bibinfo {year} {2012})}\BibitemShut {NoStop}%
\bibitem [{\citenamefont {Cirac}\ and\ \citenamefont
  {Zoller}(2012)}]{Cirac2012}%
  \BibitemOpen
  \bibfield  {author} {\bibinfo {author} {\bibfnamefont {J.~I.}\ \bibnamefont
  {Cirac}}\ and\ \bibinfo {author} {\bibfnamefont {P.}~\bibnamefont {Zoller}},\
  }\href {https://doi.org/10.1038/nphys2275} {\bibfield  {journal} {\bibinfo
  {journal} {Nature Physics}\ }\textbf {\bibinfo {volume} {8}},\ \bibinfo
  {pages} {264–266} (\bibinfo {year} {2012})}\BibitemShut {NoStop}%
\bibitem [{\citenamefont {Georgescu}\ \emph {et~al.}(2014)\citenamefont
  {Georgescu}, \citenamefont {Ashhab},\ and\ \citenamefont
  {Nori}}]{Georgescu2014}%
  \BibitemOpen
  \bibfield  {author} {\bibinfo {author} {\bibfnamefont {I.}~\bibnamefont
  {Georgescu}}, \bibinfo {author} {\bibfnamefont {S.}~\bibnamefont {Ashhab}},\
  and\ \bibinfo {author} {\bibfnamefont {F.}~\bibnamefont {Nori}},\ }\href
  {https://doi.org/10.1103/revmodphys.86.153} {\bibfield  {journal} {\bibinfo
  {journal} {Reviews of Modern Physics}\ }\textbf {\bibinfo {volume} {86}},\
  \bibinfo {pages} {153–185} (\bibinfo {year} {2014})}\BibitemShut {NoStop}%
\bibitem [{\citenamefont {Aidelsburger}\ \emph {et~al.}(2021)\citenamefont
  {Aidelsburger}, \citenamefont {Barbiero}, \citenamefont {Bermudez},
  \citenamefont {Chanda}, \citenamefont {Dauphin}, \citenamefont
  {González-Cuadra}, \citenamefont {Grzybowski}, \citenamefont {Hands},
  \citenamefont {Jendrzejewski}, \citenamefont {J\"{u}nemann}, \citenamefont
  {Juzeliūnas}, \citenamefont {Kasper}, \citenamefont {Piga}, \citenamefont
  {Ran}, \citenamefont {Rizzi}, \citenamefont {Sierra}, \citenamefont
  {Tagliacozzo}, \citenamefont {Tirrito}, \citenamefont {Zache}, \citenamefont
  {Zakrzewski}, \citenamefont {Zohar},\ and\ \citenamefont
  {Lewenstein}}]{Aidelsburger2021}%
  \BibitemOpen
  \bibfield  {author} {\bibinfo {author} {\bibfnamefont {M.}~\bibnamefont
  {Aidelsburger}}, \bibinfo {author} {\bibfnamefont {L.}~\bibnamefont
  {Barbiero}}, \bibinfo {author} {\bibfnamefont {A.}~\bibnamefont {Bermudez}},
  \bibinfo {author} {\bibfnamefont {T.}~\bibnamefont {Chanda}}, \bibinfo
  {author} {\bibfnamefont {A.}~\bibnamefont {Dauphin}}, \bibinfo {author}
  {\bibfnamefont {D.}~\bibnamefont {González-Cuadra}}, \bibinfo {author}
  {\bibfnamefont {P.~R.}\ \bibnamefont {Grzybowski}}, \bibinfo {author}
  {\bibfnamefont {S.}~\bibnamefont {Hands}}, \bibinfo {author} {\bibfnamefont
  {F.}~\bibnamefont {Jendrzejewski}}, \bibinfo {author} {\bibfnamefont
  {J.}~\bibnamefont {J\"{u}nemann}}, \bibinfo {author} {\bibfnamefont
  {G.}~\bibnamefont {Juzeliūnas}}, \bibinfo {author} {\bibfnamefont
  {V.}~\bibnamefont {Kasper}}, \bibinfo {author} {\bibfnamefont
  {A.}~\bibnamefont {Piga}}, \bibinfo {author} {\bibfnamefont {S.-J.}\
  \bibnamefont {Ran}}, \bibinfo {author} {\bibfnamefont {M.}~\bibnamefont
  {Rizzi}}, \bibinfo {author} {\bibfnamefont {G.}~\bibnamefont {Sierra}},
  \bibinfo {author} {\bibfnamefont {L.}~\bibnamefont {Tagliacozzo}}, \bibinfo
  {author} {\bibfnamefont {E.}~\bibnamefont {Tirrito}}, \bibinfo {author}
  {\bibfnamefont {T.~V.}\ \bibnamefont {Zache}}, \bibinfo {author}
  {\bibfnamefont {J.}~\bibnamefont {Zakrzewski}}, \bibinfo {author}
  {\bibfnamefont {E.}~\bibnamefont {Zohar}},\ and\ \bibinfo {author}
  {\bibfnamefont {M.}~\bibnamefont {Lewenstein}},\ }\bibfield  {journal}
  {\bibinfo  {journal} {Philosophical Transactions of the Royal Society A:
  Mathematical, Physical and Engineering Sciences}\ }\textbf {\bibinfo {volume}
  {380}},\ \href {https://doi.org/10.1098/rsta.2021.0064}
  {10.1098/rsta.2021.0064} (\bibinfo {year} {2021})\BibitemShut {NoStop}%
\bibitem [{\citenamefont {Mil}\ \emph {et~al.}(2020)\citenamefont {Mil},
  \citenamefont {Zache}, \citenamefont {Hegde}, \citenamefont {Xia},
  \citenamefont {Bhatt}, \citenamefont {Oberthaler}, \citenamefont {Hauke},
  \citenamefont {Berges},\ and\ \citenamefont {Jendrzejewski}}]{Mil2020}%
  \BibitemOpen
  \bibfield  {author} {\bibinfo {author} {\bibfnamefont {A.}~\bibnamefont
  {Mil}}, \bibinfo {author} {\bibfnamefont {T.~V.}\ \bibnamefont {Zache}},
  \bibinfo {author} {\bibfnamefont {A.}~\bibnamefont {Hegde}}, \bibinfo
  {author} {\bibfnamefont {A.}~\bibnamefont {Xia}}, \bibinfo {author}
  {\bibfnamefont {R.~P.}\ \bibnamefont {Bhatt}}, \bibinfo {author}
  {\bibfnamefont {M.~K.}\ \bibnamefont {Oberthaler}}, \bibinfo {author}
  {\bibfnamefont {P.}~\bibnamefont {Hauke}}, \bibinfo {author} {\bibfnamefont
  {J.}~\bibnamefont {Berges}},\ and\ \bibinfo {author} {\bibfnamefont
  {F.}~\bibnamefont {Jendrzejewski}},\ }\href
  {https://doi.org/10.1126/science.aaz5312} {\bibfield  {journal} {\bibinfo
  {journal} {Science}\ }\textbf {\bibinfo {volume} {367}},\ \bibinfo {pages}
  {1128–1130} (\bibinfo {year} {2020})}\BibitemShut {NoStop}%
\bibitem [{\citenamefont {Yang}\ \emph {et~al.}(2020)\citenamefont {Yang},
  \citenamefont {Sun}, \citenamefont {Ott}, \citenamefont {Wang}, \citenamefont
  {Zache}, \citenamefont {Halimeh}, \citenamefont {Yuan}, \citenamefont
  {Hauke},\ and\ \citenamefont {Pan}}]{Yang2020}%
  \BibitemOpen
  \bibfield  {author} {\bibinfo {author} {\bibfnamefont {B.}~\bibnamefont
  {Yang}}, \bibinfo {author} {\bibfnamefont {H.}~\bibnamefont {Sun}}, \bibinfo
  {author} {\bibfnamefont {R.}~\bibnamefont {Ott}}, \bibinfo {author}
  {\bibfnamefont {H.-Y.}\ \bibnamefont {Wang}}, \bibinfo {author}
  {\bibfnamefont {T.~V.}\ \bibnamefont {Zache}}, \bibinfo {author}
  {\bibfnamefont {J.~C.}\ \bibnamefont {Halimeh}}, \bibinfo {author}
  {\bibfnamefont {Z.-S.}\ \bibnamefont {Yuan}}, \bibinfo {author}
  {\bibfnamefont {P.}~\bibnamefont {Hauke}},\ and\ \bibinfo {author}
  {\bibfnamefont {J.-W.}\ \bibnamefont {Pan}},\ }\href
  {https://doi.org/10.1038/s41586-020-2910-8} {\bibfield  {journal} {\bibinfo
  {journal} {Nature}\ }\textbf {\bibinfo {volume} {587}},\ \bibinfo {pages}
  {392–396} (\bibinfo {year} {2020})}\BibitemShut {NoStop}%
\bibitem [{\citenamefont {Schweizer}\ \emph {et~al.}(2019)\citenamefont
  {Schweizer}, \citenamefont {Grusdt}, \citenamefont {Berngruber},
  \citenamefont {Barbiero}, \citenamefont {Demler}, \citenamefont {Goldman},
  \citenamefont {Bloch},\ and\ \citenamefont {Aidelsburger}}]{Schweizer2019}%
  \BibitemOpen
  \bibfield  {author} {\bibinfo {author} {\bibfnamefont {C.}~\bibnamefont
  {Schweizer}}, \bibinfo {author} {\bibfnamefont {F.}~\bibnamefont {Grusdt}},
  \bibinfo {author} {\bibfnamefont {M.}~\bibnamefont {Berngruber}}, \bibinfo
  {author} {\bibfnamefont {L.}~\bibnamefont {Barbiero}}, \bibinfo {author}
  {\bibfnamefont {E.}~\bibnamefont {Demler}}, \bibinfo {author} {\bibfnamefont
  {N.}~\bibnamefont {Goldman}}, \bibinfo {author} {\bibfnamefont
  {I.}~\bibnamefont {Bloch}},\ and\ \bibinfo {author} {\bibfnamefont
  {M.}~\bibnamefont {Aidelsburger}},\ }\href
  {https://doi.org/10.1038/s41567-019-0649-7} {\bibfield  {journal} {\bibinfo
  {journal} {Nature Physics}\ }\textbf {\bibinfo {volume} {15}},\ \bibinfo
  {pages} {1168–1173} (\bibinfo {year} {2019})}\BibitemShut {NoStop}%
\bibitem [{\citenamefont {G\"{o}rg}\ \emph {et~al.}(2019)\citenamefont
  {G\"{o}rg}, \citenamefont {Sandholzer}, \citenamefont {Minguzzi},
  \citenamefont {Desbuquois}, \citenamefont {Messer},\ and\ \citenamefont
  {Esslinger}}]{Goerg2019}%
  \BibitemOpen
  \bibfield  {author} {\bibinfo {author} {\bibfnamefont {F.}~\bibnamefont
  {G\"{o}rg}}, \bibinfo {author} {\bibfnamefont {K.}~\bibnamefont
  {Sandholzer}}, \bibinfo {author} {\bibfnamefont {J.}~\bibnamefont
  {Minguzzi}}, \bibinfo {author} {\bibfnamefont {R.}~\bibnamefont
  {Desbuquois}}, \bibinfo {author} {\bibfnamefont {M.}~\bibnamefont {Messer}},\
  and\ \bibinfo {author} {\bibfnamefont {T.}~\bibnamefont {Esslinger}},\ }\href
  {https://doi.org/10.1038/s41567-019-0615-4} {\bibfield  {journal} {\bibinfo
  {journal} {Nature Physics}\ }\textbf {\bibinfo {volume} {15}},\ \bibinfo
  {pages} {1161–1167} (\bibinfo {year} {2019})}\BibitemShut {NoStop}%
\bibitem [{\citenamefont {Barcel{\'o}}\ \emph {et~al.}(2011)\citenamefont
  {Barcel{\'o}}, \citenamefont {Liberati},\ and\ \citenamefont
  {Visser}}]{Barcelo2011}%
  \BibitemOpen
  \bibfield  {author} {\bibinfo {author} {\bibfnamefont {C.}~\bibnamefont
  {Barcel{\'o}}}, \bibinfo {author} {\bibfnamefont {S.}~\bibnamefont
  {Liberati}},\ and\ \bibinfo {author} {\bibfnamefont {M.}~\bibnamefont
  {Visser}},\ }\href {https://doi.org/10.12942/lrr-2011-3} {\bibfield
  {journal} {\bibinfo  {journal} {Living Rev. Relativ.}\ }\textbf {\bibinfo
  {volume} {14}},\ \bibinfo {pages} {3} (\bibinfo {year} {2011})}\BibitemShut
  {NoStop}%
\bibitem [{\citenamefont {Visser}\ \emph {et~al.}(2002)\citenamefont {Visser},
  \citenamefont {Barcel{\'o}},\ and\ \citenamefont {Liberati}}]{Visser2002}%
  \BibitemOpen
  \bibfield  {author} {\bibinfo {author} {\bibfnamefont {M.}~\bibnamefont
  {Visser}}, \bibinfo {author} {\bibfnamefont {C.}~\bibnamefont
  {Barcel{\'o}}},\ and\ \bibinfo {author} {\bibfnamefont {S.}~\bibnamefont
  {Liberati}},\ }\href {https://doi.org/10.1023/A:1020180409214} {\bibfield
  {journal} {\bibinfo  {journal} {Gen. Relativ. Gravit.}\ }\textbf {\bibinfo
  {volume} {34}},\ \bibinfo {pages} {1719} (\bibinfo {year}
  {2002})}\BibitemShut {NoStop}%
\bibitem [{\citenamefont {Novello}\ \emph {et~al.}(2002)\citenamefont
  {Novello}, \citenamefont {Visser},\ and\ \citenamefont
  {Volovik}}]{Novello2002}%
  \BibitemOpen
  \bibinfo {editor} {\bibfnamefont {M.}~\bibnamefont {Novello}}, \bibinfo
  {editor} {\bibfnamefont {M.}~\bibnamefont {Visser}},\ and\ \bibinfo {editor}
  {\bibfnamefont {G.~E.}\ \bibnamefont {Volovik}},\ eds.,\ \href
  {https://doi.org/10.1142/4861} {\emph {\bibinfo {title} {{A}rtificial {B}lack
  {H}oles}}}\ (\bibinfo  {publisher} {World Scientific Publishing},\ \bibinfo
  {address} {Singapore},\ \bibinfo {year} {2002})\BibitemShut {NoStop}%
\bibitem [{\citenamefont {Jacquet}\ \emph {et~al.}(2020)\citenamefont
  {Jacquet}, \citenamefont {Weinfurtner},\ and\ \citenamefont
  {K{\"o}nig}}]{Jacquet2020}%
  \BibitemOpen
  \bibfield  {author} {\bibinfo {author} {\bibfnamefont {M.~J.}\ \bibnamefont
  {Jacquet}}, \bibinfo {author} {\bibfnamefont {S.}~\bibnamefont
  {Weinfurtner}},\ and\ \bibinfo {author} {\bibfnamefont {F.}~\bibnamefont
  {K{\"o}nig}},\ }\href {https://doi.org/10.1098/rsta.2019.0239} {\bibfield
  {journal} {\bibinfo  {journal} {Philos. Trans. Royal Soc. A}\ }\textbf
  {\bibinfo {volume} {378}},\ \bibinfo {pages} {20190239} (\bibinfo {year}
  {2020})}\BibitemShut {NoStop}%
\bibitem [{\citenamefont {Barcel{\'o}}\ \emph {et~al.}(2003)\citenamefont
  {Barcel{\'o}}, \citenamefont {Liberati},\ and\ \citenamefont
  {Visser}}]{Barcelo2003c}%
  \BibitemOpen
  \bibfield  {author} {\bibinfo {author} {\bibfnamefont {C.}~\bibnamefont
  {Barcel{\'o}}}, \bibinfo {author} {\bibfnamefont {S.}~\bibnamefont
  {Liberati}},\ and\ \bibinfo {author} {\bibfnamefont {M.}~\bibnamefont
  {Visser}},\ }\href {https://doi.org/10.1103/PhysRevA.68.053613} {\bibfield
  {journal} {\bibinfo  {journal} {Phys. Rev. A}\ }\textbf {\bibinfo {volume}
  {68}},\ \bibinfo {pages} {053613} (\bibinfo {year} {2003})}\BibitemShut
  {NoStop}%
\bibitem [{\citenamefont {Fedichev}\ and\ \citenamefont
  {Fischer}(2003)}]{Fedichev2003}%
  \BibitemOpen
  \bibfield  {author} {\bibinfo {author} {\bibfnamefont {P.~O.}\ \bibnamefont
  {Fedichev}}\ and\ \bibinfo {author} {\bibfnamefont {U.~R.}\ \bibnamefont
  {Fischer}},\ }\href {https://doi.org/10.1103/PhysRevLett.91.240407}
  {\bibfield  {journal} {\bibinfo  {journal} {Phys. Rev. Lett.}\ }\textbf
  {\bibinfo {volume} {91}},\ \bibinfo {pages} {240407} (\bibinfo {year}
  {2003})}\BibitemShut {NoStop}%
\bibitem [{\citenamefont {Fedichev}\ and\ \citenamefont
  {Fischer}(2004)}]{Fedichev2004}%
  \BibitemOpen
  \bibfield  {author} {\bibinfo {author} {\bibfnamefont {P.~O.}\ \bibnamefont
  {Fedichev}}\ and\ \bibinfo {author} {\bibfnamefont {U.~R.}\ \bibnamefont
  {Fischer}},\ }\href {https://doi.org/10.1103/PhysRevA.69.033602} {\bibfield
  {journal} {\bibinfo  {journal} {Phys. Rev. A}\ }\textbf {\bibinfo {volume}
  {69}},\ \bibinfo {pages} {033602} (\bibinfo {year} {2004})}\BibitemShut
  {NoStop}%
\bibitem [{\citenamefont {Fischer}(2004)}]{Fischer2004}%
  \BibitemOpen
  \bibfield  {author} {\bibinfo {author} {\bibfnamefont {U.~R.}\ \bibnamefont
  {Fischer}},\ }\href {https://doi.org/10.1142/S0217732304015099} {\bibfield
  {journal} {\bibinfo  {journal} {Mod. Phys. Lett. A}\ }\textbf {\bibinfo
  {volume} {19}},\ \bibinfo {pages} {1789} (\bibinfo {year}
  {2004})}\BibitemShut {NoStop}%
\bibitem [{\citenamefont {Fischer}\ and\ \citenamefont
  {Sch\"utzhold}(2004)}]{Fischer2004b}%
  \BibitemOpen
  \bibfield  {author} {\bibinfo {author} {\bibfnamefont {U.~R.}\ \bibnamefont
  {Fischer}}\ and\ \bibinfo {author} {\bibfnamefont {R.}~\bibnamefont
  {Sch\"utzhold}},\ }\href {https://doi.org/10.1103/PhysRevA.70.063615}
  {\bibfield  {journal} {\bibinfo  {journal} {Phys. Rev. A}\ }\textbf {\bibinfo
  {volume} {70}},\ \bibinfo {pages} {063615} (\bibinfo {year}
  {2004})}\BibitemShut {NoStop}%
\bibitem [{\citenamefont {Uhlmann}\ \emph {et~al.}(2005)\citenamefont
  {Uhlmann}, \citenamefont {Xu},\ and\ \citenamefont
  {Sch{\"u}tzhold}}]{Uhlmann2005}%
  \BibitemOpen
  \bibfield  {author} {\bibinfo {author} {\bibfnamefont {M.}~\bibnamefont
  {Uhlmann}}, \bibinfo {author} {\bibfnamefont {Y.}~\bibnamefont {Xu}},\ and\
  \bibinfo {author} {\bibfnamefont {R.}~\bibnamefont {Sch{\"u}tzhold}},\ }\href
  {https://doi.org/10.1088/1367-2630/7/1/248} {\bibfield  {journal} {\bibinfo
  {journal} {New J. Phys.}\ }\textbf {\bibinfo {volume} {7}},\ \bibinfo {pages}
  {248} (\bibinfo {year} {2005})}\BibitemShut {NoStop}%
\bibitem [{\citenamefont {Calzetta}\ and\ \citenamefont
  {Hu}(2005)}]{Calzetta2005}%
  \BibitemOpen
  \bibfield  {author} {\bibinfo {author} {\bibfnamefont {E.~A.}\ \bibnamefont
  {Calzetta}}\ and\ \bibinfo {author} {\bibfnamefont {B.~L.}\ \bibnamefont
  {Hu}},\ }\href {https://doi.org/10.1007/s10773-005-8889-y} {\bibfield
  {journal} {\bibinfo  {journal} {Int. J. Theor. Phys.}\ }\textbf {\bibinfo
  {volume} {44}},\ \bibinfo {pages} {1691} (\bibinfo {year}
  {2005})}\BibitemShut {NoStop}%
\bibitem [{\citenamefont {Liberati}\ \emph
  {et~al.}(2006{\natexlab{a}})\citenamefont {Liberati}, \citenamefont
  {Visser},\ and\ \citenamefont {Weinfurtner}}]{Liberati2006}%
  \BibitemOpen
  \bibfield  {author} {\bibinfo {author} {\bibfnamefont {S.}~\bibnamefont
  {Liberati}}, \bibinfo {author} {\bibfnamefont {M.}~\bibnamefont {Visser}},\
  and\ \bibinfo {author} {\bibfnamefont {S.}~\bibnamefont {Weinfurtner}},\
  }\href {https://doi.org/10.1088/0264-9381/23/9/023} {\bibfield  {journal}
  {\bibinfo  {journal} {Class. Quantum Gravity}\ }\textbf {\bibinfo {volume}
  {23}},\ \bibinfo {pages} {3129} (\bibinfo {year}
  {2006}{\natexlab{a}})}\BibitemShut {NoStop}%
\bibitem [{\citenamefont {Liberati}\ \emph
  {et~al.}(2006{\natexlab{b}})\citenamefont {Liberati}, \citenamefont
  {Visser},\ and\ \citenamefont {Weinfurtner}}]{Liberati2006b}%
  \BibitemOpen
  \bibfield  {author} {\bibinfo {author} {\bibfnamefont {S.}~\bibnamefont
  {Liberati}}, \bibinfo {author} {\bibfnamefont {M.}~\bibnamefont {Visser}},\
  and\ \bibinfo {author} {\bibfnamefont {S.}~\bibnamefont {Weinfurtner}},\
  }\href {https://doi.org/10.1103/PhysRevLett.96.151301} {\bibfield  {journal}
  {\bibinfo  {journal} {Phys. Rev. Lett.}\ }\textbf {\bibinfo {volume} {96}},\
  \bibinfo {pages} {151301} (\bibinfo {year} {2006}{\natexlab{b}})}\BibitemShut
  {NoStop}%
\bibitem [{\citenamefont {Jain}\ \emph {et~al.}(2007)\citenamefont {Jain},
  \citenamefont {Weinfurtner}, \citenamefont {Visser},\ and\ \citenamefont
  {Gardiner}}]{Weinfurtner2007}%
  \BibitemOpen
  \bibfield  {author} {\bibinfo {author} {\bibfnamefont {P.}~\bibnamefont
  {Jain}}, \bibinfo {author} {\bibfnamefont {S.}~\bibnamefont {Weinfurtner}},
  \bibinfo {author} {\bibfnamefont {M.}~\bibnamefont {Visser}},\ and\ \bibinfo
  {author} {\bibfnamefont {C.~W.}\ \bibnamefont {Gardiner}},\ }\href
  {https://doi.org/10.1103/PhysRevA.76.033616} {\bibfield  {journal} {\bibinfo
  {journal} {Phys. Rev. A}\ }\textbf {\bibinfo {volume} {76}},\ \bibinfo
  {pages} {033616} (\bibinfo {year} {2007})}\BibitemShut {NoStop}%
\bibitem [{\citenamefont {Prain}\ \emph {et~al.}(2010)\citenamefont {Prain},
  \citenamefont {Fagnocchi},\ and\ \citenamefont {Liberati}}]{Prain2010}%
  \BibitemOpen
  \bibfield  {author} {\bibinfo {author} {\bibfnamefont {A.}~\bibnamefont
  {Prain}}, \bibinfo {author} {\bibfnamefont {S.}~\bibnamefont {Fagnocchi}},\
  and\ \bibinfo {author} {\bibfnamefont {S.}~\bibnamefont {Liberati}},\ }\href
  {https://doi.org/10.1103/PhysRevD.82.105018} {\bibfield  {journal} {\bibinfo
  {journal} {Phys. Rev. D}\ }\textbf {\bibinfo {volume} {82}},\ \bibinfo
  {pages} {105018} (\bibinfo {year} {2010})}\BibitemShut {NoStop}%
\bibitem [{\citenamefont {Bili\'{c}}\ and\ \citenamefont
  {Toli\'{c}}(2013)}]{Bilic2013}%
  \BibitemOpen
  \bibfield  {author} {\bibinfo {author} {\bibfnamefont {N.}~\bibnamefont
  {Bili\'{c}}}\ and\ \bibinfo {author} {\bibfnamefont {D.}~\bibnamefont
  {Toli\'{c}}},\ }\href {https://doi.org/10.1103/PhysRevD.88.105002} {\bibfield
   {journal} {\bibinfo  {journal} {Phys. Rev. D}\ }\textbf {\bibinfo {volume}
  {88}},\ \bibinfo {pages} {105002} (\bibinfo {year} {2013})}\BibitemShut
  {NoStop}%
\bibitem [{\citenamefont {Weinfurtner}\ \emph {et~al.}(2006)\citenamefont
  {Weinfurtner}, \citenamefont {Liberati},\ and\ \citenamefont
  {Visser}}]{Weinfurtner2006}%
  \BibitemOpen
  \bibfield  {author} {\bibinfo {author} {\bibfnamefont {S.}~\bibnamefont
  {Weinfurtner}}, \bibinfo {author} {\bibfnamefont {S.}~\bibnamefont
  {Liberati}},\ and\ \bibinfo {author} {\bibfnamefont {M.}~\bibnamefont
  {Visser}},\ }\href {https://doi.org/10.1088/0305-4470/39/21/s83} {\bibfield
  {journal} {\bibinfo  {journal} {J. Phys. A}\ }\textbf {\bibinfo {volume}
  {39}},\ \bibinfo {pages} {6807} (\bibinfo {year} {2006})}\BibitemShut
  {NoStop}%
\bibitem [{\citenamefont {Weinfurtner}\ \emph {et~al.}(2007)\citenamefont
  {Weinfurtner}, \citenamefont {White},\ and\ \citenamefont
  {Visser}}]{Weinfurtner2007BoseNova}%
  \BibitemOpen
  \bibfield  {author} {\bibinfo {author} {\bibfnamefont {S.}~\bibnamefont
  {Weinfurtner}}, \bibinfo {author} {\bibfnamefont {A.}~\bibnamefont {White}},\
  and\ \bibinfo {author} {\bibfnamefont {M.}~\bibnamefont {Visser}},\
  }\bibfield  {journal} {\bibinfo  {journal} {Physical Review D}\ }\textbf
  {\bibinfo {volume} {76}},\ \href {https://doi.org/10.1103/physrevd.76.124008}
  {10.1103/physrevd.76.124008} (\bibinfo {year} {2007})\BibitemShut {NoStop}%
\bibitem [{\citenamefont {Weinfurtner}\ \emph {et~al.}(2009)\citenamefont
  {Weinfurtner}, \citenamefont {Jain}, \citenamefont {Wisser},\ and\
  \citenamefont {Gardiner}}]{Weinfurtner2009}%
  \BibitemOpen
  \bibfield  {author} {\bibinfo {author} {\bibfnamefont {S.}~\bibnamefont
  {Weinfurtner}}, \bibinfo {author} {\bibfnamefont {P.}~\bibnamefont {Jain}},
  \bibinfo {author} {\bibfnamefont {M.}~\bibnamefont {Wisser}},\ and\ \bibinfo
  {author} {\bibfnamefont {C.~W.}\ \bibnamefont {Gardiner}},\ }\href
  {https://doi.org/10.1088/0264-9381/26/6/065012} {\bibfield  {journal}
  {\bibinfo  {journal} {Class. Quantum Gravity}\ }\textbf {\bibinfo {volume}
  {26}},\ \bibinfo {pages} {065012} (\bibinfo {year} {2009})}\BibitemShut
  {NoStop}%
\bibitem [{\citenamefont {Ch\"{a}}\ and\ \citenamefont
  {Fischer}(2017)}]{Fischer2017}%
  \BibitemOpen
  \bibfield  {author} {\bibinfo {author} {\bibfnamefont {S.-Y.}\ \bibnamefont
  {Ch\"{a}}}\ and\ \bibinfo {author} {\bibfnamefont {U.~R.}\ \bibnamefont
  {Fischer}},\ }\bibfield  {journal} {\bibinfo  {journal} {Physical Review
  Letters}\ }\textbf {\bibinfo {volume} {118}},\ \href
  {https://doi.org/10.1103/physrevlett.118.130404}
  {10.1103/physrevlett.118.130404} (\bibinfo {year} {2017})\BibitemShut
  {NoStop}%
\bibitem [{\citenamefont {Eckel}\ \emph {et~al.}(2018)\citenamefont {Eckel},
  \citenamefont {Kumar}, \citenamefont {Jacobson}, \citenamefont {Spielman},\
  and\ \citenamefont {Campbell}}]{Eckel2018}%
  \BibitemOpen
  \bibfield  {author} {\bibinfo {author} {\bibfnamefont {S.}~\bibnamefont
  {Eckel}}, \bibinfo {author} {\bibfnamefont {A.}~\bibnamefont {Kumar}},
  \bibinfo {author} {\bibfnamefont {T.}~\bibnamefont {Jacobson}}, \bibinfo
  {author} {\bibfnamefont {I.~B.}\ \bibnamefont {Spielman}},\ and\ \bibinfo
  {author} {\bibfnamefont {G.~K.}\ \bibnamefont {Campbell}},\ }\href
  {https://doi.org/10.1103/PhysRevX.8.021021} {\bibfield  {journal} {\bibinfo
  {journal} {Phys. Rev. X}\ }\textbf {\bibinfo {volume} {8}},\ \bibinfo {pages}
  {021021} (\bibinfo {year} {2018})}\BibitemShut {NoStop}%
\bibitem [{\citenamefont {Tajik}\ \emph {et~al.}(2023)\citenamefont {Tajik},
  \citenamefont {Gluza}, \citenamefont {Sebe}, \citenamefont {Schüttelkopf},
  \citenamefont {Cataldini}, \citenamefont {Sabino}, \citenamefont {Møller},
  \citenamefont {Ji}, \citenamefont {Erne}, \citenamefont {Guarnieri},
  \citenamefont {Sotiriadis}, \citenamefont {Eisert},\ and\ \citenamefont
  {Schmiedmayer}}]{Tajik2023}%
  \BibitemOpen
  \bibfield  {author} {\bibinfo {author} {\bibfnamefont {M.}~\bibnamefont
  {Tajik}}, \bibinfo {author} {\bibfnamefont {M.}~\bibnamefont {Gluza}},
  \bibinfo {author} {\bibfnamefont {N.}~\bibnamefont {Sebe}}, \bibinfo {author}
  {\bibfnamefont {P.}~\bibnamefont {Schüttelkopf}}, \bibinfo {author}
  {\bibfnamefont {F.}~\bibnamefont {Cataldini}}, \bibinfo {author}
  {\bibfnamefont {J.}~\bibnamefont {Sabino}}, \bibinfo {author} {\bibfnamefont
  {F.}~\bibnamefont {Møller}}, \bibinfo {author} {\bibfnamefont {S.-C.}\
  \bibnamefont {Ji}}, \bibinfo {author} {\bibfnamefont {S.}~\bibnamefont
  {Erne}}, \bibinfo {author} {\bibfnamefont {G.}~\bibnamefont {Guarnieri}},
  \bibinfo {author} {\bibfnamefont {S.}~\bibnamefont {Sotiriadis}}, \bibinfo
  {author} {\bibfnamefont {J.}~\bibnamefont {Eisert}},\ and\ \bibinfo {author}
  {\bibfnamefont {J.}~\bibnamefont {Schmiedmayer}},\ }\href
  {https://doi.org/10.1073/pnas.2301287120} {\bibfield  {journal} {\bibinfo
  {journal} {Proceedings of the National Academy of Sciences}\ }\textbf
  {\bibinfo {volume} {120}},\ \bibinfo {pages} {e2301287120} (\bibinfo {year}
  {2023})}\BibitemShut {NoStop}%
\bibitem [{\citenamefont {Hung}\ \emph {et~al.}(2013)\citenamefont {Hung},
  \citenamefont {Gurarie},\ and\ \citenamefont {Chin}}]{Hung2013}%
  \BibitemOpen
  \bibfield  {author} {\bibinfo {author} {\bibfnamefont {C.-L.}\ \bibnamefont
  {Hung}}, \bibinfo {author} {\bibfnamefont {V.}~\bibnamefont {Gurarie}},\ and\
  \bibinfo {author} {\bibfnamefont {C.}~\bibnamefont {Chin}},\ }\href
  {https://doi.org/10.1126/science.1237557} {\bibfield  {journal} {\bibinfo
  {journal} {Science}\ }\textbf {\bibinfo {volume} {341}},\ \bibinfo {pages}
  {1213–1215} (\bibinfo {year} {2013})}\BibitemShut {NoStop}%
\bibitem [{\citenamefont {Chen}\ \emph {et~al.}(2021)\citenamefont {Chen},
  \citenamefont {Khlebnikov},\ and\ \citenamefont {Hung}}]{Chen2021}%
  \BibitemOpen
  \bibfield  {author} {\bibinfo {author} {\bibfnamefont {C.-A.}\ \bibnamefont
  {Chen}}, \bibinfo {author} {\bibfnamefont {S.}~\bibnamefont {Khlebnikov}},\
  and\ \bibinfo {author} {\bibfnamefont {C.-L.}\ \bibnamefont {Hung}},\ }\href
  {https://doi.org/10.1103/PhysRevLett.127.060404} {\bibfield  {journal}
  {\bibinfo  {journal} {Phys. Rev. Lett.}\ }\textbf {\bibinfo {volume} {127}},\
  \bibinfo {pages} {060404} (\bibinfo {year} {2021})}\BibitemShut {NoStop}%
\bibitem [{\citenamefont {Wittemer}\ \emph {et~al.}(2019)\citenamefont
  {Wittemer}, \citenamefont {Hakelberg}, \citenamefont {Kiefer}, \citenamefont
  {Schr\"oder}, \citenamefont {Fey}, \citenamefont {Sch\"utzhold},
  \citenamefont {Warring},\ and\ \citenamefont {Schaetz}}]{Wittemer2019}%
  \BibitemOpen
  \bibfield  {author} {\bibinfo {author} {\bibfnamefont {M.}~\bibnamefont
  {Wittemer}}, \bibinfo {author} {\bibfnamefont {F.}~\bibnamefont {Hakelberg}},
  \bibinfo {author} {\bibfnamefont {P.}~\bibnamefont {Kiefer}}, \bibinfo
  {author} {\bibfnamefont {J.-P.}\ \bibnamefont {Schr\"oder}}, \bibinfo
  {author} {\bibfnamefont {C.}~\bibnamefont {Fey}}, \bibinfo {author}
  {\bibfnamefont {R.}~\bibnamefont {Sch\"utzhold}}, \bibinfo {author}
  {\bibfnamefont {U.}~\bibnamefont {Warring}},\ and\ \bibinfo {author}
  {\bibfnamefont {T.}~\bibnamefont {Schaetz}},\ }\href
  {https://doi.org/10.1103/PhysRevLett.123.180502} {\bibfield  {journal}
  {\bibinfo  {journal} {Phys. Rev. Lett.}\ }\textbf {\bibinfo {volume} {123}},\
  \bibinfo {pages} {180502} (\bibinfo {year} {2019})}\BibitemShut {NoStop}%
\bibitem [{\citenamefont {Steinhauer}\ \emph {et~al.}(2022)\citenamefont
  {Steinhauer}, \citenamefont {Abuzarli}, \citenamefont {Aladjidi},
  \citenamefont {Bienaimé}, \citenamefont {Piekarski}, \citenamefont {Liu},
  \citenamefont {Giacobino}, \citenamefont {Bramati},\ and\ \citenamefont
  {Glorieux}}]{Steinhauer2022}%
  \BibitemOpen
  \bibfield  {author} {\bibinfo {author} {\bibfnamefont {J.}~\bibnamefont
  {Steinhauer}}, \bibinfo {author} {\bibfnamefont {M.}~\bibnamefont
  {Abuzarli}}, \bibinfo {author} {\bibfnamefont {T.}~\bibnamefont {Aladjidi}},
  \bibinfo {author} {\bibfnamefont {T.}~\bibnamefont {Bienaimé}}, \bibinfo
  {author} {\bibfnamefont {C.}~\bibnamefont {Piekarski}}, \bibinfo {author}
  {\bibfnamefont {W.}~\bibnamefont {Liu}}, \bibinfo {author} {\bibfnamefont
  {E.}~\bibnamefont {Giacobino}}, \bibinfo {author} {\bibfnamefont
  {A.}~\bibnamefont {Bramati}},\ and\ \bibinfo {author} {\bibfnamefont
  {Q.}~\bibnamefont {Glorieux}},\ }\bibfield  {journal} {\bibinfo  {journal}
  {Nature Communications}\ }\textbf {\bibinfo {volume} {13}},\ \href
  {https://doi.org/10.1038/s41467-022-30603-1} {10.1038/s41467-022-30603-1}
  (\bibinfo {year} {2022})\BibitemShut {NoStop}%
\bibitem [{\citenamefont {Horstmann}\ \emph {et~al.}(2010)\citenamefont
  {Horstmann}, \citenamefont {Reznik}, \citenamefont {Fagnocchi},\ and\
  \citenamefont {Cirac}}]{Horstmann2010}%
  \BibitemOpen
  \bibfield  {author} {\bibinfo {author} {\bibfnamefont {B.}~\bibnamefont
  {Horstmann}}, \bibinfo {author} {\bibfnamefont {B.}~\bibnamefont {Reznik}},
  \bibinfo {author} {\bibfnamefont {S.}~\bibnamefont {Fagnocchi}},\ and\
  \bibinfo {author} {\bibfnamefont {J.~I.}\ \bibnamefont {Cirac}},\ }\href
  {https://doi.org/10.1103/PhysRevLett.104.250403} {\bibfield  {journal}
  {\bibinfo  {journal} {Phys. Rev. Lett.}\ }\textbf {\bibinfo {volume} {104}},\
  \bibinfo {pages} {250403} (\bibinfo {year} {2010})}\BibitemShut {NoStop}%
\bibitem [{\citenamefont {Weinfurtner}\ \emph {et~al.}(2011)\citenamefont
  {Weinfurtner}, \citenamefont {Tedford}, \citenamefont {Penrice},
  \citenamefont {Unruh},\ and\ \citenamefont {Lawrence}}]{Weinfurtner2011}%
  \BibitemOpen
  \bibfield  {author} {\bibinfo {author} {\bibfnamefont {S.}~\bibnamefont
  {Weinfurtner}}, \bibinfo {author} {\bibfnamefont {E.~W.}\ \bibnamefont
  {Tedford}}, \bibinfo {author} {\bibfnamefont {M.~C.~J.}\ \bibnamefont
  {Penrice}}, \bibinfo {author} {\bibfnamefont {W.~G.}\ \bibnamefont {Unruh}},\
  and\ \bibinfo {author} {\bibfnamefont {G.~A.}\ \bibnamefont {Lawrence}},\
  }\href {https://doi.org/10.1103/PhysRevLett.106.021302} {\bibfield  {journal}
  {\bibinfo  {journal} {Phys. Rev. Lett.}\ }\textbf {\bibinfo {volume} {106}},\
  \bibinfo {pages} {021302} (\bibinfo {year} {2011})}\BibitemShut {NoStop}%
\bibitem [{\citenamefont {Steinhauer}(2016)}]{Steinhauer2016}%
  \BibitemOpen
  \bibfield  {author} {\bibinfo {author} {\bibfnamefont {J.}~\bibnamefont
  {Steinhauer}},\ }\href {https://doi.org/10.1038/nphys3863} {\bibfield
  {journal} {\bibinfo  {journal} {Nat. Phys.}\ }\textbf {\bibinfo {volume}
  {12}},\ \bibinfo {pages} {959} (\bibinfo {year} {2016})}\BibitemShut
  {NoStop}%
\bibitem [{\citenamefont {Mu{\~n}oz~de Nova}\ \emph {et~al.}(2019)\citenamefont
  {Mu{\~n}oz~de Nova}, \citenamefont {Golubkov}, \citenamefont {Kolobov},\ and\
  \citenamefont {Steinhauer}}]{MunozdeNova2019}%
  \BibitemOpen
  \bibfield  {author} {\bibinfo {author} {\bibfnamefont {J.~R.}\ \bibnamefont
  {Mu{\~n}oz~de Nova}}, \bibinfo {author} {\bibfnamefont {K.}~\bibnamefont
  {Golubkov}}, \bibinfo {author} {\bibfnamefont {V.~I.}\ \bibnamefont
  {Kolobov}},\ and\ \bibinfo {author} {\bibfnamefont {J.}~\bibnamefont
  {Steinhauer}},\ }\href {https://doi.org/10.1038/s41586-019-1241-0} {\bibfield
   {journal} {\bibinfo  {journal} {Nature}\ }\textbf {\bibinfo {volume}
  {569}},\ \bibinfo {pages} {688} (\bibinfo {year} {2019})}\BibitemShut
  {NoStop}%
\bibitem [{\citenamefont {Ribeiro}\ \emph {et~al.}(2022)\citenamefont
  {Ribeiro}, \citenamefont {Baak},\ and\ \citenamefont
  {Fischer}}]{Ribeiro2022}%
  \BibitemOpen
  \bibfield  {author} {\bibinfo {author} {\bibfnamefont {C.~C.~H.}\
  \bibnamefont {Ribeiro}}, \bibinfo {author} {\bibfnamefont {S.-S.}\
  \bibnamefont {Baak}},\ and\ \bibinfo {author} {\bibfnamefont {U.~R.}\
  \bibnamefont {Fischer}},\ }\bibfield  {journal} {\bibinfo  {journal}
  {Physical Review D}\ }\textbf {\bibinfo {volume} {105}},\ \href
  {https://doi.org/10.1103/physrevd.105.124066} {10.1103/physrevd.105.124066}
  (\bibinfo {year} {2022})\BibitemShut {NoStop}%
\bibitem [{\citenamefont {Jaskula}\ \emph {et~al.}(2012)\citenamefont
  {Jaskula}, \citenamefont {Partridge}, \citenamefont {Bonneau}, \citenamefont
  {Lopes}, \citenamefont {Ruaudel}, \citenamefont {Boiron},\ and\ \citenamefont
  {Westbrook}}]{Jaskula2012}%
  \BibitemOpen
  \bibfield  {author} {\bibinfo {author} {\bibfnamefont {J.-C.}\ \bibnamefont
  {Jaskula}}, \bibinfo {author} {\bibfnamefont {G.~B.}\ \bibnamefont
  {Partridge}}, \bibinfo {author} {\bibfnamefont {M.}~\bibnamefont {Bonneau}},
  \bibinfo {author} {\bibfnamefont {R.}~\bibnamefont {Lopes}}, \bibinfo
  {author} {\bibfnamefont {J.}~\bibnamefont {Ruaudel}}, \bibinfo {author}
  {\bibfnamefont {D.}~\bibnamefont {Boiron}},\ and\ \bibinfo {author}
  {\bibfnamefont {C.~I.}\ \bibnamefont {Westbrook}},\ }\href
  {https://doi.org/10.1103/PhysRevLett.109.220401} {\bibfield  {journal}
  {\bibinfo  {journal} {Phys. Rev. Lett.}\ }\textbf {\bibinfo {volume} {109}},\
  \bibinfo {pages} {220401} (\bibinfo {year} {2012})}\BibitemShut {NoStop}%
\bibitem [{\citenamefont {Rodriguez-Laguna}\ \emph {et~al.}(2017)\citenamefont
  {Rodriguez-Laguna}, \citenamefont {Tarruell}, \citenamefont {Lewenstein},\
  and\ \citenamefont {Celi}}]{Rodriguez-Laguna2017}%
  \BibitemOpen
  \bibfield  {author} {\bibinfo {author} {\bibfnamefont {J.}~\bibnamefont
  {Rodriguez-Laguna}}, \bibinfo {author} {\bibfnamefont {L.}~\bibnamefont
  {Tarruell}}, \bibinfo {author} {\bibfnamefont {M.}~\bibnamefont
  {Lewenstein}},\ and\ \bibinfo {author} {\bibfnamefont {A.}~\bibnamefont
  {Celi}},\ }\href {https://doi.org/10.1103/PhysRevA.95.013627} {\bibfield
  {journal} {\bibinfo  {journal} {Phys. Rev. A}\ }\textbf {\bibinfo {volume}
  {95}},\ \bibinfo {pages} {013627} (\bibinfo {year} {2017})}\BibitemShut
  {NoStop}%
\bibitem [{\citenamefont {Hu}\ \emph {et~al.}(2019)\citenamefont {Hu},
  \citenamefont {Feng}, \citenamefont {Zhang},\ and\ \citenamefont
  {Chin}}]{Hu2019}%
  \BibitemOpen
  \bibfield  {author} {\bibinfo {author} {\bibfnamefont {J.}~\bibnamefont
  {Hu}}, \bibinfo {author} {\bibfnamefont {L.}~\bibnamefont {Feng}}, \bibinfo
  {author} {\bibfnamefont {Z.}~\bibnamefont {Zhang}},\ and\ \bibinfo {author}
  {\bibfnamefont {C.}~\bibnamefont {Chin}},\ }\href
  {https://doi.org/10.1038/s41567-019-0537-1} {\bibfield  {journal} {\bibinfo
  {journal} {Nat. Phys.}\ }\textbf {\bibinfo {volume} {15}},\ \bibinfo {pages}
  {785} (\bibinfo {year} {2019})}\BibitemShut {NoStop}%
\bibitem [{\citenamefont {Gooding}\ \emph {et~al.}(2020)\citenamefont
  {Gooding}, \citenamefont {Biermann}, \citenamefont {Erne}, \citenamefont
  {Louko}, \citenamefont {Unruh}, \citenamefont {Schmiedmayer},\ and\
  \citenamefont {Weinfurtner}}]{Gooding2020}%
  \BibitemOpen
  \bibfield  {author} {\bibinfo {author} {\bibfnamefont {C.}~\bibnamefont
  {Gooding}}, \bibinfo {author} {\bibfnamefont {S.}~\bibnamefont {Biermann}},
  \bibinfo {author} {\bibfnamefont {S.}~\bibnamefont {Erne}}, \bibinfo {author}
  {\bibfnamefont {J.}~\bibnamefont {Louko}}, \bibinfo {author} {\bibfnamefont
  {W.~G.}\ \bibnamefont {Unruh}}, \bibinfo {author} {\bibfnamefont
  {J.}~\bibnamefont {Schmiedmayer}},\ and\ \bibinfo {author} {\bibfnamefont
  {S.}~\bibnamefont {Weinfurtner}},\ }\href
  {https://doi.org/10.1103/PhysRevLett.125.213603} {\bibfield  {journal}
  {\bibinfo  {journal} {Phys. Rev. Lett.}\ }\textbf {\bibinfo {volume} {125}},\
  \bibinfo {pages} {213603} (\bibinfo {year} {2020})}\BibitemShut {NoStop}%
\bibitem [{\citenamefont {Zenesini}\ \emph {et~al.}(2024)\citenamefont
  {Zenesini}, \citenamefont {Berti}, \citenamefont {Cominotti}, \citenamefont
  {Rogora}, \citenamefont {Moss}, \citenamefont {Billam}, \citenamefont
  {Carusotto}, \citenamefont {Lamporesi}, \citenamefont {Recati},\ and\
  \citenamefont {Ferrari}}]{Zenesini2024}%
  \BibitemOpen
  \bibfield  {author} {\bibinfo {author} {\bibfnamefont {A.}~\bibnamefont
  {Zenesini}}, \bibinfo {author} {\bibfnamefont {A.}~\bibnamefont {Berti}},
  \bibinfo {author} {\bibfnamefont {R.}~\bibnamefont {Cominotti}}, \bibinfo
  {author} {\bibfnamefont {C.}~\bibnamefont {Rogora}}, \bibinfo {author}
  {\bibfnamefont {I.~G.}\ \bibnamefont {Moss}}, \bibinfo {author}
  {\bibfnamefont {T.~P.}\ \bibnamefont {Billam}}, \bibinfo {author}
  {\bibfnamefont {I.}~\bibnamefont {Carusotto}}, \bibinfo {author}
  {\bibfnamefont {G.}~\bibnamefont {Lamporesi}}, \bibinfo {author}
  {\bibfnamefont {A.}~\bibnamefont {Recati}},\ and\ \bibinfo {author}
  {\bibfnamefont {G.}~\bibnamefont {Ferrari}},\ }\href
  {https://doi.org/10.1038/s41567-023-02345-4} {\bibfield  {journal} {\bibinfo
  {journal} {Nature Physics}\ }\textbf {\bibinfo {volume} {20}},\ \bibinfo
  {pages} {558–563} (\bibinfo {year} {2024})}\BibitemShut {NoStop}%
\bibitem [{\citenamefont {Jenkins}\ \emph {et~al.}(2024)\citenamefont
  {Jenkins}, \citenamefont {Braden}, \citenamefont {Peiris}, \citenamefont
  {Pontzen}, \citenamefont {Johnson},\ and\ \citenamefont
  {Weinfurtner}}]{Jenkins2024}%
  \BibitemOpen
  \bibfield  {author} {\bibinfo {author} {\bibfnamefont {A.~C.}\ \bibnamefont
  {Jenkins}}, \bibinfo {author} {\bibfnamefont {J.}~\bibnamefont {Braden}},
  \bibinfo {author} {\bibfnamefont {H.~V.}\ \bibnamefont {Peiris}}, \bibinfo
  {author} {\bibfnamefont {A.}~\bibnamefont {Pontzen}}, \bibinfo {author}
  {\bibfnamefont {M.~C.}\ \bibnamefont {Johnson}},\ and\ \bibinfo {author}
  {\bibfnamefont {S.}~\bibnamefont {Weinfurtner}},\ }\href
  {https://doi.org/10.1103/PhysRevD.109.023506} {\bibfield  {journal} {\bibinfo
   {journal} {Phys. Rev. D}\ }\textbf {\bibinfo {volume} {109}},\ \bibinfo
  {pages} {023506} (\bibinfo {year} {2024})}\BibitemShut {NoStop}%
\bibitem [{\citenamefont {Tolosa-Simeón}\ \emph {et~al.}(2023)\citenamefont
  {Tolosa-Simeón}, \citenamefont {Scherer},\ and\ \citenamefont
  {Floerchinger}}]{Tolosa2023}%
  \BibitemOpen
  \bibfield  {author} {\bibinfo {author} {\bibfnamefont {M.}~\bibnamefont
  {Tolosa-Simeón}}, \bibinfo {author} {\bibfnamefont {M.~M.}\ \bibnamefont
  {Scherer}},\ and\ \bibinfo {author} {\bibfnamefont {S.}~\bibnamefont
  {Floerchinger}},\ }\href@noop {} {\bibinfo {title} {Analog of cosmological
  particle production in moir\'e dirac materials}} (\bibinfo {year} {2023}),\
  \Eprint {https://arxiv.org/abs/2307.09299} {arXiv:2307.09299
  [cond-mat.mes-hall]} \BibitemShut {NoStop}%
\bibitem [{\citenamefont {Haller}\ \emph {et~al.}(2023)\citenamefont {Haller},
  \citenamefont {Hegde}, \citenamefont {Xu}, \citenamefont {De~Beule},
  \citenamefont {Schmidt},\ and\ \citenamefont {Meng}}]{HallerMeng2023}%
  \BibitemOpen
  \bibfield  {author} {\bibinfo {author} {\bibfnamefont {A.}~\bibnamefont
  {Haller}}, \bibinfo {author} {\bibfnamefont {S.}~\bibnamefont {Hegde}},
  \bibinfo {author} {\bibfnamefont {C.}~\bibnamefont {Xu}}, \bibinfo {author}
  {\bibfnamefont {C.}~\bibnamefont {De~Beule}}, \bibinfo {author}
  {\bibfnamefont {T.~L.}\ \bibnamefont {Schmidt}},\ and\ \bibinfo {author}
  {\bibfnamefont {T.}~\bibnamefont {Meng}},\ }\href
  {https://doi.org/10.21468/SciPostPhys.14.5.119} {\bibfield  {journal}
  {\bibinfo  {journal} {SciPost Physics}\ }\textbf {\bibinfo {volume} {14}},\
  \bibinfo {pages} {119} (\bibinfo {year} {2023})}\BibitemShut {NoStop}%
\bibitem [{\citenamefont {Falque}\ \emph {et~al.}(2023)\citenamefont {Falque},
  \citenamefont {Glorieux}, \citenamefont {Giacobino}, \citenamefont
  {Bramati},\ and\ \citenamefont {Jacquet}}]{Falque2023}%
  \BibitemOpen
  \bibfield  {author} {\bibinfo {author} {\bibfnamefont {K.}~\bibnamefont
  {Falque}}, \bibinfo {author} {\bibfnamefont {Q.}~\bibnamefont {Glorieux}},
  \bibinfo {author} {\bibfnamefont {E.}~\bibnamefont {Giacobino}}, \bibinfo
  {author} {\bibfnamefont {A.}~\bibnamefont {Bramati}},\ and\ \bibinfo {author}
  {\bibfnamefont {M.~J.}\ \bibnamefont {Jacquet}},\ }\href@noop {} {\bibinfo
  {title} {Spectroscopic measurement of the excitation spectrum on effectively
  curved spacetimes in a polaritonic fluid of light}} (\bibinfo {year}
  {2023}),\ \Eprint {https://arxiv.org/abs/2311.01392} {arXiv:2311.01392
  [cond-mat.quant-gas]} \BibitemShut {NoStop}%
\bibitem [{\citenamefont {Jacquet}\ \emph {et~al.}(2022)\citenamefont
  {Jacquet}, \citenamefont {Joly}, \citenamefont {Claude}, \citenamefont
  {Giacomelli}, \citenamefont {Glorieux}, \citenamefont {Bramati},
  \citenamefont {Carusotto},\ and\ \citenamefont {Giacobino}}]{Jacquet2022}%
  \BibitemOpen
  \bibfield  {author} {\bibinfo {author} {\bibfnamefont {M.}~\bibnamefont
  {Jacquet}}, \bibinfo {author} {\bibfnamefont {M.}~\bibnamefont {Joly}},
  \bibinfo {author} {\bibfnamefont {F.}~\bibnamefont {Claude}}, \bibinfo
  {author} {\bibfnamefont {L.}~\bibnamefont {Giacomelli}}, \bibinfo {author}
  {\bibfnamefont {Q.}~\bibnamefont {Glorieux}}, \bibinfo {author}
  {\bibfnamefont {A.}~\bibnamefont {Bramati}}, \bibinfo {author} {\bibfnamefont
  {I.}~\bibnamefont {Carusotto}},\ and\ \bibinfo {author} {\bibfnamefont
  {E.}~\bibnamefont {Giacobino}},\ }\bibfield  {journal} {\bibinfo  {journal}
  {The European Physical Journal D}\ }\textbf {\bibinfo {volume} {76}},\ \href
  {https://doi.org/10.1140/epjd/s10053-022-00477-5}
  {10.1140/epjd/s10053-022-00477-5} (\bibinfo {year} {2022})\BibitemShut
  {NoStop}%
\bibitem [{\citenamefont {Jacquet}\ \emph {et~al.}(2023)\citenamefont
  {Jacquet}, \citenamefont {Giacomelli}, \citenamefont {Valnais}, \citenamefont
  {Joly}, \citenamefont {Claude}, \citenamefont {Giacobino}, \citenamefont
  {Glorieux}, \citenamefont {Carusotto},\ and\ \citenamefont
  {Bramati}}]{Jacquet2023}%
  \BibitemOpen
  \bibfield  {author} {\bibinfo {author} {\bibfnamefont {M.}~\bibnamefont
  {Jacquet}}, \bibinfo {author} {\bibfnamefont {L.}~\bibnamefont {Giacomelli}},
  \bibinfo {author} {\bibfnamefont {Q.}~\bibnamefont {Valnais}}, \bibinfo
  {author} {\bibfnamefont {M.}~\bibnamefont {Joly}}, \bibinfo {author}
  {\bibfnamefont {F.}~\bibnamefont {Claude}}, \bibinfo {author} {\bibfnamefont
  {E.}~\bibnamefont {Giacobino}}, \bibinfo {author} {\bibfnamefont
  {Q.}~\bibnamefont {Glorieux}}, \bibinfo {author} {\bibfnamefont
  {I.}~\bibnamefont {Carusotto}},\ and\ \bibinfo {author} {\bibfnamefont
  {A.}~\bibnamefont {Bramati}},\ }\bibfield  {journal} {\bibinfo  {journal}
  {Physical Review Letters}\ }\textbf {\bibinfo {volume} {130}},\ \href
  {https://doi.org/10.1103/physrevlett.130.111501}
  {10.1103/physrevlett.130.111501} (\bibinfo {year} {2023})\BibitemShut
  {NoStop}%
\bibitem [{\citenamefont {Švančara}\ \emph {et~al.}(2024)\citenamefont
  {Švančara}, \citenamefont {Smaniotto}, \citenamefont {Solidoro},
  \citenamefont {MacDonald}, \citenamefont {Patrick}, \citenamefont {Gregory},
  \citenamefont {Barenghi},\ and\ \citenamefont {Weinfurtner}}]{Svancara2024}%
  \BibitemOpen
  \bibfield  {author} {\bibinfo {author} {\bibfnamefont {P.}~\bibnamefont
  {Švančara}}, \bibinfo {author} {\bibfnamefont {P.}~\bibnamefont
  {Smaniotto}}, \bibinfo {author} {\bibfnamefont {L.}~\bibnamefont {Solidoro}},
  \bibinfo {author} {\bibfnamefont {J.~F.}\ \bibnamefont {MacDonald}}, \bibinfo
  {author} {\bibfnamefont {S.}~\bibnamefont {Patrick}}, \bibinfo {author}
  {\bibfnamefont {R.}~\bibnamefont {Gregory}}, \bibinfo {author} {\bibfnamefont
  {C.~F.}\ \bibnamefont {Barenghi}},\ and\ \bibinfo {author} {\bibfnamefont
  {S.}~\bibnamefont {Weinfurtner}},\ }\href
  {https://doi.org/10.1038/s41586-024-07176-8} {\bibfield  {journal} {\bibinfo
  {journal} {Nature}\ }\textbf {\bibinfo {volume} {628}},\ \bibinfo {pages}
  {66–70} (\bibinfo {year} {2024})}\BibitemShut {NoStop}%
\bibitem [{\citenamefont {Giacomelli}\ and\ \citenamefont
  {Carusotto}(2021)}]{Giacomelli2021}%
  \BibitemOpen
  \bibfield  {author} {\bibinfo {author} {\bibfnamefont {L.}~\bibnamefont
  {Giacomelli}}\ and\ \bibinfo {author} {\bibfnamefont {I.}~\bibnamefont
  {Carusotto}},\ }\bibfield  {journal} {\bibinfo  {journal} {Physical Review
  A}\ }\textbf {\bibinfo {volume} {103}},\ \href
  {https://doi.org/10.1103/physreva.103.043309} {10.1103/physreva.103.043309}
  (\bibinfo {year} {2021})\BibitemShut {NoStop}%
\bibitem [{\citenamefont {Delhom}\ \emph {et~al.}(2024)\citenamefont {Delhom},
  \citenamefont {Guerrero}, \citenamefont {Calizaya}, \citenamefont {Falque},
  \citenamefont {Bramati}, \citenamefont {Brady}, \citenamefont {Jacquet},\
  and\ \citenamefont {Agullo}}]{Delhom2024}%
  \BibitemOpen
  \bibfield  {author} {\bibinfo {author} {\bibfnamefont {A.}~\bibnamefont
  {Delhom}}, \bibinfo {author} {\bibfnamefont {K.}~\bibnamefont {Guerrero}},
  \bibinfo {author} {\bibfnamefont {P.}~\bibnamefont {Calizaya}}, \bibinfo
  {author} {\bibfnamefont {K.}~\bibnamefont {Falque}}, \bibinfo {author}
  {\bibfnamefont {A.}~\bibnamefont {Bramati}}, \bibinfo {author} {\bibfnamefont
  {A.~J.}\ \bibnamefont {Brady}}, \bibinfo {author} {\bibfnamefont {M.~J.}\
  \bibnamefont {Jacquet}},\ and\ \bibinfo {author} {\bibfnamefont
  {I.}~\bibnamefont {Agullo}},\ }\href@noop {} {\bibinfo {title} {Entanglement
  from superradiance and rotating quantum fluids of light}} (\bibinfo {year}
  {2024}),\ \Eprint {https://arxiv.org/abs/2310.16031} {arXiv:2310.16031
  [gr-qc]} \BibitemShut {NoStop}%
\bibitem [{\citenamefont {Floerchinger}\ and\ \citenamefont
  {Wetterich}(2008)}]{Floerchinger2008}%
  \BibitemOpen
  \bibfield  {author} {\bibinfo {author} {\bibfnamefont {S.}~\bibnamefont
  {Floerchinger}}\ and\ \bibinfo {author} {\bibfnamefont {C.}~\bibnamefont
  {Wetterich}},\ }\href {https://doi.org/10.1103/PhysRevA.77.053603} {\bibfield
   {journal} {\bibinfo  {journal} {Phys. Rev. A}\ }\textbf {\bibinfo {volume}
  {77}},\ \bibinfo {pages} {053603} (\bibinfo {year} {2008})}\BibitemShut
  {NoStop}%
\bibitem [{\citenamefont {Pitaevskii}\ and\ \citenamefont
  {Stringari}(2016)}]{Pitaevskii2016}%
  \BibitemOpen
  \bibfield  {author} {\bibinfo {author} {\bibfnamefont {L.}~\bibnamefont
  {Pitaevskii}}\ and\ \bibinfo {author} {\bibfnamefont {S.}~\bibnamefont
  {Stringari}},\ }\href@noop {} {\emph {\bibinfo {title} {Bose-Einstein
  Condensation and Superfluidity}}}\ (\bibinfo  {publisher} {Oxford University
  PressOxford},\ \bibinfo {year} {2016})\BibitemShut {NoStop}%
\bibitem [{\citenamefont {Feshbach}(1958)}]{Feshbach1958}%
  \BibitemOpen
  \bibfield  {author} {\bibinfo {author} {\bibfnamefont {H.}~\bibnamefont
  {Feshbach}},\ }\href {https://doi.org/10.1016/0003-4916(58)90007-1}
  {\bibfield  {journal} {\bibinfo  {journal} {Annals of Physics}\ }\textbf
  {\bibinfo {volume} {5}},\ \bibinfo {pages} {357–390} (\bibinfo {year}
  {1958})}\BibitemShut {NoStop}%
\bibitem [{\citenamefont {Heyen}\ and\ \citenamefont
  {Floerchinger}(2020)}]{Heyen2020}%
  \BibitemOpen
  \bibfield  {author} {\bibinfo {author} {\bibfnamefont {L.~H.}\ \bibnamefont
  {Heyen}}\ and\ \bibinfo {author} {\bibfnamefont {S.}~\bibnamefont
  {Floerchinger}},\ }\href {https://doi.org/10.1103/PhysRevD.102.036024}
  {\bibfield  {journal} {\bibinfo  {journal} {Phys. Rev. D}\ }\textbf {\bibinfo
  {volume} {102}},\ \bibinfo {pages} {036024} (\bibinfo {year}
  {2020})}\BibitemShut {NoStop}%
\bibitem [{\citenamefont {Cohl}(2011)}]{Cohl2011}%
  \BibitemOpen
  \bibfield  {author} {\bibinfo {author} {\bibfnamefont {H.~S.}\ \bibnamefont
  {Cohl}},\ }\bibfield  {journal} {\bibinfo  {journal} {Symmetry, Integrability
  and Geometry: Methods and Applications}\ }\href
  {https://doi.org/10.3842/sigma.2011.108} {10.3842/sigma.2011.108} (\bibinfo
  {year} {2011})\BibitemShut {NoStop}%
\bibitem [{\citenamefont {Donnelly}(2002)}]{donnelly2002}%
  \BibitemOpen
  \bibfield  {author} {\bibinfo {author} {\bibfnamefont {H.}~\bibnamefont
  {Donnelly}},\ }in\ \href@noop {} {\emph {\bibinfo {booktitle} {Minimal
  Surfaces, Geometric Analysis and Symplectic Geometry}}},\ Vol.~\bibinfo
  {volume} {34}\ (\bibinfo  {publisher} {Mathematical Society of Japan},\
  \bibinfo {year} {2002})\ pp.\ \bibinfo {pages} {15--30}\BibitemShut {NoStop}%
\bibitem [{\citenamefont {Carroll}(2019)}]{Carroll2019}%
  \BibitemOpen
  \bibfield  {author} {\bibinfo {author} {\bibfnamefont {S.~M.}\ \bibnamefont
  {Carroll}},\ }\href {https://doi.org/10.1017/9781108770385} {\emph {\bibinfo
  {title} {Spacetime and Geometry}}}\ (\bibinfo  {publisher} {Cambridge
  University Press},\ \bibinfo {year} {2019})\BibitemShut {NoStop}%
\bibitem [{\citenamefont {Visser}(1999)}]{Visser1999}%
  \BibitemOpen
  \bibfield  {author} {\bibinfo {author} {\bibfnamefont {M.}~\bibnamefont
  {Visser}},\ }\href {https://doi.org/10.1103/physreva.59.427} {\bibfield
  {journal} {\bibinfo  {journal} {Physical Review A}\ }\textbf {\bibinfo
  {volume} {59}},\ \bibinfo {pages} {427–438} (\bibinfo {year}
  {1999})}\BibitemShut {NoStop}%
\bibitem [{\citenamefont {Dodelson}(2003)}]{Dodelson2003}%
  \BibitemOpen
  \bibfield  {author} {\bibinfo {author} {\bibfnamefont {S.}~\bibnamefont
  {Dodelson}},\ }\href@noop {} {\emph {\bibinfo {title} {{Modern Cosmology}}}}\
  (\bibinfo  {publisher} {Academic Press},\ \bibinfo {address} {Amsterdam},\
  \bibinfo {year} {2003})\BibitemShut {NoStop}%
\bibitem [{\citenamefont {Sakharov}(1966)}]{Sakharov1966}%
  \BibitemOpen
  \bibfield  {author} {\bibinfo {author} {\bibfnamefont {A.~D.}\ \bibnamefont
  {Sakharov}},\ }\href@noop {} {\bibfield  {journal} {\bibinfo  {journal} {Sov.
  Phys. JETP}\ }\textbf {\bibinfo {volume} {22}},\ \bibinfo {pages} {241}
  (\bibinfo {year} {1966})}\BibitemShut {NoStop}%
\bibitem [{\citenamefont {Hobson}\ \emph {et~al.}(2006)\citenamefont {Hobson},
  \citenamefont {Efstathiou},\ and\ \citenamefont {Lasenby}}]{Hobson2006}%
  \BibitemOpen
  \bibfield  {author} {\bibinfo {author} {\bibfnamefont {M.~P.}\ \bibnamefont
  {Hobson}}, \bibinfo {author} {\bibfnamefont {G.~P.}\ \bibnamefont
  {Efstathiou}},\ and\ \bibinfo {author} {\bibfnamefont {A.~N.}\ \bibnamefont
  {Lasenby}},\ }\href {https://doi.org/10.1017/cbo9780511790904} {\emph
  {\bibinfo {title} {General Relativity: An Introduction for Physicists}}}\
  (\bibinfo  {publisher} {Cambridge University Press},\ \bibinfo {year}
  {2006})\BibitemShut {NoStop}%
\bibitem [{\citenamefont {Weinberg}(2008)}]{Weinberg2008}%
  \BibitemOpen
  \bibfield  {author} {\bibinfo {author} {\bibfnamefont {S.}~\bibnamefont
  {Weinberg}},\ }\href
  {https://global.oup.com/academic/product/cosmology-9780198526827?cc=es&lang=en&}
  {\emph {\bibinfo {title} {{Cosmology}}}}\ (\bibinfo  {publisher} {Oxford
  University Press},\ \bibinfo {address} {Oxford},\ \bibinfo {year}
  {2008})\BibitemShut {NoStop}%
\bibitem [{\citenamefont {Chen}\ \emph {et~al.}(2014)\citenamefont {Chen},
  \citenamefont {Gibbons}, \citenamefont {Li},\ and\ \citenamefont
  {Yang}}]{Chen2014}%
  \BibitemOpen
  \bibfield  {author} {\bibinfo {author} {\bibfnamefont {S.}~\bibnamefont
  {Chen}}, \bibinfo {author} {\bibfnamefont {G.~W.}\ \bibnamefont {Gibbons}},
  \bibinfo {author} {\bibfnamefont {Y.}~\bibnamefont {Li}},\ and\ \bibinfo
  {author} {\bibfnamefont {Y.}~\bibnamefont {Yang}},\ }\href
  {https://doi.org/10.1088/1475-7516/2014/12/035} {\bibfield  {journal}
  {\bibinfo  {journal} {Journal of Cosmology and Astroparticle Physics}\
  }\textbf {\bibinfo {volume} {2014}}\bibinfo  {number} { (12)},\ \bibinfo
  {pages} {035}}\BibitemShut {NoStop}%
\bibitem [{\citenamefont {Grishchuk}\ and\ \citenamefont
  {Sidorov}(1990)}]{Grishchuk1990}%
  \BibitemOpen
\bibfield  {number} {  }\bibfield  {author} {\bibinfo {author} {\bibfnamefont
  {L.~P.}\ \bibnamefont {Grishchuk}}\ and\ \bibinfo {author} {\bibfnamefont
  {Y.~V.}\ \bibnamefont {Sidorov}},\ }\href
  {https://doi.org/10.1103/PhysRevD.42.3413} {\bibfield  {journal} {\bibinfo
  {journal} {Phys. Rev. D}\ }\textbf {\bibinfo {volume} {42}},\ \bibinfo
  {pages} {3413} (\bibinfo {year} {1990})}\BibitemShut {NoStop}%
\bibitem [{\citenamefont {Albrecht}\ \emph {et~al.}(1994)\citenamefont
  {Albrecht}, \citenamefont {Ferreira}, \citenamefont {Joyce},\ and\
  \citenamefont {Prokopec}}]{Albrecht1994}%
  \BibitemOpen
  \bibfield  {author} {\bibinfo {author} {\bibfnamefont {A.}~\bibnamefont
  {Albrecht}}, \bibinfo {author} {\bibfnamefont {P.}~\bibnamefont {Ferreira}},
  \bibinfo {author} {\bibfnamefont {M.}~\bibnamefont {Joyce}},\ and\ \bibinfo
  {author} {\bibfnamefont {T.}~\bibnamefont {Prokopec}},\ }\href
  {https://doi.org/10.1103/PhysRevD.50.4807} {\bibfield  {journal} {\bibinfo
  {journal} {Phys. Rev. D}\ }\textbf {\bibinfo {volume} {50}},\ \bibinfo
  {pages} {4807} (\bibinfo {year} {1994})}\BibitemShut {NoStop}%
\bibitem [{\citenamefont {Martin}\ \emph {et~al.}(2022)\citenamefont {Martin},
  \citenamefont {Micheli},\ and\ \citenamefont {Vennin}}]{Martin2022}%
  \BibitemOpen
  \bibfield  {author} {\bibinfo {author} {\bibfnamefont {J.}~\bibnamefont
  {Martin}}, \bibinfo {author} {\bibfnamefont {A.}~\bibnamefont {Micheli}},\
  and\ \bibinfo {author} {\bibfnamefont {V.}~\bibnamefont {Vennin}},\ }\href
  {https://doi.org/10.1088/1475-7516/2022/04/051} {\bibfield  {journal}
  {\bibinfo  {journal} {Journal of Cosmology and Astroparticle Physics}\
  }\textbf {\bibinfo {volume} {2022}}\bibinfo  {number} { (04)},\ \bibinfo
  {pages} {051}}\BibitemShut {NoStop}%
\bibitem [{\citenamefont {Brady}\ \emph {et~al.}(2022)\citenamefont {Brady},
  \citenamefont {Agullo},\ and\ \citenamefont {Kranas}}]{Brady2022}%
  \BibitemOpen
\bibfield  {number} {  }\bibfield  {author} {\bibinfo {author} {\bibfnamefont
  {A.~J.}\ \bibnamefont {Brady}}, \bibinfo {author} {\bibfnamefont
  {I.}~\bibnamefont {Agullo}},\ and\ \bibinfo {author} {\bibfnamefont
  {D.}~\bibnamefont {Kranas}},\ }\href
  {https://doi.org/10.1103/PhysRevD.106.105021} {\bibfield  {journal} {\bibinfo
   {journal} {Phys. Rev. D}\ }\textbf {\bibinfo {volume} {106}},\ \bibinfo
  {pages} {105021} (\bibinfo {year} {2022})}\BibitemShut {NoStop}%
\bibitem [{\citenamefont {G.Grosso}\ and\ \citenamefont
  {G.P.Parravicini}(2000)}]{Grosso2000}%
  \BibitemOpen
  \bibfield  {author} {\bibinfo {author} {\bibnamefont {G.Grosso}}\ and\
  \bibinfo {author} {\bibnamefont {G.P.Parravicini}},\ }\bibinfo {title}
  {Electrons in one-dimensional periodic potentials},\ in\ \href
  {https://doi.org/10.1016/b978-012304460-0/50001-3} {\emph {\bibinfo
  {booktitle} {Solid State Physics}}}\ (\bibinfo  {publisher} {Elsevier},\
  \bibinfo {year} {2000})\ p.\ \bibinfo {pages} {1–36}\BibitemShut {NoStop}%
\bibitem [{\citenamefont {Markos}\ and\ \citenamefont
  {Soukoulis}(2008)}]{Markos2008}%
  \BibitemOpen
  \bibfield  {author} {\bibinfo {author} {\bibfnamefont {P.}~\bibnamefont
  {Markos}}\ and\ \bibinfo {author} {\bibfnamefont {C.~M.}\ \bibnamefont
  {Soukoulis}},\ }\href {https://doi.org/10.1515/9781400835676} {\emph
  {\bibinfo {title} {Wave Propagation: From Electrons to Photonic Crystals and
  Left-Handed Materials}}}\ (\bibinfo  {publisher} {Princeton University
  Press},\ \bibinfo {year} {2008})\BibitemShut {NoStop}%
\bibitem [{\citenamefont {Griffiths}\ and\ \citenamefont
  {Taussig}(1992)}]{Griffiths1992}%
  \BibitemOpen
  \bibfield  {author} {\bibinfo {author} {\bibfnamefont {D.~J.}\ \bibnamefont
  {Griffiths}}\ and\ \bibinfo {author} {\bibfnamefont {N.~F.}\ \bibnamefont
  {Taussig}},\ }\href {https://doi.org/10.1119/1.17008} {\bibfield  {journal}
  {\bibinfo  {journal} {American Journal of Physics}\ }\textbf {\bibinfo
  {volume} {60}},\ \bibinfo {pages} {883–888} (\bibinfo {year}
  {1992})}\BibitemShut {NoStop}%
\bibitem [{\citenamefont {Kronig}\ \emph {et~al.}(1931)\citenamefont {Kronig},
  \citenamefont {Penney},\ and\ \citenamefont {Fowler}}]{KronigPenney1931}%
  \BibitemOpen
  \bibfield  {author} {\bibinfo {author} {\bibfnamefont {R.~D.~L.}\
  \bibnamefont {Kronig}}, \bibinfo {author} {\bibfnamefont {W.~G.}\
  \bibnamefont {Penney}},\ and\ \bibinfo {author} {\bibfnamefont {R.~H.}\
  \bibnamefont {Fowler}},\ }\href@noop {} {\bibfield  {journal} {\bibinfo
  {journal} {Proceedings of the Royal Society of London. Series A, Containing
  Papers of a Mathematical and Physical Character}\ }\textbf {\bibinfo {volume}
  {130}},\ \bibinfo {pages} {499} (\bibinfo {year} {1931})}\BibitemShut
  {NoStop}%
\bibitem [{\citenamefont {Kalotas}\ and\ \citenamefont
  {Lee}(1991)}]{Kalotas1991}%
  \BibitemOpen
  \bibfield  {author} {\bibinfo {author} {\bibfnamefont {T.~M.}\ \bibnamefont
  {Kalotas}}\ and\ \bibinfo {author} {\bibfnamefont {A.~R.}\ \bibnamefont
  {Lee}},\ }\href {https://doi.org/10.1088/0143-0807/12/6/006} {\bibfield
  {journal} {\bibinfo  {journal} {European Journal of Physics}\ }\textbf
  {\bibinfo {volume} {12}},\ \bibinfo {pages} {275–282} (\bibinfo {year}
  {1991})}\BibitemShut {NoStop}%
\bibitem [{\citenamefont {Senn}(1988)}]{Senn1988}%
  \BibitemOpen
  \bibfield  {author} {\bibinfo {author} {\bibfnamefont {P.}~\bibnamefont
  {Senn}},\ }\href {https://doi.org/10.1119/1.15359} {\bibfield  {journal}
  {\bibinfo  {journal} {American Journal of Physics}\ }\textbf {\bibinfo
  {volume} {56}},\ \bibinfo {pages} {916} (\bibinfo {year} {1988})}\BibitemShut
  {NoStop}%
\bibitem [{\citenamefont {Wigner}(1948)}]{Wigner1948}%
  \BibitemOpen
  \bibfield  {author} {\bibinfo {author} {\bibfnamefont {E.~P.}\ \bibnamefont
  {Wigner}},\ }\href {https://doi.org/10.1103/PhysRev.73.1002} {\bibfield
  {journal} {\bibinfo  {journal} {Phys. Rev.}\ }\textbf {\bibinfo {volume}
  {73}},\ \bibinfo {pages} {1002} (\bibinfo {year} {1948})}\BibitemShut
  {NoStop}%
\bibitem [{\citenamefont {Weinberg}(2003)}]{Weinberg2003}%
  \BibitemOpen
  \bibfield  {author} {\bibinfo {author} {\bibfnamefont {S.}~\bibnamefont
  {Weinberg}},\ }\href {https://doi.org/10.1103/PhysRevD.67.123504} {\bibfield
  {journal} {\bibinfo  {journal} {Phys. Rev. D}\ }\textbf {\bibinfo {volume}
  {67}},\ \bibinfo {pages} {123504} (\bibinfo {year} {2003})}\BibitemShut
  {NoStop}%
\bibitem [{\citenamefont {Mukhanov}\ \emph {et~al.}(1992)\citenamefont
  {Mukhanov}, \citenamefont {Feldman},\ and\ \citenamefont
  {Brandenberger}}]{MukhanovFeldmanBrandenberger1992}%
  \BibitemOpen
  \bibfield  {author} {\bibinfo {author} {\bibfnamefont {V.}~\bibnamefont
  {Mukhanov}}, \bibinfo {author} {\bibfnamefont {H.}~\bibnamefont {Feldman}},\
  and\ \bibinfo {author} {\bibfnamefont {R.}~\bibnamefont {Brandenberger}},\
  }\href@noop {} {\bibfield  {journal} {\bibinfo  {journal} {Physics Reports}\
  }\textbf {\bibinfo {volume} {215}},\ \bibinfo {pages} {203} (\bibinfo {year}
  {1992})}\BibitemShut {NoStop}%
\bibitem [{\citenamefont {Lyth}(1985)}]{Lyth1984}%
  \BibitemOpen
  \bibfield  {author} {\bibinfo {author} {\bibfnamefont {D.~H.}\ \bibnamefont
  {Lyth}},\ }\href {https://doi.org/10.1103/PhysRevD.31.1792} {\bibfield
  {journal} {\bibinfo  {journal} {Phys. Rev. D}\ }\textbf {\bibinfo {volume}
  {31}},\ \bibinfo {pages} {1792} (\bibinfo {year} {1985})}\BibitemShut
  {NoStop}%
\bibitem [{\citenamefont {Martin}(2005)}]{Martin2005}%
  \BibitemOpen
  \bibfield  {author} {\bibinfo {author} {\bibfnamefont {J.}~\bibnamefont
  {Martin}},\ }\bibinfo {title} {Inflationary cosmological perturbations of
  quantum-mechanical origin},\ in\ \href {https://doi.org/10.1007/11377306_7}
  {\emph {\bibinfo {booktitle} {Lecture Notes in Physics}}}\ (\bibinfo
  {publisher} {Springer-Verlag},\ \bibinfo {year} {2005})\ p.\ \bibinfo {pages}
  {199–244}\BibitemShut {NoStop}%
\bibitem [{\citenamefont {Wellner}(1964)}]{Wellner1964}%
  \BibitemOpen
  \bibfield  {author} {\bibinfo {author} {\bibfnamefont {M.}~\bibnamefont
  {Wellner}},\ }\href {https://doi.org/10.1119/1.1969857} {\bibfield  {journal}
  {\bibinfo  {journal} {American Journal of Physics}\ }\textbf {\bibinfo
  {volume} {32}},\ \bibinfo {pages} {787–789} (\bibinfo {year}
  {1964})}\BibitemShut {NoStop}%
\bibitem [{\citenamefont {Boya}\ and\ \citenamefont
  {Casahorr\'{a}n}(2007)}]{Boya2007}%
  \BibitemOpen
  \bibfield  {author} {\bibinfo {author} {\bibfnamefont {L.~J.}\ \bibnamefont
  {Boya}}\ and\ \bibinfo {author} {\bibfnamefont {J.}~\bibnamefont
  {Casahorr\'{a}n}},\ }\href {https://doi.org/10.1007/s10773-006-9321-y}
  {\bibfield  {journal} {\bibinfo  {journal} {Internat. J. Theoret. Phys.}\
  }\textbf {\bibinfo {volume} {46}},\ \bibinfo {pages} {1998} (\bibinfo {year}
  {2007})}\BibitemShut {NoStop}%
\bibitem [{\citenamefont {Glenz}\ and\ \citenamefont
  {Parker}(2009)}]{Glenz2009}%
  \BibitemOpen
  \bibfield  {author} {\bibinfo {author} {\bibfnamefont {M.~M.}\ \bibnamefont
  {Glenz}}\ and\ \bibinfo {author} {\bibfnamefont {L.}~\bibnamefont {Parker}},\
  }\bibfield  {journal} {\bibinfo  {journal} {Physical Review D}\ }\textbf
  {\bibinfo {volume} {80}},\ \href {https://doi.org/10.1103/physrevd.80.063534}
  {10.1103/physrevd.80.063534} (\bibinfo {year} {2009})\BibitemShut {NoStop}%
\bibitem [{\citenamefont {\'Alvarez-Dom\'{\i}nguez}\ \emph
  {et~al.}(2023)\citenamefont {\'Alvarez-Dom\'{\i}nguez}, \citenamefont
  {Cembranos}, \citenamefont {Garay}, \citenamefont {Mart\'{\i}n-Benito},
  \citenamefont {Parra-L\'opez},\ and\ \citenamefont
  {Vel\'azquez}}]{Dominguez2023}%
  \BibitemOpen
  \bibfield  {author} {\bibinfo {author} {\bibfnamefont {A.}~\bibnamefont
  {\'Alvarez-Dom\'{\i}nguez}}, \bibinfo {author} {\bibfnamefont {J.~A.~R.}\
  \bibnamefont {Cembranos}}, \bibinfo {author} {\bibfnamefont {L.~J.}\
  \bibnamefont {Garay}}, \bibinfo {author} {\bibfnamefont {M.}~\bibnamefont
  {Mart\'{\i}n-Benito}}, \bibinfo {author} {\bibfnamefont {A.}~\bibnamefont
  {Parra-L\'opez}},\ and\ \bibinfo {author} {\bibfnamefont {J.~M.~S.}\
  \bibnamefont {Vel\'azquez}},\ }\href
  {https://doi.org/10.1103/PhysRevD.108.065008} {\bibfield  {journal} {\bibinfo
   {journal} {Phys. Rev. D}\ }\textbf {\bibinfo {volume} {108}},\ \bibinfo
  {pages} {065008} (\bibinfo {year} {2023})}\BibitemShut {NoStop}%
\bibitem [{\citenamefont {Gel'fand}\ and\ \citenamefont
  {Levitan}(1955)}]{GelfandLevitan1955}%
  \BibitemOpen
  \bibfield  {author} {\bibinfo {author} {\bibfnamefont {I.~M.}\ \bibnamefont
  {Gel'fand}}\ and\ \bibinfo {author} {\bibfnamefont {B.~M.}\ \bibnamefont
  {Levitan}}\ }(\bibinfo {year} {1955})\BibitemShut {NoStop}%
\bibitem [{\citenamefont {Marchenko}(1955)}]{Marchenko1955}%
  \BibitemOpen
  \bibfield  {author} {\bibinfo {author} {\bibfnamefont {V.}~\bibnamefont
  {Marchenko}},\ }\href@noop {} {\bibfield  {journal} {\bibinfo  {journal}
  {Dokl. Akad. Nauk SSSR}\ }\textbf {\bibinfo {volume} {104}},\ \bibinfo
  {pages} {695 } (\bibinfo {year} {1955})}\BibitemShut {NoStop}%
\bibitem [{\citenamefont {Chadan}\ \emph {et~al.}(2013)\citenamefont {Chadan},
  \citenamefont {Newton},\ and\ \citenamefont {Sabatier}}]{Chadan2013}%
  \BibitemOpen
  \bibfield  {author} {\bibinfo {author} {\bibfnamefont {K.}~\bibnamefont
  {Chadan}}, \bibinfo {author} {\bibfnamefont {R.}~\bibnamefont {Newton}},\
  and\ \bibinfo {author} {\bibfnamefont {P.}~\bibnamefont {Sabatier}},\ }\href
  {https://books.google.de/books?id=RMb3CAAAQBAJ} {\emph {\bibinfo {title}
  {Inverse problems in quantum scattering theory}}},\ Theoretical and
  mathematical physics\ (\bibinfo  {publisher} {Springer Berlin Heidelberg},\
  \bibinfo {year} {2013})\BibitemShut {NoStop}%
\bibitem [{\citenamefont {Bogolyubov}(1947)}]{Bogolyubov:1947zz}%
  \BibitemOpen
  \bibfield  {author} {\bibinfo {author} {\bibfnamefont {N.~N.}\ \bibnamefont
  {Bogolyubov}},\ }\href@noop {} {\bibfield  {journal} {\bibinfo  {journal} {J.
  Phys. (USSR)}\ }\textbf {\bibinfo {volume} {11}},\ \bibinfo {pages} {23}
  (\bibinfo {year} {1947})}\BibitemShut {NoStop}%
\bibitem [{\citenamefont {Micheli}\ and\ \citenamefont
  {Robertson}(2022)}]{MicheliRobertson2022}%
  \BibitemOpen
  \bibfield  {author} {\bibinfo {author} {\bibfnamefont {A.}~\bibnamefont
  {Micheli}}\ and\ \bibinfo {author} {\bibfnamefont {S.}~\bibnamefont
  {Robertson}},\ }\href {https://doi.org/10.1103/PhysRevB.106.214528}
  {\bibfield  {journal} {\bibinfo  {journal} {Phys. Rev. B}\ }\textbf {\bibinfo
  {volume} {106}},\ \bibinfo {pages} {214528} (\bibinfo {year} {2022})},\
  \Eprint {https://arxiv.org/abs/2205.15826} {arXiv:2205.15826
  [cond-mat.quant-gas]} \BibitemShut {NoStop}%
\bibitem [{\citenamefont {Wands}(1999)}]{Wands1998}%
  \BibitemOpen
  \bibfield  {author} {\bibinfo {author} {\bibfnamefont {D.}~\bibnamefont
  {Wands}},\ }\href {https://doi.org/10.1103/PhysRevD.60.023507} {\bibfield
  {journal} {\bibinfo  {journal} {Phys. Rev. D}\ }\textbf {\bibinfo {volume}
  {60}},\ \bibinfo {pages} {023507} (\bibinfo {year} {1999})}\BibitemShut
  {NoStop}%
\bibitem [{\citenamefont {Finelli}\ and\ \citenamefont
  {Brandenberger}(2002)}]{FinelliBrandenberger2002}%
  \BibitemOpen
  \bibfield  {author} {\bibinfo {author} {\bibfnamefont {F.}~\bibnamefont
  {Finelli}}\ and\ \bibinfo {author} {\bibfnamefont {R.}~\bibnamefont
  {Brandenberger}},\ }\href {https://doi.org/10.1103/PhysRevD.65.103522}
  {\bibfield  {journal} {\bibinfo  {journal} {Phys. Rev. D}\ }\textbf {\bibinfo
  {volume} {65}},\ \bibinfo {pages} {103522} (\bibinfo {year}
  {2002})}\BibitemShut {NoStop}%
\bibitem [{\citenamefont {Shaikh}\ \emph {et~al.}(2022)\citenamefont {Shaikh},
  \citenamefont {Tarai}, \citenamefont {Tripathy},\ and\ \citenamefont
  {Mishra}}]{Shaikh2022}%
  \BibitemOpen
  \bibfield  {author} {\bibinfo {author} {\bibfnamefont {A.~Y.}\ \bibnamefont
  {Shaikh}}, \bibinfo {author} {\bibfnamefont {S.}~\bibnamefont {Tarai}},
  \bibinfo {author} {\bibfnamefont {S.~K.}\ \bibnamefont {Tripathy}},\ and\
  \bibinfo {author} {\bibfnamefont {B.}~\bibnamefont {Mishra}},\ }\bibfield
  {journal} {\bibinfo  {journal} {International Journal of Geometric Methods in
  Modern Physics}\ }\textbf {\bibinfo {volume} {19}},\ \href
  {https://doi.org/10.1142/s0219887822501936} {10.1142/s0219887822501936}
  (\bibinfo {year} {2022})\BibitemShut {NoStop}%
\end{thebibliography}%
\bibliographystyle{apsrev4-2} 

\end{document}